\documentclass{aa}
\usepackage[varg]{txfonts}
\usepackage{graphicx}
\usepackage{natbib}
\usepackage{appendix}
\usepackage{listings}
\usepackage{amsmath}
\usepackage{mathtools}
\usepackage{enumitem} 
\usepackage{hyperref}
\usepackage{pdflscape}
\usepackage{caption}
\usepackage{rotating}
\usepackage{float}
\usepackage{subfig}
\usepackage{stfloats}
\usepackage{chngcntr}

\usepackage{listings}
\usepackage{xcolor}

\hypersetup{
     colorlinks   = true,
     allcolors    = blue
}
\begin{document}

\counterwithin*{figure}{section}

\titlerunning{The $^{12}$CO gas structures of protoplanetary disks in the Upper Scorpius region}
\title{The $^{12}$CO gas structures of protoplanetary disks in the Upper Scorpius region
  } 

\author{Luigi Zallio\inst{1,2}
  \and Giovanni P. Rosotti\inst{1} 
  \and Miguel Vioque\inst{2}
  \and Anna Miotello\inst{2}
  \and Sean M. Andrews\inst{3}
  \and Carlo F. Manara\inst{2}
  \and John M. Carpenter\inst{4}
  \and Aaron Empey\inst{5}
  \and Nicolás T. Kurtovic\inst{6,7}
  \and Charles J. Law\inst{8}\thanks{NASA Hubble Fellowship Program Sagan Fellow}
  \and Cristiano Longarini\inst{9}
  \and Teresa Paneque-Carreño\inst{10}
  \and Richard Teague\inst{11}
  \and Marion Villenave\inst{12}
  \and Hsi-Wei Yen\inst{13}
  \and Francesco Zagaria\inst{7}
  }

\offprints{L. Zallio, \email{\href{mailto:luigi.zallio@unimi.it}{luigi.zallio@unimi.it}}}

\institute{Dipartimento di Fisica ‘Aldo Pontremoli’, Università degli Studi di Milano, via G. Celoria 16, I-20133 Milano, Italy.
  \and European Southern Observatory, Karl-Schwarzschild-Strasse 2, D-85748 Garching bei München, Germany.
  \and Center for Astrophysics \textbar\ Harvard \& Smithsonian, 60 Garden Street, Cambridge, MA 02138, USA
  \and Joint ALMA Observatory, Avenida Alonso de Córdova 3107, Vitacura, Santiago, Chile
  \and University College Dublin (UCD), Department of Physics, Belfield, Dublin 4, Ireland
  \and Max Planck Institute for Extraterrestrial Physics, Giessenbachstrasse 1, D-85748 Garching, Germany
  \and Max-Planck-Institut fur Astronomie (MPIA), Konigstuhl 17, 69117 Heidelberg, Germany
  \and Department of Astronomy, University of Virginia, Charlottesville, VA 22904, USA
  \and Institute of Astronomy, University of Cambridge, Madingley Road, Cambridge, CB3 0HA, UK 
  \and Department of Astronomy, University of Michigan, 1085 South University Avenue, Ann Arbor, MI 48109, USA
  \and Department of Earth, Atmospheric, and Planetary Sciences, Massachusetts Institute of Technology, Cambridge, MA 02139, USA
  \and Univ. Grenoble Alpes, CNRS, IPAG, F-38000 Grenoble, France
  \and Academia Sinica Institute of Astronomy and Astrophysics, 11F of Astronomy-Mathematics Building, AS/NTU, No.1, Sec 4, Roosevelt Rd, Taipei 106216, Taiwan}
  
\date{Received ... / Accepted ...}

\abstract{We present measurements of key protoplanetary disk properties inferred from parametric models of ALMA $^{12}$CO spectral line visibilities. We derived gas-disk radii, integrated fluxes, optically thick emission layers, and brightness temperature profiles for the disk population of the old ($4 - 14$ Myr) Upper Scorpius star-forming region. We measured CO emission sizes for 37 disks with bright CO $J=3-2$ emission (S/N > 10 on the integrated flux; out of the 83 disks with CO detections), finding that the median radius containing 90\% of the flux is $\sim82$ au, with radii spanning from $22$ up to $247$ au.
We report a correlation between the $^{12}$CO brightness temperatures and stellar luminosities, with a Pearson coefficient of 0.6, which we used to prove that the $^{12}$CO optically thick emission layer primarily emanates from a region below the superheated dust, which is optically thin to the stellar irradiation. Moreover, we derive 33 CO emission-surface height profiles, finding a median aspect ratio of $\langle z/r \rangle \sim 0.16$ in a range from $\sim0.01$ up to $\sim0.45$ over the sample. Finally, we comment on the multiple systems in our sample, of which only some were already known. 
These results confirm that it is possible to derive bulk disk properties by modeling moderate-angular-resolution ALMA visibilities.}

\keywords{protoplanetary disks, protoplanetary disk populations, protoplanetary gas-disk radii, disk vertical structure, line brightness temperature.}
\maketitle

\section{Introduction}
\label{sec:introduction}

Thousands of exoplanets have been discovered in the last 20 years\footnote{\url{https://exoplanetarchive.ipac.caltech.edu/}}, but many fundamental questions remain about how they were born. Our picture of planet formation is incomplete, mainly because we still do not understand how the protoplanetary disks that orbit young stars evolve and eventually form planets (\citealt{Morbidelli_2016}). Surveys with the Atacama Large Millimeter/submillimeter Array (ALMA) have greatly improved our knowledge regarding protoplanetary disk evolution, enabling population studies to investigate how gas disks evolve in different star-forming regions (SFRs; e.g., \citealt{Ansdell_2018}, \citealt{Sanchis2021}, \citealt{Long2022}, and \citealt{Zhang_2025}).

The typical observables for gas-disk evolution studies are line fluxes and radii. In particular, the latter can be used to constrain the dominant mechanisms controlling the evolution of protoplanetary disks (for a review, see \citealt{Manara_2023}). For example, \cite{Tabone_2025} recently used population-synthesis models to show that magnetohydrodynamic (MHD) wind-driven evolution best reproduces the typical masses and sizes of protoplanetary disks. The biggest challenge for measuring precise bulk gas-disk properties (such as disk radii and line-emission surfaces) for large samples is that most surveys are observed with moderate angular resolution and modest signal-to-noise ratios (S/N); this is a necessary trade-off to observe a large number of disks. For this reason, ALMA shallow surveys are often used as precursors for dedicated high-resolution observations of particularly bright and extended disks (see, e.g., \citealt{Oberg_2021} and \citealt{Teague_2025}). 

One possibility for deriving bulk disk properties from moderate-angular-resolution data is offered by modeling spectral-line visibilities. So far, this technique has functioned via radiative-transfer codes (e.g., \citealt{Guilloteau_1998}, \citealt{Simon_2000}, \citealt{Dartois_2003}, \citealt{Pietu_2007}, \citealt{Rosenfeld_2012}, \citealt{Flaherty_2015}) or parametric emission models (\citealt{Kurtovic2024}) on a handful of resolved protoplanetary disks, but it has never been applied on a large survey: in this work, we filled this gap.

We implemented two new simple gas-disk models and used the software \verb|csalt|\footnote{\url{https://github.com/seanandrews/csalt}} (Andrews et al., in prep.) to infer important quantities of protoplanetary disks, such as radius, line-emission layer, and integrated flux. We conducted our analysis on the $^{12}$CO $J=3-2$ data presented by \cite{Carpenter2025} for the Upper Scorpius old SFR ($4 - 14$ Myr, \citealt{Ratzenbock_2023a}, \citealt{Ratzenbock_2023b}).

This paper is structured as follows. In Sect. \ref{sec:sample}, we present the sample over which we conducted our analysis. In Sect. \ref{sec:analysis}, we show the parametric models adopted in this work and discuss the details of our modeling technique. In Sect. \ref{sec:results}, we show an overview of our results and provide a comparison with literature values. In Sect. \ref{sec:discussion}, we discuss our results and highlight confirmed and candidate multiple systems identified in our sample. Finally, in Sect. \ref{sec:conclusions} we draw our conclusions. 

In this paper, we do not comment on the stellar masses we derived, even though we report their values for completeness. This specific topic will be covered by \cite{2026A&A...708L...1Z}.

\vspace{-0.3cm}
\section{Sample selection}
\label{sec:sample}
The sample was assembled from the work of \cite{Carpenter2025}, which presented ALMA continuum and $^{12}$CO observations (project codes 2011.0.00526.S, 2012.1.00688.S, 2013.1.00395.S, and 2018.1.00564.S) for a total of 284 Class II objects. The observations were taken with the 12-m array using between 43 and 50 antennas, providing an angular resolution from $0.1^{\prime\prime}-0.3^{\prime\prime}$, with a typical on-source integration time of 2.5 minutes. Among this sample, \cite{Carpenter2025} identified 202 sources associated with the Upper Scorpius SFR. The observations were obtained in Band 7, with spectral windows centered at 334.2, 336.1, 346.2, and 348.1 GHz. The window centered at 346.2 GHz included the $^{12}$CO $J=3 - 2$ transition, with a channel width of 0.488 MHz (0.429 kms$^{-1}$ for $^{12}$CO $J=3 - 2$).
The data were calibrated with the ALMA calibration pipeline. Among the 202 sources, \cite{Carpenter2025} identified 83 $^{12}$CO $J=3 - 2$ detections with S/N $\geq3$; these are the sources we analyzed.

In addition to the ALMA data of protoplanetary disks, we used the Very Large Telescope / X-Shooter spectra presented in \cite{Manara_2020} (project codes 097.C-0378, 0101.C-0866) and new data that will be published in Empey et al., in prep. (project codes 105.2082.003, 113.26NN.001, 113.26NN.003, and 115.27XL.001) to derive the stellar properties of the disks that overlap with our ALMA sample.

\vspace{-0.3cm}
\section{Analysis}
\label{sec:analysis}
In this section, we present our analysis technique to constrain protoplanetary-disk parameters. Choosing a proper parametric model, it is possible to reproduce the channel maps (or their integrated maps) with high fidelity, and to interpret the best-fit parameters that maximize the specified likelihood (\citealt{Teague2018}, \citealt{Pinte2018}, \citealt{Casassus2019}, \citealt{Izquierdo2021}). 

We performed our analysis using the software \verb|csalt| (Andrews et al., in prep.), which allows robust inference on model parameters using visibility fitting. In this work, we present the first large-scale application of \verb|csalt| to fit and extract information from moderate-angular-resolution ALMA observations. For each model calculation, we generated a set of model visibility spectra in the kinematic Local Standard of Rest (LSRK) frequencies at each timestamp, convolved it with the spectral response function (SRF) of the ALMA correlator, and then calculated a likelihood that explicitly treats the spectral covariance following \cite{Loomis_2018}. Finally, for user-specified priors, we sampled the posterior space using \verb|emcee| (\citealt{Foreman_2013}).

\vspace{-0.3cm}
\subsection{Model description}
\label{sec:model}
In our sample, we identified two main groups of disks we could fit using \verb|csalt|: "resolved" and "marginally resolved". In the next sections, we present two different parametric prescriptions used to fit these two groups.

\vspace{-0.3cm}
\subsubsection{Parametric model A - detailed disk description}
\label{sec:model}
The first parametric model used in this work was designed to be simple and to mimic a zeroth-order radiative-transfer model. In particular, this parametric model is conceptually similar to the one implemented in \verb|DISCMINER| (\citealt{Izquierdo2021}), but it uses a different parametrization for the line width. With this prescription, we fit the 33 "resolved" disks of our sample. We divided the model parameters into the four groups listed below.
\begin{enumerate}
    \item Geometrical parameters. We fit for the protoplanetary disk inclination, PA, and offsets dRA and dDEC. The PA notation was used to represent the angle to the redshifted major axis of the emission at the midplane when moving E of N in the plane of the sky.
    \item Vertical height. We assumed that all the disk emission comes from two thin vertical layers, with the back surface beyond the front one from the perspective of the observer. We fit the height of the layer (vertical height) through an exponentially tapered power law. For simplicity, as the sources of our sample are observed at moderate-to-low angular resolution, and therefore we are not able to distinguish between the front and back surface layers, we set the back emission surface $z_b$ equal to the front surface $z_f$: $z_b = -z_f$.\footnote{This choice was adopted as the optimal compromise after testing different parametric prescriptions where the back surface was fixed at the midplane or completely removed.}
    The vertical height equation is 
    \begin{equation}
        z(r) = z_{1} \> \biggl(\frac{r}{1''}\biggr)^{\psi_z} \exp\biggl(-\biggl(\frac{r}{r_{z}} \biggr)^{\phi_z} \biggr).
        \label{eq:z(r)}
    \end{equation}
    \item Velocity. The model takes into account pure Keplerian rotation $v_{kep}$, which is projected into the observer frame. Then, the systemic velocity $v_{sys}$ is added to obtain the velocity on the plane of the sky $v_{sky}$.
    The equations are
    \begin{equation}
        v_{kep} (r,z) = \sqrt{\frac{GM_* r}{(r^2+z^2)^{3/2}}}, 
    \end{equation}
    \begin{equation}
        v_{sky} (r,z) = v_{kep} \cos(\theta) \sin\bigl(|\text{inc}|\bigr) + v_{sys},
    \end{equation}
    where $G$ is the gravitational constant, $M_*$ is the stellar mass, $\theta$ is the azimuthal angle in cylindrical coordinates, and inc is the disk inclination.
    \item Line parameters. The model uses a power-law prescription for the brightness temperature, similarly to other studies (see \citealt{Law2021}). We then adopted an exponentially tapered power law to spatially characterize the optical depth. This choice has the advantage of enabling a more accurate treatment of optically thin lines and line wings. Both temperature and optical depth profiles are assumed to be equal for the front and back surfaces. The equation for temperature reads
    \begin{equation}
        T(r) = T_{10}\> \biggl(\frac{r}{10 \> \text{au}} \biggr)^{q},
        \label{eq:T(r)}
    \end{equation},
    while the equation describing the optical depth is 
    \begin{equation}
    \label{eq:tau(r)}
        \tau(r) = {\tau_{10}} \> \biggl(\frac{r}{10 \>\text{au}}\biggr)^{\psi_{\tau}} \exp\biggl(-\biggl(\frac{r}{r_{\tau}} \biggr)^{\phi_{\tau}} \biggr),
    \end{equation}
    where we considered $\log_{10}(\tau_{10})$ when sampling the posteriors.
    The line-profile shape is characterized by a Gaussian, such that
    \begin{equation}
        \tau_{\nu}(r, z) = \tau(r) \> \exp\biggl(- \> \Biggl( \frac{v-v_{sky}(r,z)}{\Delta v} \Biggr)^2 \> \biggr),
        \label{eq:tau(r,z)}
    \end{equation}
    where the line width, $\Delta v$, is purely thermal and described through 
    \begin{equation}
        \Delta v (r)= \sqrt{\frac{2 k_B T(r)}{\mu \> m_H}}.
        \label{eq:vkep}
    \end{equation}
Here, $k_B$ is the Boltzmann constant, $\mu$ is the molecular mass of the emitting molecule, and $m_H$ is the hydrogen mass.
\end{enumerate}
Finally, we computed the surface-brightness profile for the front surface as 
\begin{equation}
    I_{\nu, f} (T, r, z) = B_{\nu} (T) \> (1-\exp(-\tau_{\nu} (r, z))),
    \label{eq:Inu_f}
\end{equation}
while the one for the back surface is 
\begin{equation}
    I_{\nu, b}(T,r, z) = B_{\nu}(T)\> (1-\exp(-\tau_{\nu}(r, z)))\> \exp(-\tau_{\nu}(r, z)),
    \label{eq:Inu_b}
\end{equation}
where $B_{\nu}$ is the Planck function.
Finally, the total surface brightness profile is computed as 
\begin{equation}
    I_{\nu} (T,r,z )= I_{\nu, f} (T,r, z)+ I_{\nu, b}(T,r, z).
\end{equation}
We summarize the equations of prescription A in Table \ref{Table:params}, dividing them by the physical disk attribute they describe.
\begin{table}[h!]
\centering
\caption{Full model A parametrization prescription.}
\vspace{-0.3cm}
\resizebox{0.9\linewidth}{!}{
\begin{tabular}{l l} 
 \hline
 \hline
 \noalign{\vskip 0.03in} 
 Group & Prescription A\\ [0.2ex]
 \hline
 \hline
 \noalign{\vskip 0.03in} 
 Geometry & inc, PA, dRA, dDEC \\ [0.2ex]
 \hline
 \noalign{\vskip 0.03in} 
 Front surface & $z_f = z_1 (r/1'')^{\psi_z}\> \exp(-(r/r_z)^{\phi_z})$ \\ [0.2ex]
 Back surface & $z_b = -z_f$ \\ [0.2ex]
 \hline
 \noalign{\vskip 0.01in} 
 Keplerian velocity & $v_{kep} = \sqrt{GM_*r / (r^2+z^2)^{3/2}}$ \\ [0.2ex]
 Sky velocity & $v_{sky} (r,z) = v_{kep} \cos(\theta) \sin\bigl(|\text{inc}|\bigr) + v_{sys}$ \\ [0.2ex]
 \hline
 \noalign{\vskip 0.03in} 
 Brightness temperature & $T = T_{10} \> (r/10 \text{ au})^{q}$\\ [0.2ex]
 Line width & $\Delta v = \sqrt{2k_BT / \mu m_H}$ \\ [0.2ex]
 \hline
 \noalign{\vskip 0.02in} 
  Optical depth & $\tau = \tau_{10} (r/10 \text{ au})^{{\psi}_{\tau}} \exp(-(r/r_\tau)^{\phi_\tau})$ \\ [0.2ex]
 Line profile & $\tau_{\nu} = \tau \exp\bigl(- ((v-v_{sky})/\Delta v)^2\bigr)$ \\ [0.2ex]
 \hline
 \noalign{\vskip 0.03in} 
 Front surface brightness & $I_{\nu, f} = B_{\nu} (1-\exp(-\tau_{\nu}))$\\ [0.2ex]
 Back surface brightness & $I_{\nu, b} = B_{\nu} (1-\exp(-\tau_{\nu})) \exp(-\tau_{\nu})$\\ [0.2ex]
 Surface brightness & $I_{\nu} = I_{\nu,f} + I_{\nu, b}$\\[0.2ex]
 \hline
 \hline
\end{tabular}
}
\label{Table:params}
\end{table}

We stress that even though model A provides a full description of a protoplanetary disk relying on the assumptions discussed above, this is still very simplistic in comparison with the real complexity of the disk. However, given the limited quality of the data used in our study, this approximation is adequate.

\vspace{-0.3cm}
\subsubsection{Parametric model B - minimal disk description}
To extract information from the four marginally resolved disks we have in the sample, we built a secondary disk prescription, which is more suited to describing their faint and compact line emission. The model B case assumes the emission is generated in the midplane (i.e., there is effectively no back surface, and the emission surface has $z=0$ at all radii) and has a constant optical depth $\tau_0$ out to radius $R_{out}$. We made this choice because the poor angular resolution did not allow us to resolve the optically thin outer disk, so we could not put any constraints on the disk's vertical structure.
In Table \ref{Table:params_B}, we summarize the resulting equations for parametrization B.
\begin{table}[h!]
\centering
\caption{Full model B parametrization prescription.}
\vspace{-0.3cm}
\resizebox{0.9\linewidth}{!}{
\begin{tabular}{l l} 
 \hline
 \hline
 \noalign{\vskip 0.03in} 
 Group & Prescription B\\ [0.2ex]
 \hline
 \hline
 \noalign{\vskip 0.03in} 
 Geometry & inc, PA, dRA, dDEC \\ [0.2ex]
 \hline
 \noalign{\vskip 0.01in} 
 Keplerian velocity & $v_{kep} = \sqrt{GM_*/r^2} $ \\ [0.2ex]
 Sky velocity & $v_{sky} (r) = v_{kep} \cos(\theta) \sin\bigl(|\text{inc}|\bigr) + v_{sys}$ \\ [0.2ex]
 \hline
 \noalign{\vskip 0.03in} 
 Brightness temperature & $T = T_{10}(r/10 \text{ au})^{q}$\\ [0.2ex]
 Line width & $\Delta v = \sqrt{2k_BT /\mu m_H}$ \\ [0.2ex]
 \hline
 \noalign{\vskip 0.02in} 
  Optical depth  & $\tau(r) =
        \begin{cases}
        \tau_0, & \text{if } r \leq R_{out} \\
        0,  & \text{if } r > R_{out}
        \end{cases}$ \\ [0.2ex]
 Line profile & $\tau_{\nu} = \tau \exp\bigl(- ((v-v_{sky})/\Delta v)^2\bigr)$ \\ [0.2ex]
 \hline
 \noalign{\vskip 0.03in} 
 Surface brightness & $I_{\nu} = B_{\nu}(1-\exp(-\tau_{\nu}))$\\[0.2ex]
 \hline
 \hline
\end{tabular}
}
\label{Table:params_B}
\end{table}

\vspace{-0.3cm}
\subsection{Best-fit models, residuals, and gas-disk size extraction}
In this section, we discuss the fitting procedure in detail, using the J16213469-2612269 disk as an example. This disk is compact, but still resolved in the image plane, and thus demonstrates the advantages of our fitting approach. The disk parametrization we adopted to fit this disk is prescription A (see Table \ref{Table:Fits_A} for the collection of best-fit parameters). 

\vspace{-0.3cm}
\subsubsection{Fitting procedure and residual plots}
We used Markov chain Monte Carlo (MCMC) algorithms to model the posterior distribution of our parametric models, and we sampled it using \verb|emcee| (\citealt{Foreman_2013}), which is implemented inside \verb|csalt|. Inspecting the channel maps, we chose priors for inc, PA, $v_{sys}$, $M_*$, $r_{\tau}$ and $r_{z}$ for each disk. We chose a conservative number of burn-in steps (250), we let 90 walkers evolve for 1000 steps afterward, and we then checked the results. If the relative change in the median value of each parameter was below $5\%$ over the last 500 steps, the fit was stopped; otherwise, the walkers were allowed to evolve for additional steps until this condition was met.
The results that we report in Appendix \ref{appendix:best_fit} (Tables \ref{Table:Fits_A}, \ref{Table:Fits_B}) are the median as the best fit and the uncertainties as its $\pm 68\%$ confidence intervals.
In Table \ref{Table:priors_summary}, we summarize the priors we chose for the fit on J16213469-2612269 performed with model A.
\begin{table}[h!]
\centering
\caption{Reference table for priors used in fitting J16213469-2612269 with model A.}
\vspace{-0.3cm}
\resizebox{0.7\linewidth}{!}{
\begin{tabular}{l c c} 
 \hline
 \hline
 \noalign{\vskip 0.03in} 
 Disk parameter & Prior & Values provided \\ [0.2ex]
 \hline
 \hline
 \noalign{\vskip 0.03in} 
  & Gaussian & [mean, std] \\[0.2ex]
   & Uniform & [low, high] \\[0.2ex]
 \hline
 \noalign{\vskip 0.03in} 
 inc [°] & Gaussian & $[50, 3]$ \\[0.2ex]
 PA [°] & Gaussian & $[44, 5]$ \\[0.2ex]
 ${M}_* \> [M_{\odot}]$ & Uniform & $[0.6, 1.4]$ \\[0.2ex]
 $r_{z}$ [au] & Gaussian & $[110, 10]$ \\[0.2ex]
 $z_{1}\>[^{\prime\prime}]$ & Uniform & $[0, 0.5]$ \\[0.2ex]
 $\psi_{z}$ & Gaussian & $[1.25, 0.15]$ \\[0.2ex]
 $T_{10}$ [K] & Gaussian & $[55, 5]$ \\[0.2ex]
 $q$ & Gaussian & $[-0.4, 0.1]$ \\[0.2ex]
 $r_{\tau}$ [au] & Gaussian & $[100, 10]$ \\[0.2ex]
 $\phi_{\tau}$ & Uniform & $[0, 50]$ \\[0.2ex]
 $\tau_{10}$ & Uniform & $[1, 40]$ \\[0.2ex]
 $\psi_{\tau}$ & Uniform & $[-4, 1]$ \\[0.2ex]
 $\phi_{z}$ & Uniform & $[0, 50]$ \\[0.2ex]
 $v_{sys}$ [m s$^{-1}$] & Gaussian & $[5400, 150]$ \\[0.2ex]
 $\text{d}_{\text{RA}} \>[^{\prime\prime}]$ & Gaussian & $[0, 0.05]$ \\[0.2ex]
 $\text{d}_{\text{DEC}} \>[^{\prime\prime}]$ & Gaussian & $[0, 0.05]$ \\[0.2ex]
 \hline
 \hline
\end{tabular}
}
\tablefoot{The full set of priors used in this work and the scripts to generate them for both parametrizations are available on \texttt{GitHub}\protect\hyperref[footnote:github]{\textsuperscript{\ref*{footnote:github}}}.}
\vspace{-1.2cm}
\label{Table:priors_summary}
\end{table}
The setup for the priors of each fit on the full sample for both parametrizations can be found on \verb|GitHub|\footnote{\label{footnote:github}\url{https://github.com/lzallio/USco_gas_2025}}.

\begin{figure*}[t!]
    \centering
    \includegraphics[width=\linewidth]{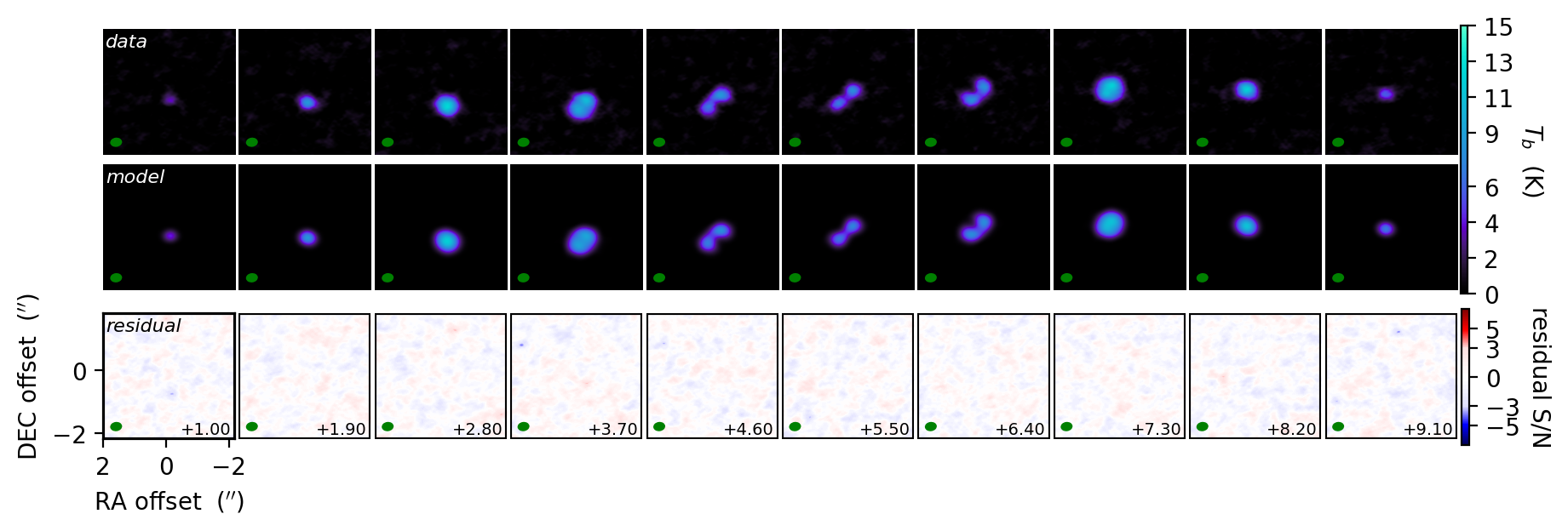}
    \caption{Residual plot for J16213469-2612269. In the first row, we show the channel maps of the data; in the second row, we show the channel maps of the best-fit model; in the third row, we show the channel maps of the residuals (data - model). The residual plots for all the disks analyzed in this work can be found in Appendix \ref{appendix:residuals} and on \texttt{GitHub}\protect\hyperref[footnote:github]{\textsuperscript{\ref*{footnote:github}}}.}
    \vspace{-0.4cm}
    \label{fig:residuals_example}
\end{figure*}

After convergence was met, we imaged the best-fit model and residual visibilities in the same way as the data: in Fig. \ref{fig:residuals_example} we show the residual plot for J16213469-2612269 (the full collection of residual plots can be found in Appendix \ref{appendix:residuals}, while their corresponding corner plots can be found on \verb|GitHub|\hyperref[footnote:github]{\textsuperscript{\ref{footnote:github}}}.). 
Here, in the first row we show the channel maps of the data. To produce these channel maps, we used the tclean routine in Common Astronomy Software Applications (CASA, \citealt{CASA_2007}) version 6.6.0.20. We chose a different velocity range for each disk so that it was possible to visualize the full-disk emission in ten channels. 

For cleaning, we used an image size of 1500 pix, with a cell of 0.01$^{\prime\prime}$. We set four different scales (0, 10, 30, 60), with a Briggs weighting of robust=1, a threshold of 20 mJy per beam. We also specified a Keplerian mask based on the best-fit parameters for each specific disk.
In the second row, we show the channel maps of the best-fit results: to clean the best-fit visibilities we used the same cleaning parameters as above.
In the third row, we show the channel maps of residuals, which were obtained by cleaning the residual visibilities (data - model). These channel maps are particularly useful, because if any non-Keplerian motion is found in the data, it manifests as either a red or blue blob in the residual maps. In the case of J16213469-2612269, we only see residual maps of noise.

\subsubsection{Gas-disk radii extraction and comparison with image-plane analysis}
Once we derived the best-fit model, we computed the gas-disk radii. This is usually done in the literature by first computing velocity-integrated (moment 0) maps, extracting the midplane disk coordinates from the sky projection, and azimuthally averaging to obtain the radial profile, which is then integrated to extract the radius that encompasses $68\%$ and $90\%$ of the disk flux ($R_{68\%}$ and $R_{90\%}$, respectively; see \citealt{Ansdell_2018}, \citealt{Sanchis2021}, \citealt{Galloway2025}, and \citealt{Trapman_2025}).
However, this methodology has an intrinsic problem, which is that the radius of the disk depends on the inclination and on the projected velocity field of the disk even after the disk was deprojected from the plane of the sky, as we discuss in Appendix \ref{appendix:R_vs_inc}. To avoid this intrinsic effect, to determine the gas-disk radii, we first computed the peak surface-brightness profile of the disk front surface at the $^{12}$CO $J=3 - 2$ rest frequency $(\nu=345.7959$ GHz) using Eq. \ref{eq:Inu_f}, rather than calculating the moment 0 map. Secondly, we integrated the peak surface-brightness profile in cylindrical coordinates and extracted the radius that encompasses $68\%$ and $90\%$ of the line emission. 

To determine the uncertainties of the gas-disk radii, we selected all random draws from the MCMC chain parameters needed in Eq. \ref{eq:Inu_f} and computed $R_{68\%}$ and $R_{90\%}$ as above. We then took the median value of the measured radii, $R_{90\%}$, and the 16$^{\text{th}}$ and 84$^{\text{th}}$ percentiles to derive their upper and lower errors, respectively.

\begin{figure}[t]
    \centering
    \includegraphics[width=\linewidth]{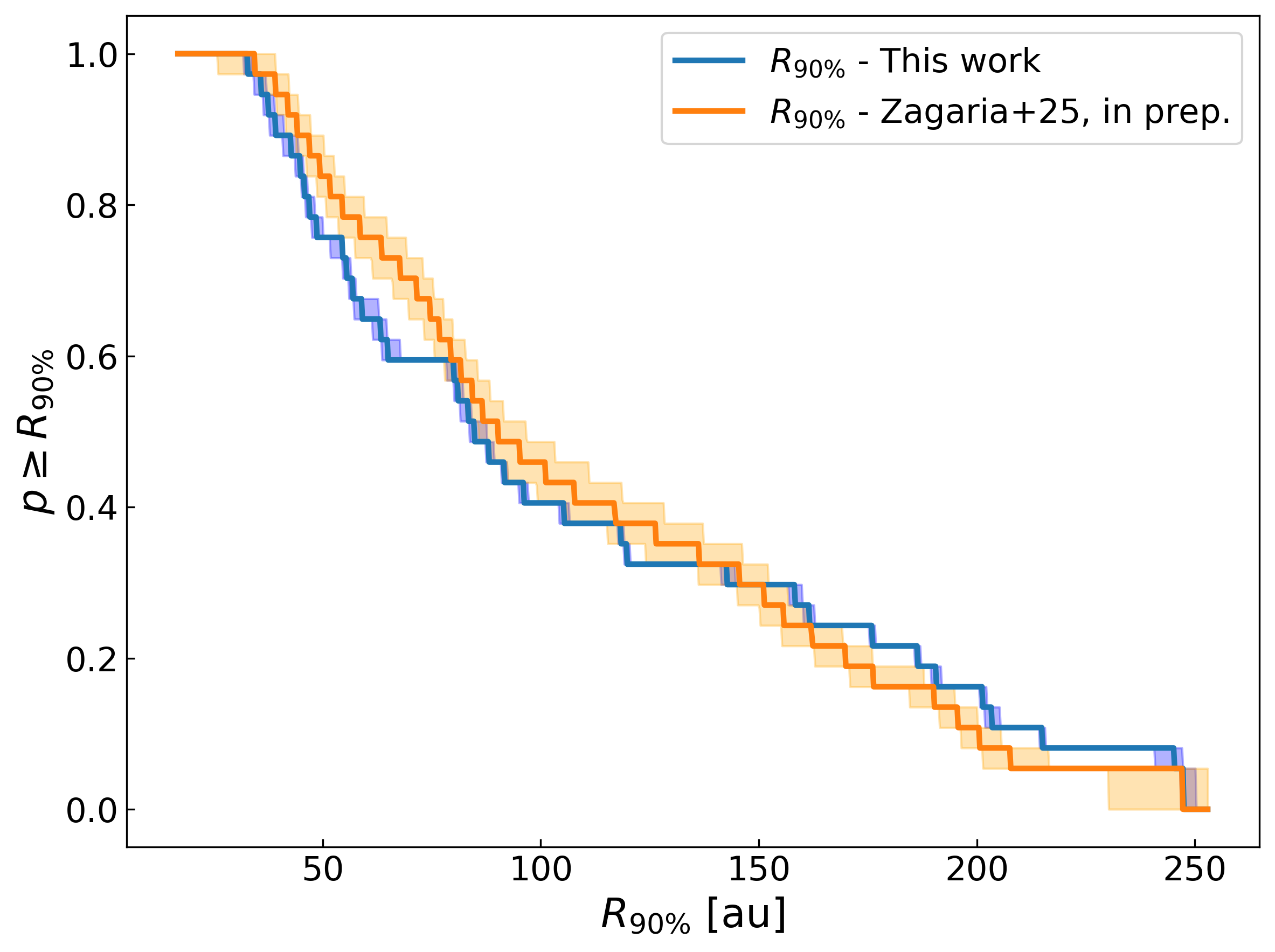}
    \caption{CCDF of $^{12}$CO $R_{90\%}$ computed for this work (in blue) and of the radii evaluated using image-plane analysis from Zagaria et al., in prep. (in orange).}
    \vspace{-0.5cm}
    \label{fig:R90_csalt_vs_Zagaria}
\end{figure}

We are aware that the radius definition adopted in this work intrinsically differs from the one usually adopted in literature, as we integrated the peak line flux rather than the spectrally integrated line flux. For this reason, we stress that our results are not directly comparable with what has been done in the literature so far. To provide a comparison with what is usually done in the literature, in Fig. \ref{fig:R90_csalt_vs_Zagaria} we show the complementary cumulative distribution function (CCDF\footnote{The CCDF of $R_{90\%}$ is defined as the probability $p\geq R_{90\%}$; i.e., the fraction of disks with a gas radius larger than or equal to a given value.}) of $R_{90\%}$ of our sample and compare it with that of Zagaria et al., in prep., which analyzed the same sources of our sample, but using an image-plane analysis as described above. 
After running the Anderson-Darling two-sample test (AD; as implemented in \verb|scipy|\footnote{\url{https://scipy.org}}; see \citealt{Scipy}), we obtained an AD $p$-value $p=0.8$ in the range of $[0.6, 0.9]$, which means that the two distributions cannot be distinguished within uncertainties. This check shows that the methodology for extracting gas-disk radii presented in this paper provides results that are comparable to what is usually done in the literature when working on the same sample.

\vspace{-0.3cm}
\subsubsection{Fitting small disks with prescription A}
\label{sec:compact_disks}
\begin{figure*}[t!]
    \centering
    \includegraphics[width=\linewidth]{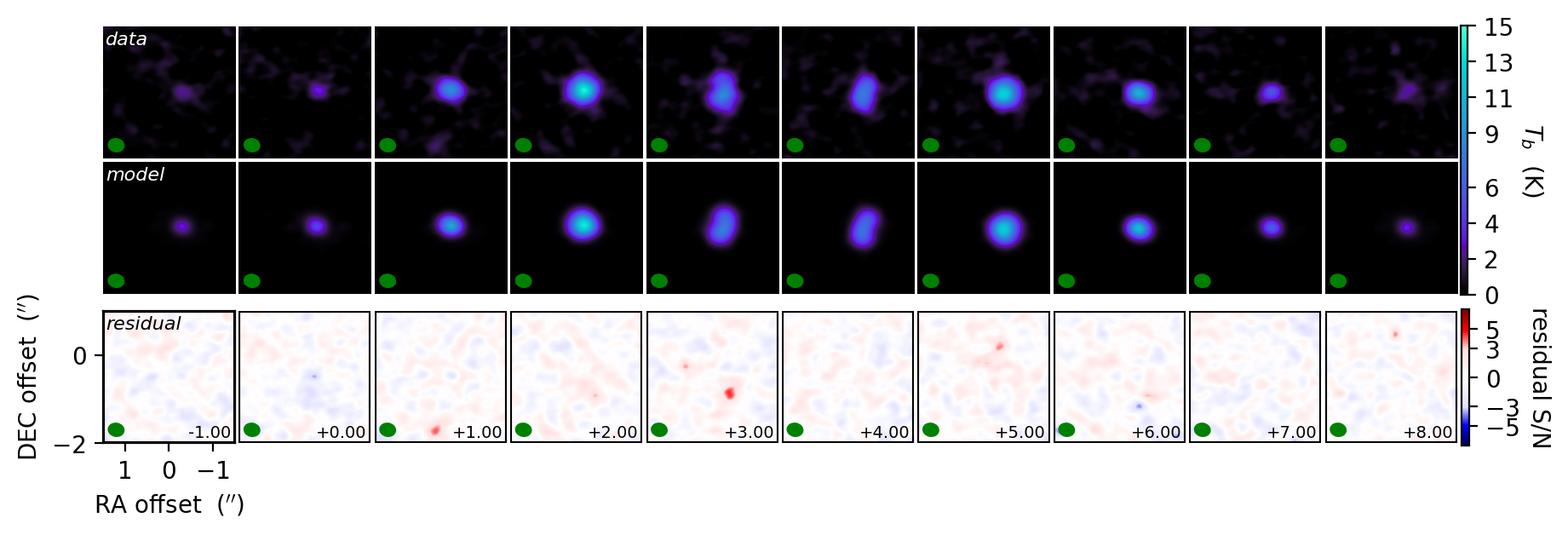}
    \caption{Residual plot for J16062861-2121297, shown in the same way as in Fig. \ref{fig:residuals_example}. The residual plots for all the disks analyzed in this work can be found in Appendix \ref{appendix:residuals} and on \texttt{GitHub}\protect\hyperref[footnote:github]{\textsuperscript{\ref*{footnote:github}}}.}
    \vspace{-0.3cm}
    \label{fig:residuals_compact}
\end{figure*}

Using \verb|csalt|, we can also retrieve bulk properties for small protoplanetary disks. In Fig. \ref{fig:residuals_compact}, we show the residual plot of J16062861-2121297, a compact disk (gas-disk size $R_{90\%} = 56\pm1$ au, with a beam of $\sim25$ au).
Looking at Fig. \ref{fig:residuals_compact}, we see how \verb|csalt| reproduces the disk morphology well using the parametric prescription A. We are also able to infer the vertical structure of the thick $^{12}$CO emitting layer from this fit. The channel maps of the residuals are almost completely clean, showing that an optimal fit was found for this source.

\vspace{-0.3cm}
\subsubsection{Dealing with cloud-absorbed channels}
\label{sec:cloud_absorption}
\begin{figure*}[t!]
    \centering
    \includegraphics[width=\linewidth]{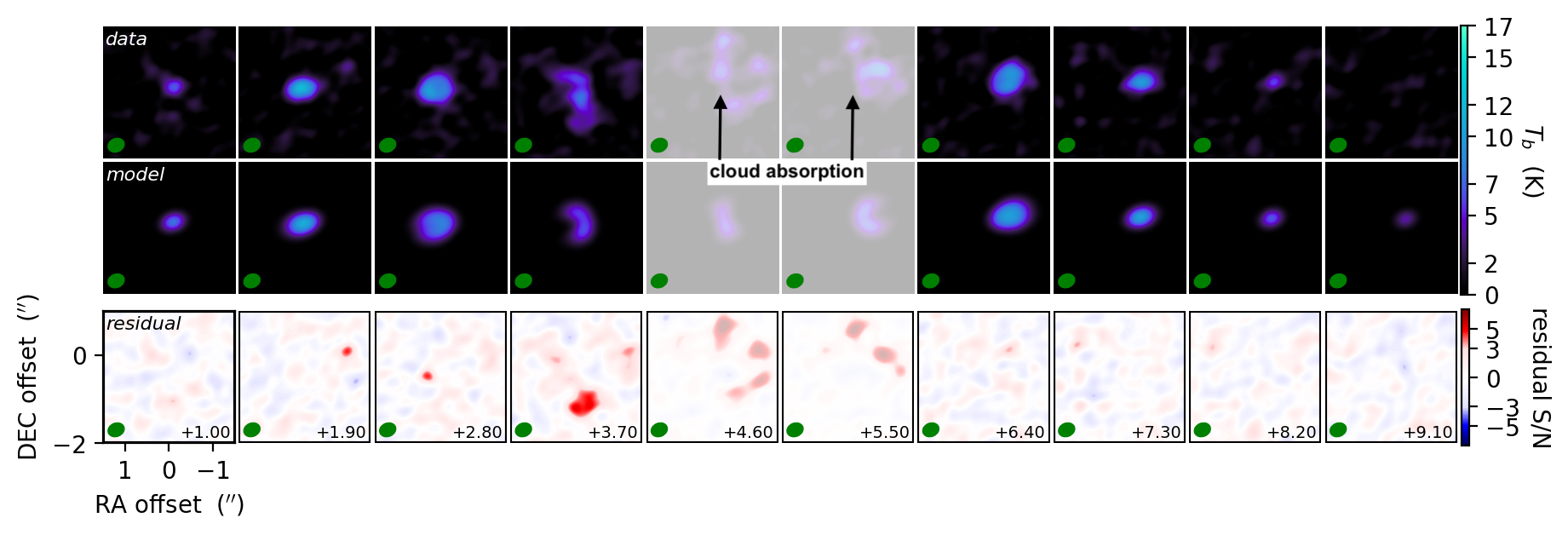}
    \caption{Residual plot for J16123916-1859284, shown in the same way as in Fig. \ref{fig:residuals_example}. We highlight the residuals in the channel maps at $v=4.6, 5.5$ km s$^{-1}$, which are those affected by cloud absorption. Moreover, in the channel map at $v=3.7$ km s$^{-1}$, we show how \texttt{csalt} is able to separate the diffuse cloud emission (the localized red emission) and the Keplerian disk emission. The residual plots for all the disks analyzed in this work can be found in Appendix \ref{appendix:residuals} and on \texttt{GitHub}\protect\hyperref[footnote:github]{\textsuperscript{\ref*{footnote:github}}}.}
    \vspace{-0.3cm}
    \label{fig:residuals_cloud}
\end{figure*}

Some disks in our sample showed signs of foreground cloud absorption: the disk emission is either obscured in some channels or completely embedded in more diffuse cloud emission. To deal with such problematic features, we first checked the channel maps and the spectra of each of the disks in our sample and identified the frequency ranges that showed clear cloud absorption. Then, we used an option implemented in \verb|csalt| that allows one to exclude specific frequency ranges from the MCMC likelihood evaluation; this allowed us to consider only clean disk emission in our fits. In Fig. \ref{fig:residuals_cloud}, we show the residual plots we obtained for the cloud-contaminated disk J16123916-1859284.
This plot shows clear 3 and 5$\sigma$ residuals in the channel maps at $v=4.6, 5.5$ km s$^{-1}$, which are the channels affected by cloud absorption. Moreover, in the channel map at $v=3.7$ km s$^{-1}$ we show how \verb|csalt| is able to separate the diffuse cloud emission, which results in red residuals, and Keplerian disk emission. 

\vspace{-0.3cm}
\section{Results}
\label{sec:results}
In this section, we present our results and provide median values of the relevant physical quantities of the disks in Upper Scorpius. We were able to fit 37 sources out of a sample of 83, 33 with prescription A and four with prescription B. We present the best-fit results in Tables \ref{Table:Fits_A} and \ref{Table:Fits_B}, respectively.

We attempted to perform our analysis on all 83 $^{12}$CO $J=3-2$ detections, but we were unable to achieve any acceptable convergence for our parametric models on the remaining 46 disks for a variety of reasons. The first is the poor spectral sampling that we have for these sources: if $^{12}$CO is only weakly detected in a few channels (a bandwidth of $\lesssim2$ km s$^{-1}$), \verb|csalt| cannot fit a Keplerian rotation spectrum due to lack of information from the data. The second is very low S/N; by looking at the integrated flux reported in \cite{Carpenter2025}, most of these sources have integrated S/Ns of $\lesssim 10$. We therefore empirically report that visibility fitting becomes difficult if the S/N of the integrated flux is $\lesssim 10$. The third reason is poor angular resolution (the disk size is comparable with the size of the beam): if a disk is both spectrally and spatially unresolved, disk fitting fails even if visibilities are used. We report that a combination of these three effects is causing the fits to fail for 44 of these remaining objects. Finally, two peculiar objects that we were unable to fit are J16113134-1838259, which is a potential fly-by, and J16193570-1950426, which may hide at least one companion, as further detailed in Sect. \ref{sec:discussion}.

\vspace{-0.3cm}
\subsection{Gas-disk radii}
We find a median gas-disk size of $R_{90\%} \sim 82$ au, with the largest disk being J16042165-2130284, a well-known face-on disk (see, e.g., \citealt{Stadler_2023} and references therein) for which we derive $R_{90\%}=247\pm1$ au. The most compact disk is J16095933-1800090, for which we derive $R_{90\%}=22\pm3$ au. However, as said before, our results are sensitivity-limited. Since there is a tight correlation between the gas-disk radius and the line flux (\citealt{Long2022}, \citealt{Zagaria2023}, and discussed later on in this section), we conclude that the 44 disks that could not be fit (we excluded J16113134-1838259 and J16193570-1950426 as discussed before) are smaller than or of comparable size to J16095933-1800090. As a consequence, the real median gas-disk size for the $^{12}$CO detections should be, at the zeroth order, $R_{90\%}\lesssim 22$ au. A detailed discussion on this topic will be presented in Zagaria et al., in prep.

In Fig. \ref{fig:R_90_CDF}, we report the CCDF of $R_{90\%}$ of our sample and compare it with that of \cite{Trapman_2025} for the AGE-PRO Upper Scorpius sample.
\setcounter{figure}{4}
\begin{figure}
    \centering
    \includegraphics[width=\linewidth]{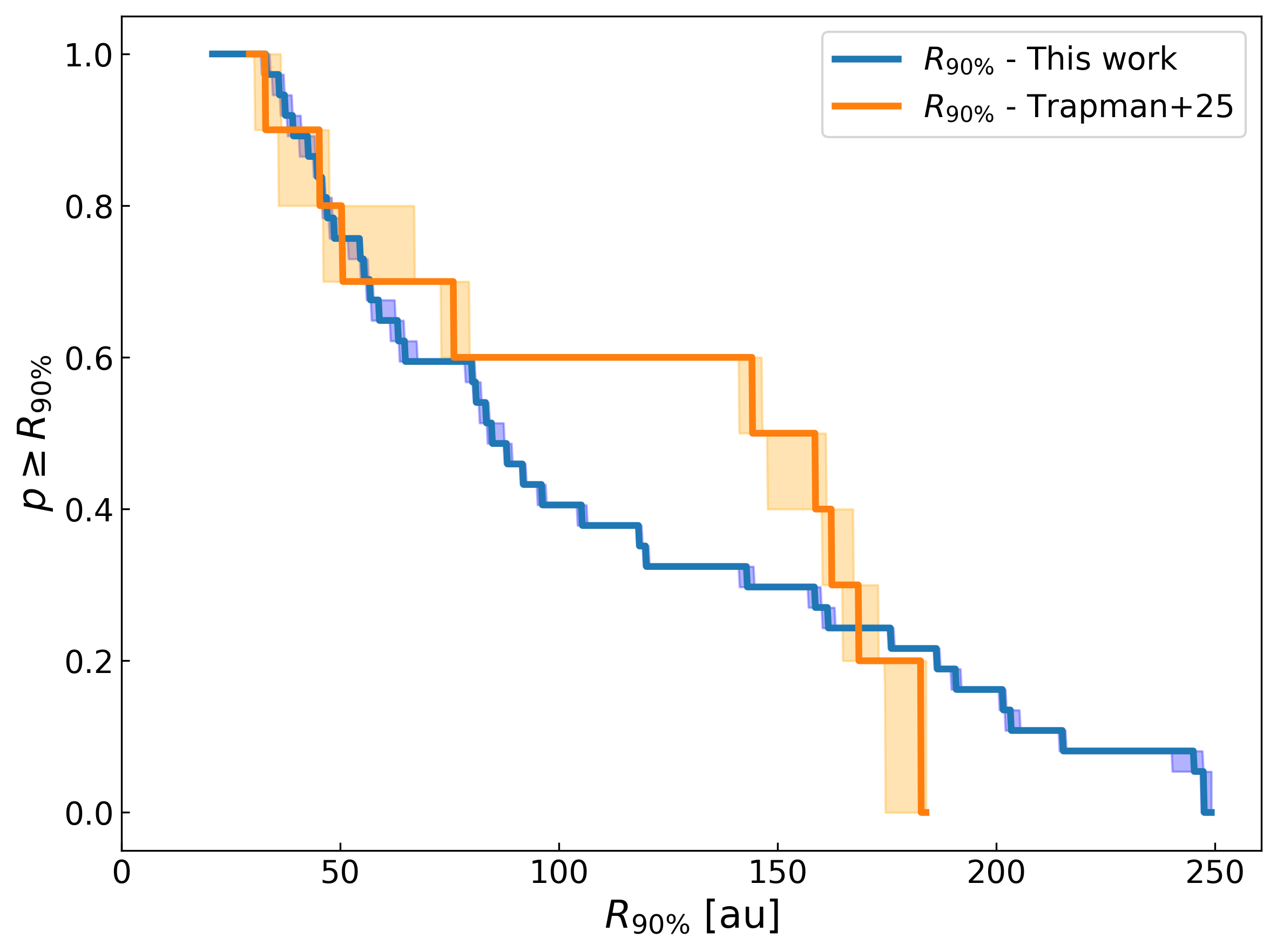}
    \caption{CCDF of $^{12}$CO $R_{90\%}$ for our sample (in blue) and of the sources in Upper Scorpius reported from \cite{Trapman_2025} (in orange). The gas-disk sizes reported in this figure are reported in Tables \ref{Table:Fits_A} and \ref{Table:Fits_B}.}
    \label{fig:R_90_CDF}
\end{figure}
Our median gas-disk size is somewhat lower than the value $R_{\text{CO}} \sim 109.9$ au reported in \cite{Agurto-Gangas_2025} and \cite{Zhang_2025}. To determine if our results are consistent, we ran the AD two-sample test, and we obtained a $p$-value $p=0.7$ in the range of $[0.5, 0.8]$. This means that the two distributions of $R_{90\%}$ cannot be distinguished within the uncertainties. We again emphasize the different techniques used in \cite{Trapman_2025} and in our work to determine the gas-disk radii, and we note that the two studies involve different sample sizes and ALMA observations (with AGE-PRO observations conducted using ALMA $^{12}$CO $J=2-1$ data with an integration time $\sim1$ hour on source).

\vspace{-0.3cm}
\subsection{Integrated fluxes}
In Appendix \ref{appendix:fluxes}, we explain how we extracted the fluxes both from data and best-fit models, while in Table \ref{Table:Fluxes} we report them together with their associated uncertainties. We also include a column that shows the velocity ranges per source for which we identified foreground cloud absorption, and a column that reports whether after the fit any 3 or 5$\sigma$ residuals are visible in the residual plots.
\begin{table*}[!t]
\centering
\caption{Fluxes from data and best-fit models, together with foreground absorption-velocity ranges and a keyword for residuals.}
\vspace{-0.3cm}
\resizebox{0.67\linewidth}{!}{
\centering
\begin{tabular}{l c c c c} 
 \hline
 \hline
 \noalign{\vskip 0.03in} 
 Source (2MASS) & Flux $[\text{Jy}\> \text{m s}^{-1}]$ & Flux models $[\text{Jy}\> \text{m s}^{-1}]$ & Absorption $[\text{m s}^{-1}]$ & Residuals \\ [0.2ex]
 \hline
 \hline
 \noalign{\vskip 0.03in} 
J15583620-1946135 & $633.5 \pm 90.1$ & $747.0 \pm 10.2$ & - & Y \\ [0.4ex]
J15583692-2257153 & $6040.2 \pm 239.5$ & $5792.6 \pm 52.0$ & - & Y \\ [0.4ex]
J16035793-1942108 & $989.2 \pm 70.8$ & $1725.0 \pm 45.4$ & - & Y \\ [0.4ex]
J16052157-1821412 & $7460.1 \pm 157.2$ & $7987.3 \pm 29.6$ & - & N \\ [0.4ex]
J16062383-1807183 & $818.4 \pm 69.7$ & $632.8 \pm 9.8$ & - & Y \\ [0.4ex]
J16062861-2121297 & $1700.8 \pm 78.3$ & $1718.9 \pm 14.7$ & - & Y \\ [0.4ex]
J16064794-1841437 & $8819.5 \pm 195.6$ & $8515.9 \pm 26.1$ & - & N \\ [0.4ex]
J16095933-1800090 & $301.0 \pm 54.5$ & $634.6 \pm 69.8$ & - & Y \\ [0.4ex]
J16101264-2104446 & $8529.1 \pm 251.9$ & $8891.2 \pm 23.3$ & - & Y \\ [0.4ex]
J16120668-3010270 & $11240.8 \pm 2155.1$ & $10086.0 \pm 76.9$ & - & Y \\ [0.4ex]
J16123916-1859284 & $1770.8 \pm 207.5$ & $2450.3 \pm 2.0$ & 3900-5600 & Y \\ [0.4ex]
J16132190-2136136 & $556.1 \pm 63.3$ & $615.5 \pm 28.1$ & - & Y \\ [0.4ex]
J16140792-1938292 & $3022.6 \pm 78.5$ & $3059.7 \pm 13.6$ & - & N \\ [0.4ex]
J16142091-1906051 & $7186.1 \pm 339.9$ & $5581.5 \pm 17.2$ & - & Y \\ [0.4ex]
J16145024-2100599 & $2107.5 \pm 126.4$ & $2379.4 \pm 29.1$ & - & N \\ [0.4ex]
J16152752-1847097 & $7860.8 \pm 254.1$ & $7576.6 \pm 21.5$ & - & Y \\ [0.4ex]
J16181445-2319251 & $268.9 \pm 70.4$ & $395.2 \pm 26.3$ & - & Y \\ [0.4ex]
J16185382-2053182 & $1188.6 \pm 47.6$ & $1406.0 \pm 23.9$ & - & Y \\ [0.4ex]
J16202291-2227041 & $308.2 \pm 57.8$ & $429.1 \pm 17.5$ & 3400-4400 & Y \\ [0.4ex]
J16202863-2442087 & $3914.2 \pm 141.9$ & $3525.0 \pm 13.8$ & - & Y \\ [0.4ex]
J16203960-2634284 & $6868.9 \pm 213.1$ & $7008.4 \pm 16.4$ & - & Y \\ [0.4ex]
J16213469-2612269 & $2615.7 \pm 134.2$ & $2411.1 \pm 13.2$ & - & N \\ [0.4ex]
J16215472-2752053 & $3213.8 \pm 61.6$ & $3307.3 \pm 21.3$ & - & N \\ [0.4ex]
J16221532-2511349 & $1973.1 \pm 80.9$ & $1911.2 \pm 14.0$ & - & N \\ [0.4ex]
J16222982-2002472 & $4564.5 \pm 53.4$ & $5434.3 \pm 22.4$ & 700-3000 & N \\ [0.4ex]
J16230761-2516339 & $819.3 \pm 31.6$ & $921.5 \pm 12.3$ & - & Y \\ [0.4ex]
J16253798-1943162 & $3854.7 \pm 198.3$ & $2902.6 \pm 17.4$ & - & Y \\ [0.4ex]
J16271273-2504017 & $662.7 \pm 66.7$ & $673.5 \pm 34.6$ & 4000-4600 & Y \\ [0.4ex]
J16274905-2602437 & $2459.1 \pm 89.4$ & $2385.3 \pm 32.2$ & - & Y \\ [0.4ex]
J16293267-2543291 & $1459.8 \pm 47.1$ & $1322.0 \pm 11.8$ & - & Y \\ [0.4ex]
J16395577-2347355 & $865.3 \pm 61.2$ & $698.3 \pm 14.1$ & - & Y \\ [0.4ex]
J16142029-1906481 & $3286.3 \pm 52.7$ & $4362.3 \pm 43.8$ & - & N \\ [0.4ex]
J16042165-2130284 & $19049.3 \pm 1596.0$ & $19654.6 \pm 49.9$ & - & N \\ [0.4ex]
J16012268-2408003 & $3494.2 \pm 256.3$ & $1535.4 \pm 35.1$ & - & Y \\ [0.4ex]
J16145026-2332397 & $833.1 \pm 63.1$ & $763.0 \pm 9.3$ & - & Y \\ [0.4ex]
J16020757-2257467 & $708.9 \pm 74.6$ & $931.5 \pm 54.4$ & - & Y \\ [0.4ex]
J16121242-1907191 & $322.3 \pm 23.5$ & $300.1 \pm 6.5$ & - & Y \\ [0.4ex]

 \hline
 \hline
\end{tabular}
}
\tablefoot{The table is also available on \texttt{GitHub}\protect\hyperref[footnote:github]{\textsuperscript{\ref*{footnote:github}}}.}
\vspace{-0.6cm}
\label{Table:Fluxes}
\end{table*}

\vspace{-0.3cm}
\subsubsection*{$^{12}$CO flux - size correlation}
\begin{figure}
    \centering
    \includegraphics[width=\linewidth]{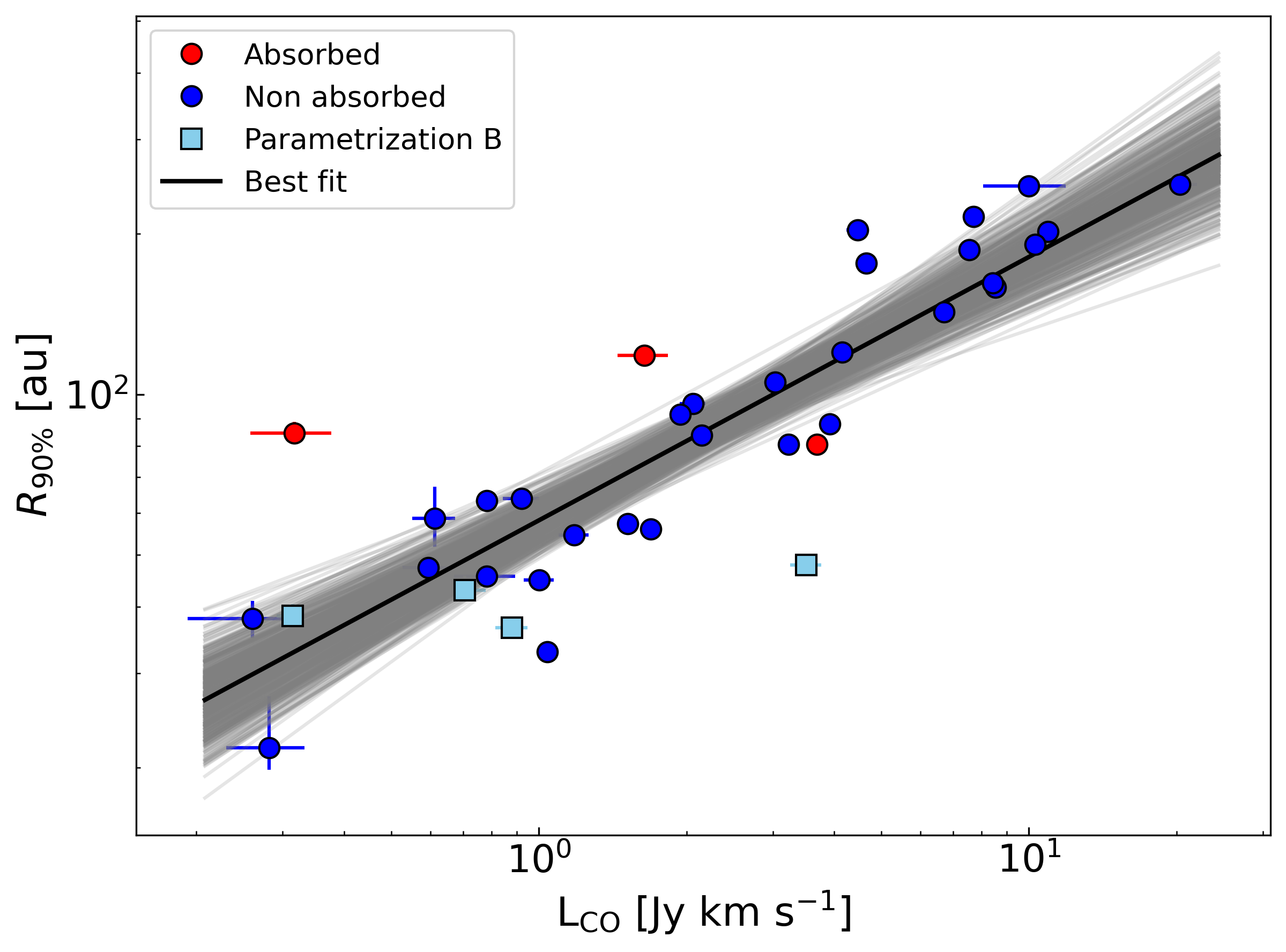}
    \caption{Correlation between $^{12}$CO gas-disk size and luminosity. The values for $R_{90\%}$ and the disk fluxes, together with the line absorption ranges, can be found in Tables \ref{Table:Fits_A}, \ref{Table:Fits_B}, and \ref{Table:Fluxes}, respectively.}
    \vspace{-0.5cm}
    \label{fig:R_L_corr}
\end{figure}

In Fig. \ref{fig:R_L_corr}, we show the correlation between the $^{12}$CO sizes and the $^{12}$CO disk luminosity $\text{L}_{\text{CO}}= \text{F}_{\text{CO}} \times (\text{d} \>[\text{pc}]\>/\>140 \>\text{pc})^2$ Jy km s$^{-1}$, where we highlight, with different markers and colors, the disks we fit with different parametrizations and those that showed clear absorption. We fit the correlation using the \verb|linmix|\footnote{\url{https://linmix.readthedocs.io/en/latest/}} package and find the best-fit correlation: $\log_{10}(R_{\text{90\%}}) = (1.77\pm0.03) + (0.49\pm0.05)\> \log_{10}\bigl(\text{L}_{\text{CO}}\bigr)$. The slope we derived is comparable to that found in \cite{Long2022}, which reported the relationship $\log_{10}(\text{R}_{\text{CO}}) = (0.52\pm0.05)\>\log_{10}\bigl(\text{L}_{\text{CO}}\bigr) + (2.07\pm0.03)$ using $^{12}$CO $J=2 - 1$ data.
The slope we measured is in agreement with the fact that the $^{12}$CO layer is optically thick\footnote{The expected correlation in a regime of optically thick emission is $R_{\text{CO}}\propto L_{\text{CO}}^{0.5}$, since in this case $L_{\text{CO}}\sim B_{\nu}(T) \times \pi R_{\text{CO}}^2$.} and therefore can be used as a proxy for the gas temperature of the emitting layer (e.g., \citealt{Sanchis2021}, \citealt{Long2022}), as further discussed in Sect. \ref{sec:discussion}. However, the intercept found in this work is different from the one reported by \cite{Long2022}, with a difference of $\sim 0.3$ in the intercept and a significance of $\sim 7 \sigma$; this is likely the result of using different techniques to extract gas-disk radii, combined with having different data qualities in different samples. 

\vspace{-0.3cm}
\subsection{Vertical heights}
Our first disk model includes information on the vertical location of the line-emitting layer. We derived a collection of $^{12}$CO $J=3 - 2$ line-emission surfaces of 33 compact disks. We created the profiles according to Eq. \ref{eq:z(r)} using random draws from the MCMC chains of $z_1$, ${\psi}_z$, $r_z$, and $\phi_z$, evaluating the associated $z(r)$ profiles, and taking their median value and the $16^{\text{th}}$ and $84^{\text{th}}$ percentiles for the associated uncertainties. In Fig. \ref{fig:all_disks_zr} of Appendix \ref{appendix:vertical_heights}, we show the vertical-height profiles as a function of radius. 

Through vertical-height profiles, we computed the values of the aspect ratio $ \langle z/r \rangle $ in the same way as \cite{Law2021} and \cite{Paneque_2023} to homogenize our analysis. This was first done by defining $r_{cut}$ as the value of the radius for which the vertical-height profile stops increasing. At this point, we defined $r_{taper} = 0.8 \times r_{cut}$, and we evaluated $ \langle z/r \rangle $, with $0\leq r \leq r_{taper}$.

\begin{figure}
    \centering
    \includegraphics[width=\linewidth]{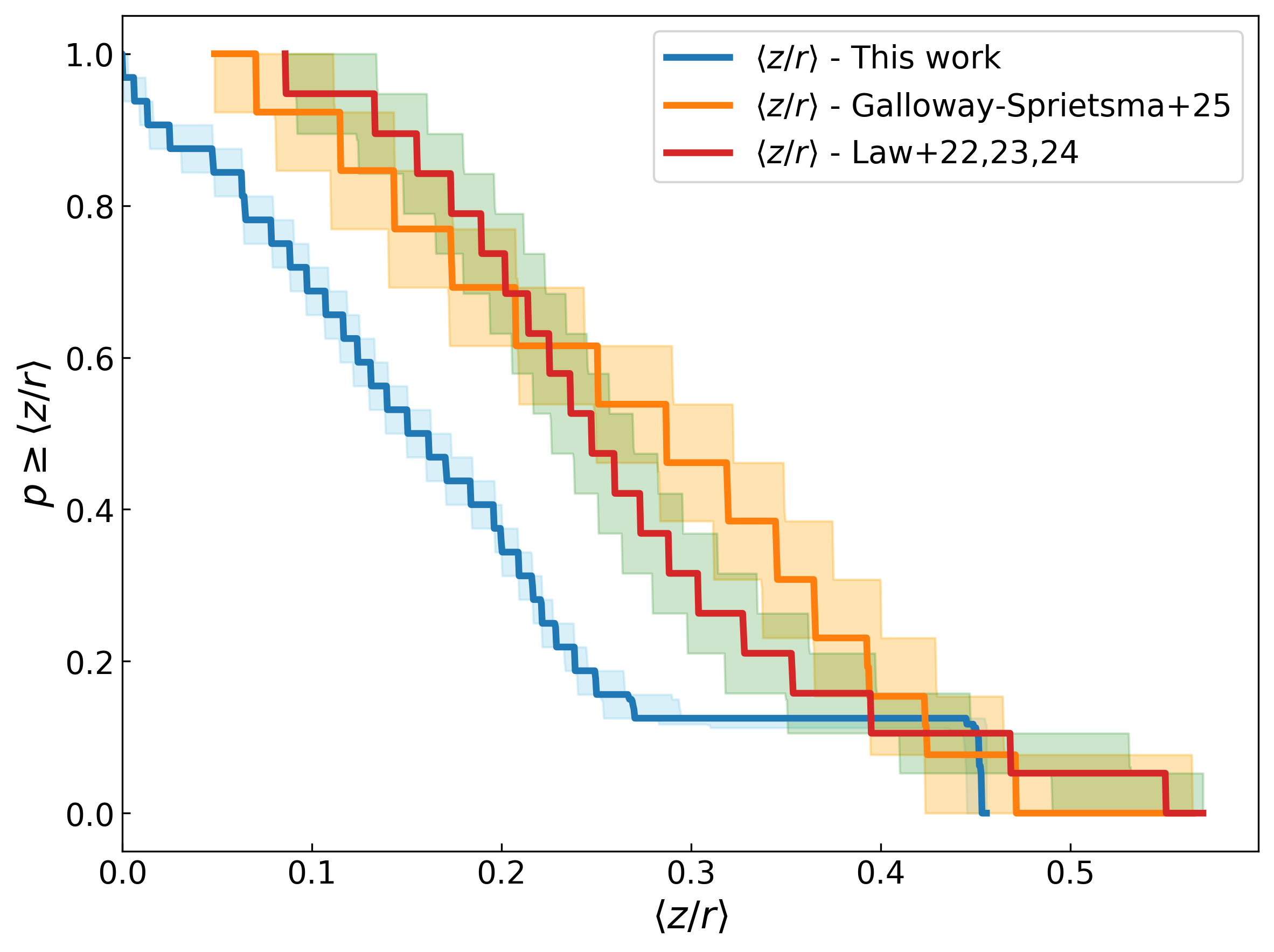}
    \caption{Kaplan-Meier estimator of radially averaged $^{12}$CO $ \langle z/r \rangle $ values for our sample (in blue), compared with those of \cite{Galloway2025} (in orange), and the combined works of \cite{Law_2022, Law_2023, Law_2024} (in red).}
    \vspace{-0.5cm}
    \label{fig:z_over_r_histo}
\end{figure}

We find a median value of $ \langle z/r \rangle \sim0.16$, which is lower than the one reported in \cite{Galloway2025} $( \langle z/r \rangle \sim0.28)$. This is likely because we analyzed a very different sample. The exoALMA sample (\citealt{Teague_2025}) is composed of bright and extended protoplanetary disks; the same applies for the disks considered in the works of \cite{Law_2022, Law_2023, Law_2024}. Instead, we considered the disk population of Upper Scorpius, which is composed of both compact and extended disks. In Fig. \ref{fig:z_over_r_histo}, we show the Kaplan-Meier estimator (as implemented in \verb|lifelines|\footnote{\url{https://lifelines.readthedocs.io/en/latest/index.html}}; see \citealt{linfelines_2019}) of the radially averaged $ \langle z/r \rangle $ values for our sample and compare it with those of \cite{Galloway2025}, and the combined works of \cite{Law_2022, Law_2023, Law_2024}. 
We ran the log-rank test (again, as implemented in \verb|lifelines|), which accounts for upper limits via the Kaplan-Meier estimator, finding that the distribution of $ \langle z/r \rangle $ derived for this work differs significantly from the one of \cite{Galloway2025} ($p=0.05$ in a range $[0.02, 0.12]$) and the one of \cite{Law_2022, Law_2023, Law_2024} ($p=0.03$ in a range $[0.01, 0.08]$). 
A caveat of our analysis is that the $\langle z/r \rangle$ values derived for our disk sample are based on observations with significantly lower angular resolution compared to those used in \cite{Galloway2025} and \cite{Law_2022, Law_2023, Law_2024}. Therefore, the differences observed in the $\langle z/r \rangle$ distributions cannot be solely attributed to the compactness of the disks considered in our analysis.

In Table \ref{Table:z/r_values}, we report all the values of $ \langle z/r \rangle $ used in Fig. \ref{fig:z_over_r_histo}.
\begin{table*}[h!]
\centering
\caption{Values of $ \langle z/r \rangle $ derived in this work, together with values compiled from the literature.}
\vspace{-0.3cm}
\resizebox{\linewidth}{!}{
\begin{tabular}{l c c | l c c | l c c} 
 \hline
 \hline
 \noalign{\vskip 0.03in} 
 2MASS - This work & $ \langle z/r \rangle $ & Line ($^{12}$CO) & Source - Law+22,23,24 & $ \langle z/r \rangle $ & Line ($^{12}$CO) & Source - Galloway-Sprietsma+25 & $ \langle z/r \rangle $ & Line ($^{12}$CO) \\ [0.4ex]
 \hline
 \hline
 \noalign{\vskip 0.03in} 
 J15583620-1946135 & $<0.39$ & $J=3-2$ & HD 142666 & $0.20_{-0.05}^{+0.02}$ & $J=2-1$ &  DM Tau & $0.35\pm{0.07}$ & $J=3-2$ \\ [0.4ex]
J15583692-2257153 & $0.25_{-0.02}^{+0.01}$ & $J=3-2$ & MY Lup & $0.16_{-0.06}^{+0.01}$ & $J=2-1$ & AA Tau & $0.42\pm0.05$ & $J=3-2$ \\ [0.4ex]
J16035793-1942108 & $0.014_{-0.003}^{+0.002}$ & $J=3-2$ & V4046 Sgr & $0.24_{-0.07}^{+0.12}$ & $J=3-2$ & LkCa 15 & $0.34\pm0.04$ & $J=3-2$ \\ [0.4ex]
J16052157-1821412 & $0.13_{-0.01}^{+0.01}$ & $J=3-2$ & HD 100546 & $0.24_{-0.05}^{+0.07}$ & $J=2-1$ & HD 34282 & $0.41\pm0.04$ & $J=3-2$\\ [0.4ex]
J16062383-1807183 & $0.16_{-0.03}^{+0.02}$ & $J=3-2$ & GW Lup & $0.21_{-0.02}^{+0.13}$ & $J=2-1$ & MWC 758 & $0.29\pm0.13$ & $J=3-2$ \\ [0.4ex]
J16062861-2121297 & $0.20_{-0.02}^{+0.02}$ & $J=3-2$ & WaOph 6 & $0.19_{-0.05}^{+0.03}$ & $J=2-1$ & CQ Tau & $0.17\pm0.09$ & $J=3-2$\\ [0.4ex]
J16064794-1841437 & $0.221_{-0.003}^{+0.003}$ & $J=3-2$ & DoAr 25 & $0.25_{-0.06}^{+0.04}$ & $J=2-1$ & SY Cha & $0.29\pm0.06$ & $J=3-2$ \\ [0.4ex]
J16095933-1800090 & $0.20_{-0.06}^{+0.05}$ & $J=3-2$ & Sz 91 & $0.22_{-0.1}^{+0.05}$ & $J=3-2$ & PDS 66 & $0.12\pm0.05$ & $J=3-2$ \\ [0.4ex]
J16101264-2104446 & $0.22_{-0.02}^{+0.01}$ & $J=3-2$ & CI Tau & $0.24_{-0.04}^{+0.07}$ & $J=3-2$ & HD 143006 & $0.40\pm0.16$ & $J=3-2$\\ [0.4ex]
J16120668-3010270 & $0.45_{-0.01}^{+0.01}$ & $J=3-2$ & DM Tau & $0.53_{-0.09}^{+0.04}$ & $J=2-1$ & RXJ1615.3-3255 & $0.34\pm0.06$ & $J=3-2$\\ [0.4ex]
J16123916-1859284 & $0.063_{-0.003}^{+0.003}$ & $J=3-2$ & IM Lup & $0.34_{-0.11}^{+0.08}$ & $J=2-1$ & V4046 Sgr & $0.13\pm0.05$ & $J=3-2$ \\ [0.4ex]
J16132190-2136136 & $<0.29$ & $J=3-2$ & GM Aur & $0.35_{-0.11}^{+0.06}$ & $J=2-1$ & RXJ1842.9-3532 & $0.20\pm0.05$ & $J=3-2$ \\ [0.4ex]
J16140792-1938292 & $0.16_{-0.02}^{+0.01}$ & $J=3-2$ & AS 209 & $0.17_{-0.07}^{+0.04}$ & $J=2-1$ & RXJ1852.3-3700 & $0.15\pm0.05$ & $J=3-2$ \\ [0.4ex]
\cline{7-9}
J16142091-1906051 & $<0.30$ & $J=3-2$ & HD 163296 & $0.24_{-0.09}^{+0.06}$ & $J=2-1$  \\ [0.4ex]
J16145024-2100599 & $0.09_{-0.01}^{+0.01}$ & $J=3-2$ & MWC 480 & $0.22_{-0.11}^{+0.05}$ & $J=2-1$ \\ [0.4ex]
J16152752-1847097 & $0.199_{-0.001}^{+0.001}$ & $J=3-2$ & HD 97048 & $0.26_{-0.03}^{+0.05}$ & $J=2-1$ \\ [0.4ex]
J16181445-2319251 & $0.08_{-0.03}^{+0.06}$ & $J=3-2$ & LkCa 15 & $0.26_{-0.03}^{+0.04}$ & $J=2-1$  \\ [0.4ex]
J16185382-2053182 & $0.11_{-0.03}^{+0.04}$ & $J=3-2$ & HD 34282 & $0.46_{-0.14}^{+0.04}$ & $J=2-1$ \\ [0.4ex]
J16202291-2227041 & $0.05_{-0.02}^{+0.03}$ & $J=3-2$ & PDS 70 & $0.32_{-0.06}^{+0.01}$ & $J=2-1$ \\ [0.4ex]
\cline{4-6}
J16202863-2442087 & $0.21_{-0.01}^{+0.01}$ & $J=3-2$  \\ [0.4ex]
J16203960-2634284 & $0.239_{-0.003}^{+0.003}$ & $J=3-2$ \\ [0.4ex]
J16213469-2612269 & $0.14_{-0.01}^{+0.01}$ & $J=3-2$ \\ [0.4ex]
J16215472-2752053 & $0.12_{-0.01}^{+0.02}$ & $J=3-2$ \\ [0.4ex]
J16221532-2511349 & $0.18_{-0.01}^{+0.02}$ & $J=3-2$ \\ [0.4ex]
J16222982-2002472 & $0.13_{-0.05}^{+0.03}$ & $J=3-2$ \\ [0.4ex]
J16230761-2516339 & $0.26_{-0.03}^{+0.02}$ & $J=3-2$ \\ [0.4ex]
J16253798-1943162 & $0.12_{-0.01}^{+0.01}$ & $J=3-2$ \\ [0.4ex]
J16271273-2504017 & $0.07_{-0.06}^{+0.05}$ & $J=3-2$ \\ [0.4ex]
J16274905-2602437 & $0.002_{-0.001}^{+0.004}$ & $J=3-2$ \\ [0.4ex]
J16293267-2543291 & $0.18_{-0.03}^{+0.03}$ & $J=3-2$ \\ [0.4ex]
J16395577-2347355 & $0.05_{-0.01}^{+0.02}$ & $J=3-2$ \\ [0.4ex]
J16142029-1906481 & $0.13_{-0.03}^{+0.02}$ & $J=3-2$ \\ [0.4ex]
J16042165-2130284 & $0.03_{-0.02}^{+0.07}$ & $J=3-2$ \\ [0.4ex]
 \hline
 \hline
\end{tabular}
}
\vspace{-0.3cm}
\tablefoot{$\langle z/r \rangle$ values were collected from the works of \citet{Law_2022,Law_2023,Law_2024}, \cite{Galloway2025}, and references therein.}
\vspace{-0.9cm}
\label{Table:z/r_values}
\end{table*}
In Sect. \ref{sec:discussion}, we show and discuss the correlation between $ \langle z/r \rangle $ and $R_{90\%}$.

\vspace{-0.3cm}
\section{Discussion}
\label{sec:discussion}
\subsection{Brightness temperature - stellar luminosity correlation}
We report a correlation between the brightness temperature of the $^{12}$CO layer at a given radius and the stellar luminosity. In the following, we chose a distance of 10 au from the central star to illustrate the results, as shown in Fig. \ref{fig:Tb_L*_corr}. Since we are not directly constraining the $^{12}$CO temperature at 10 au, but rather inferring its value from a power-law parametric fit, we stress that the correlation should be interpreted as tracing how stellar luminosity affects the CO temperature structure in the outer disk, rather than a direct measurement of the temperature at 10 au. While here we choose to illustrate the results at 10 au because it is the quantity directly fit by the model, we verified that the correlation remains present when evaluating the temperature at larger radii (50–100 au).
\setcounter{figure}{7}
\begin{figure}
    \centering
    \includegraphics[width=\linewidth]{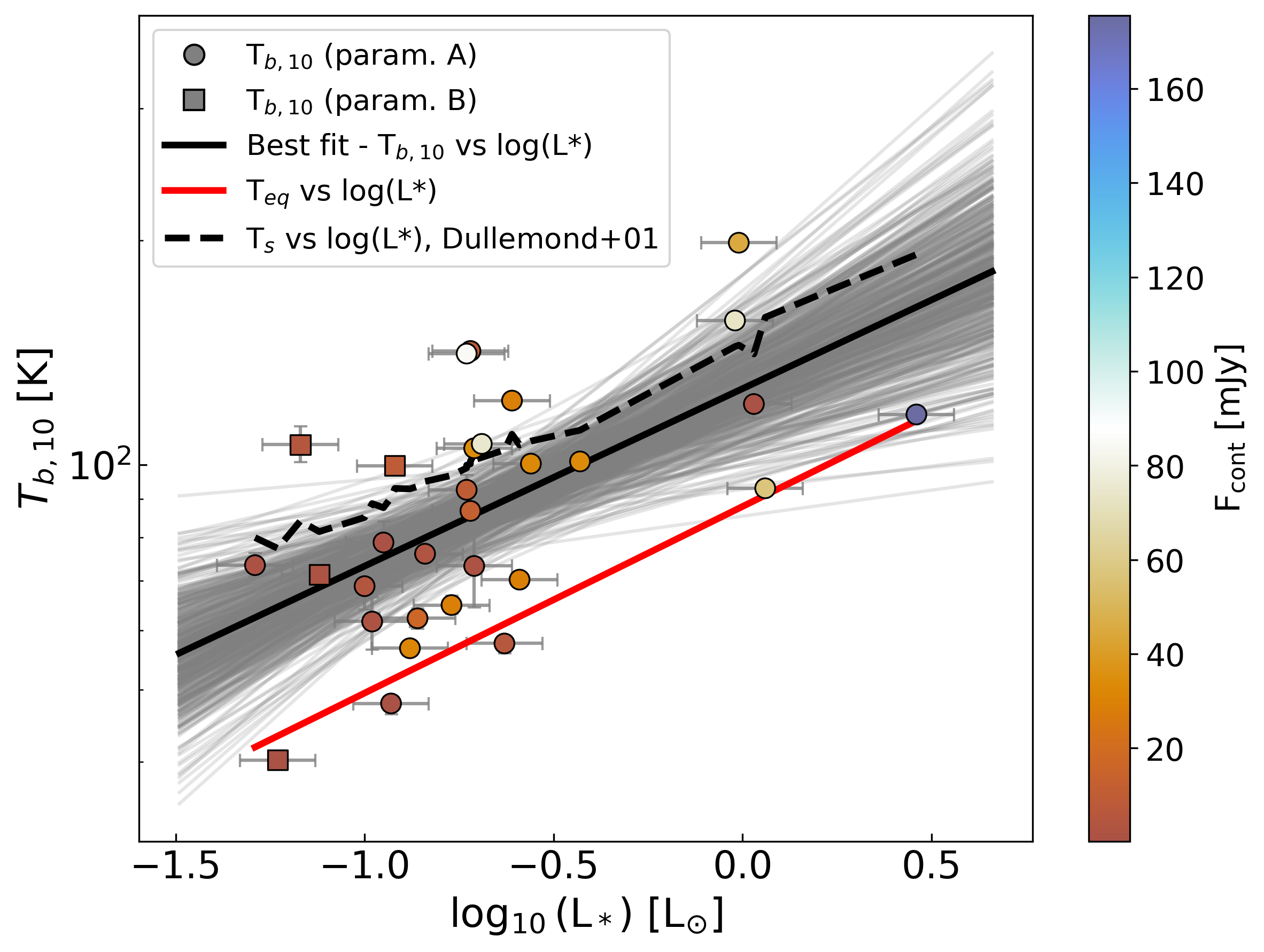}
    \caption{Correlation between $^{12}$CO brightness temperature at a distance of 10 au from the central star and stellar luminosity. The values of $T_{b,10}$ and $\log_{10}\bigl(\text{L}_*\bigr)$ were taken from Tables \ref{Table:Fits_A} and \ref{Table:Fits_B} and Empey et al., in prep., respectively. The values of $\text{T}_{b,10}$ were converted from cylindrical coordinates to spherical coordinates using the vertical-height-profile values reported in Table \ref{Table:Fits_A}. The values of F$_{\text{cont}}$ were taken from \cite{Carpenter2025}.}
    \vspace{-0.5cm}
    \label{fig:Tb_L*_corr}
\end{figure}
The correlation is $\log(T_{b,10}) = (2.10 \pm 0.05) + (0.24 \pm 0.07) \log(\text{L}_*)$, with a Pearson correlation coefficient of 0.6 in a range $[0.4, 0.7]$. This correlation is expected because more luminous stars irradiate the nearby circumstellar disk more strongly and therefore make it hotter. However, its slope and intercept have not been reported in the literature so far due to the lack of spatially resolved brightness temperature estimates on a large enough sample. We also compared these results to the total continuum flux analyzed in \cite{Carpenter2025}, but found no obvious trend. 

Additionally, we checked the presence of a correlation between $T_{b,10}$ and $R_{90\%}$ and found a positive correlation of $T_{b,10} = (70\pm12) + (0.2\pm0.1)\> R_{90\%}$ at a $3\sigma$ confidence level, and a Pearson coefficient of $r=0.4$. This hints toward a dependence of the CO brightness temperature on disk size, possibly reflecting underlying variations in disk mass and offering an alternative explanation in which the $T_{b,10}-L_*$ correlation reflects the correlation between the stellar luminosity and the stellar mass, which is in turn correlated to the disk mass. However, we found no empirical evidence in the literature for the presence of a correlation between gas disk size and gas disk mass. Pending a more detailed investigation on a larger sample that can disentangle the two explanations, in the following we proceed with the former since it provides a physical reason for the correlation we report.

The temperature values of the $^{12}$CO optically thick layer at a distance of 10 au from the central star were then compared to the temperature values of different dust grains at the same distance. First, as a reference, the red line shown in Fig. \ref{fig:Tb_L*_corr} represents the expected theoretical correlation of the equilibrium temperature, $T_{\text{eq}}$, at 10 au against the stellar luminosity. The equilibrium temperature is the temperature of a perfect blackbody dust grain irradiated by the central star: in this case, the slope of the correlation is 0.25. However, it is well-known that dust grains are not perfect blackbodies and that they absorb more efficiently than they re-emit (e.g., \citealt{Draine_2003}). For this reason, they undergo so-called superheating (i.e., not perfectly gray grains; see \citealt{Chiang_1997}).
Second, we evaluated the expected temperature of the optically thin-to-stellar irradiation-superheated dust layer reported in Eq. 7 of \cite{Dullemond2001}. We adopted a pure silicate dust composition for the Planck mean opacities\footnote{The Planck mean opacity $\kappa_P(T)$ is the average opacity of a material, weighted by the Planck blackbody function at a given temperature, $T$.}, evaluated using the \verb|dsharp_opac|\footnote{\url{https://github.com/birnstiel/dsharp_opac}} package presented in \cite{Birnstiel2018}: the result is shown as a dashed black line in Fig. \ref{fig:Tb_L*_corr}. 

There are two considerations we can make. The first is that the dashed black line could easily be interpreted as one of the gray best-fit lines in the background; this means that the dust at the optically thin-to-stellar-irradiation layer at 10 au has a similar temperature with respect to the optically thick $^{12}$CO layer that we extracted. The second is that the dust temperatures evaluated as a function of stellar luminosity are, on average, higher with respect to the best-fit correlation (solid black line). Since protoplanetary disks have a stratified temperature structure (\citealt{Chiang_1997}, \citealt{Oberg_2021}), our results empirically imply that, on average, the optically thin-to-stellar-irradiation dust layer lies higher up than the $^{12}$CO optically thick emission layer. A similar result was reported by \cite{Law_2024} and references therein, where a comparison between scattered-light images and the $^{12}$CO and $^{13}$CO layers is presented. We emphasize that, a priori, the stellar-irradiated layer and the CO emitting height could be totally different, given that they have vastly different opacities and the former is set by the optical depth from the star (e.g., \citealt{d'alessio_1998}, Bolchini et al., in prep.), while the latter comes from the observer (e.g., \citealt{Rosotti_2025}). The fact that we find they are similar (albeit with a slightly lower $^{12}$CO emitting layer) is an empirical finding. Anecdotally, this is often found in radiative transfer calculations, but we are not aware of theoretical works that explicitly looked at the relative heights of these layers.

Finally, the slope of the observed correlation can be used to empirically measure how the Planck mean opacities scale as a function of temperature. If we suppose that the Planck mean opacities scale as a linear function of the temperature ($\kappa_P \propto \alpha T^{\beta}$), after some simple algebra, Eq. 7 of \cite{Dullemond2001} reads 
\begin{equation}
    T_s^{4+\beta} = T_*^{\beta} L_* \frac{1}{16\pi\sigma R^2},
    \label{eq:Dullemond}
\end{equation}
where $T_s$ is the temperature of the optically thin-to-stellar-irradiation layer, $T_*$ is the stellar surface temperature, $\sigma$ is the Stefan-Boltzmann constant, and $R$ is the distance from the stellar surface. From Eq. \ref{eq:Dullemond}, we evaluated the exponent, $\beta$, which allowed us to retrieve the slope we evaluated for the correlation shown in Fig. \ref{fig:Tb_L*_corr}; from this we found $\beta\sim0.2$. As shown in Fig. 8 of \cite{Birnstiel2018}, the Planck mean opacities are expected to increase as a function of temperature: therefore, a slope $\beta \sim0.2$ is allowed. Moreover, in the temperature range of $10-200$ K, Fig. 8 of \cite{Birnstiel2018} shows that the mean dependence of the Planck mean opacities from the temperature is $\kappa_P \propto T^{0.45}$, the exponent of which is comparable to our $\beta\sim0.2$, even though it is not exactly equal. This result confirms the fact that the $^{12}$CO optically thick emission layer is more deeply embedded in the protoplanetary disk structure with respect to the optically thin to stellar irradiation dust layer. Finally, considering Fig. \ref{fig:Tb_L*_corr}, we report that the median slope of the black dashed line is comparable with that of the solid black line (slope $\sim0.22$). Although not particularly surprising, the existence of this correlation offers empirical evidence that the basic dust models we used are reasonable.

\vspace{-0.3cm}
\subsection{ $\langle z/r \rangle$  - disk size correlation}
\label{sec:z/r_corr}
\begin{figure}
    \centering
    \includegraphics[width=\linewidth]{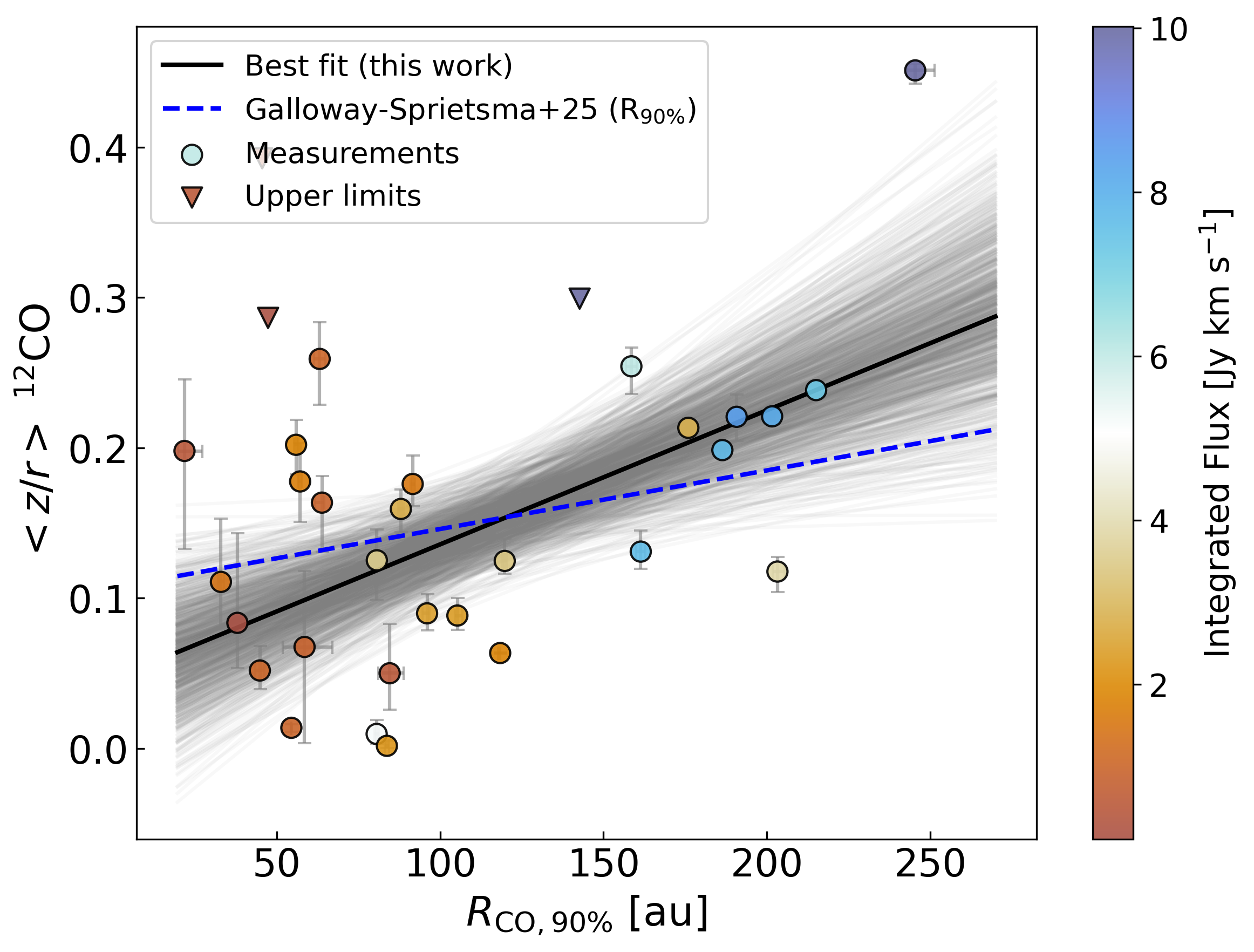}
    \vspace{-0.5cm}
    \caption{Correlation between $^{12}$CO $ \langle z/r \rangle $ and $R_{90\%}$. The values of $ \langle z/r \rangle $ were evaluated as discussed in Sect. \ref{sec:analysis}, while the values of R$_{90\%}$ were taken from Table \ref{Table:Fits_A}, and the integrated fluxes from Table \ref{Table:Fluxes}.}
    \vspace{-0.7cm}
    \label{fig:z/r_R90}
\end{figure}

\cite{Law2021} reported a correlation between $ \langle z/r \rangle $ and the gas-disk size $R_{90\%}$. We find that this still holds for our sample of compact disks orbiting T Tauri stars, as shown in Fig. \ref{fig:z/r_R90}.
Here, we show $ \langle z/r \rangle $ of our sample of disks as a function of $R_{90\%}$. The solid black curve represents the best fit, and in blue we over-plot the best-fit curve found by \cite{Galloway2025} for comparison. We also color-code the data points by their integrated fluxes to show how the most vertically extended disks are, on average, also the brightest. We removed J16042165-2130284 from the sample, as it is found to be a face-on disk (inc $\sim 6$°), for which the vertical-height extraction is very uncertain. We used J16142091-1906051, which is a known binary system, and J15583620-1946135 and J16132190-2136136, for which the resolution of the disk limits the derivation of reliable $ \langle z/r \rangle $ values, as upper limits. We stress that the correlation would still be significant if we had not made such arrangements.
The correlation is $ \langle z/r \rangle  = (0.05 \pm 0.03) + (0.0009 \pm 0.0003)$ $R_{90\%}$, and it is found with a Pearson correlation coefficient of $0.6$. Moreover, the fraction of posterior samples with an intercept $> 0$ is $\sim 0.999$, indicating a significance of $\gtrsim3\sigma$. This result is in agreement within the 68th percentile with the results shown in \cite{Galloway2025}, and it shows that this correlation, to date found only for extended, massive, and bright protoplanetary disks, holds for fainter and more compact disks. Therefore, we have reason to think that the correlation between $ \langle z/r \rangle $ and $R_{90\%}$ suggests something fundamental about all protoplanetary disks. However, it is not yet clear how this correlation is intimately related to the disk surface density. \cite{Paneque_2025} showed how the vertical structure of the disk as traced by CO is very sensitive to variations in disk mass by running thermochemical DALI models (\citealt{Bruderer_2013}) and accounting for hydrostatic equilibrium. More massive disks would therefore appear to be more extended both radially and vertically, and this would explain the observed correlation. However, CO depletion from the canonical ISM value $(\sim10^{-4})$ can also affect the vertical extent of the emitting layer (\citealt{Paneque_2025}, \citealt{Rosotti_2025}), and the radial cutoff may be affected by the surface-density profile (\citealt{Trapman_2020}). 

\vspace{-0.3cm}
\subsection{Multiple systems}
\label{sec:binaries}
\begin{figure}[t]
  \includegraphics[width=\linewidth]{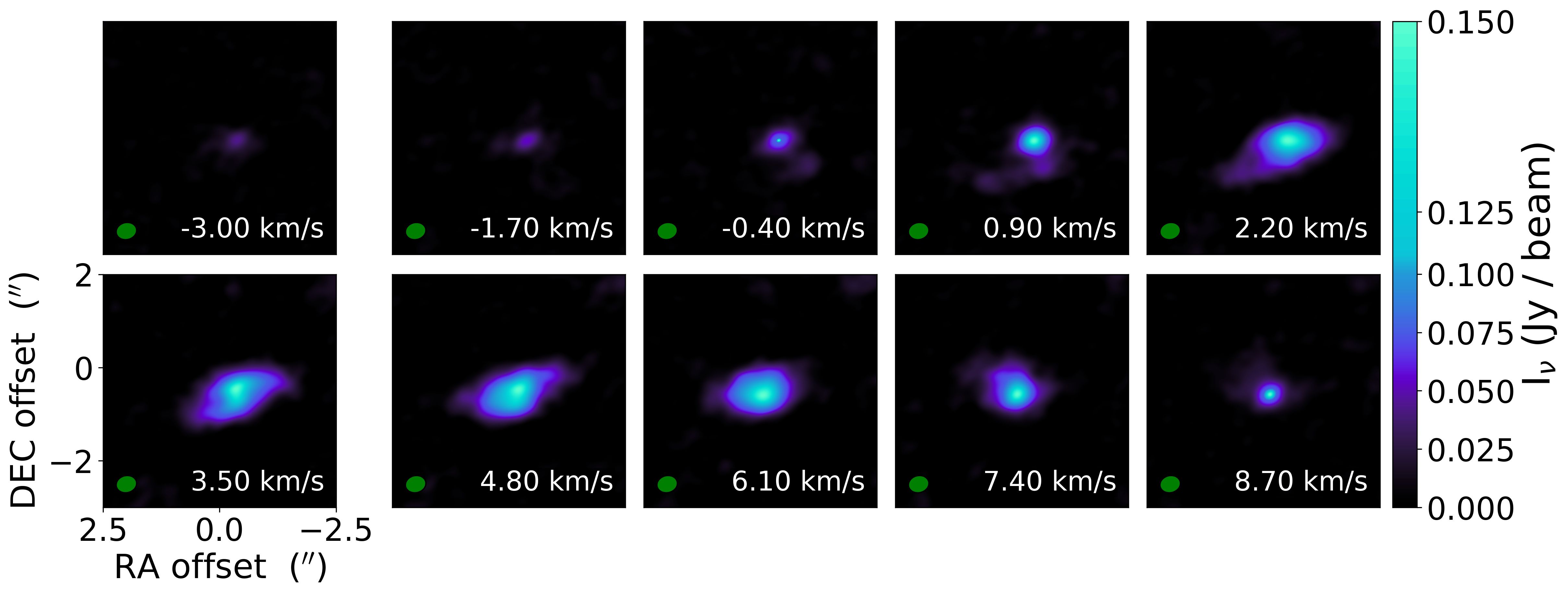}
  \vspace{-0.5cm}
  \caption{Channel maps of the binary system J16142091-1906051. The binary companion is placed at a RA offset of $\sim+1^{\prime\prime}$.}
  \label{fig:2MASS_J16142091-1906051}
  \vspace{-0.4cm}
\end{figure}

Among the sources we selected, we identified a group of multiple systems that show different characteristics. J16142091-1906051 shows a faint companion marginally visible in the continuum emission (see \citealt{Carpenter2025} for the continuum gallery), while in the gas a spiral feature is clearly visible (see Fig. \ref{fig:2MASS_J16142091-1906051}).
However, the keplerian velocity pattern for the main source is perfectly visible and still well behaved: therefore, we were able to successfully run our analysis for it, and we list the results in Table \ref{Table:Fits_A}, and show the residuals in Fig. \ref{fig:residuals_2} (f).

\begin{figure}[t]
  \includegraphics[width=\linewidth]{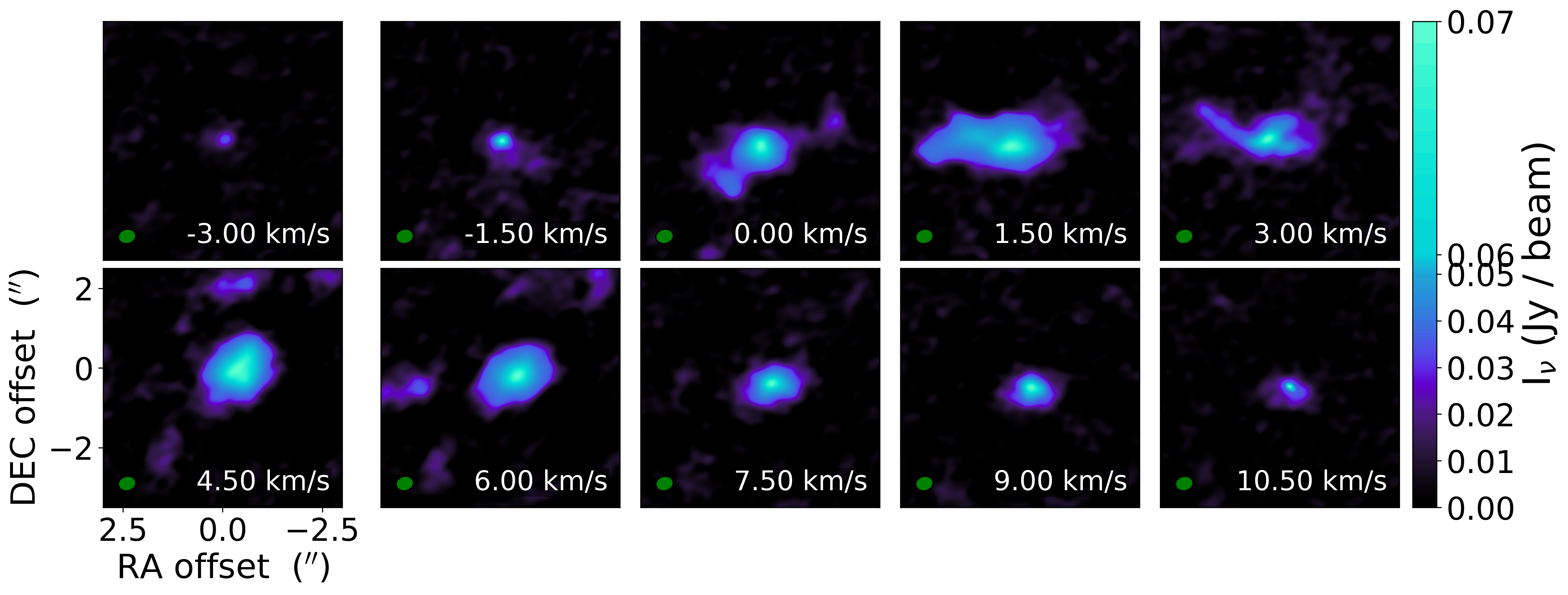}
  \vspace{-0.5cm}
  \caption{Channel maps of J16193570-1950426. A bridge of extended emission is visible in the channel maps at $v=1.5$ km s$^{-1}$, and a clear spiral pattern is visible from the channel map at $v=-0.4$ km s$^{-1}$ to the one at $v=6.1$ km s$^{-1}$.}
  \vspace{-0.6cm}
  \label{fig:2MASS_J16193570-1950426}
\end{figure}

J16193570-1950426, shows evidence for what is likely to be an emission inflow/outflow ($v=1.5$ km s$^{-1}$), which made it impossible to fit this source with the prescriptions adopted in this work (see Fig. \ref{fig:2MASS_J16193570-1950426}). For this reason, we report the possibility that this source hides at least one hidden wide separated companion, which might be connected to the main star through the diffuse material. This source was not listed as a binary in \cite{Carpenter2025} or anywhere else, but was reported as a continuum burst-accreting source from \cite{Cody_2017}. 

\begin{figure}[t]
  \includegraphics[width=\linewidth]{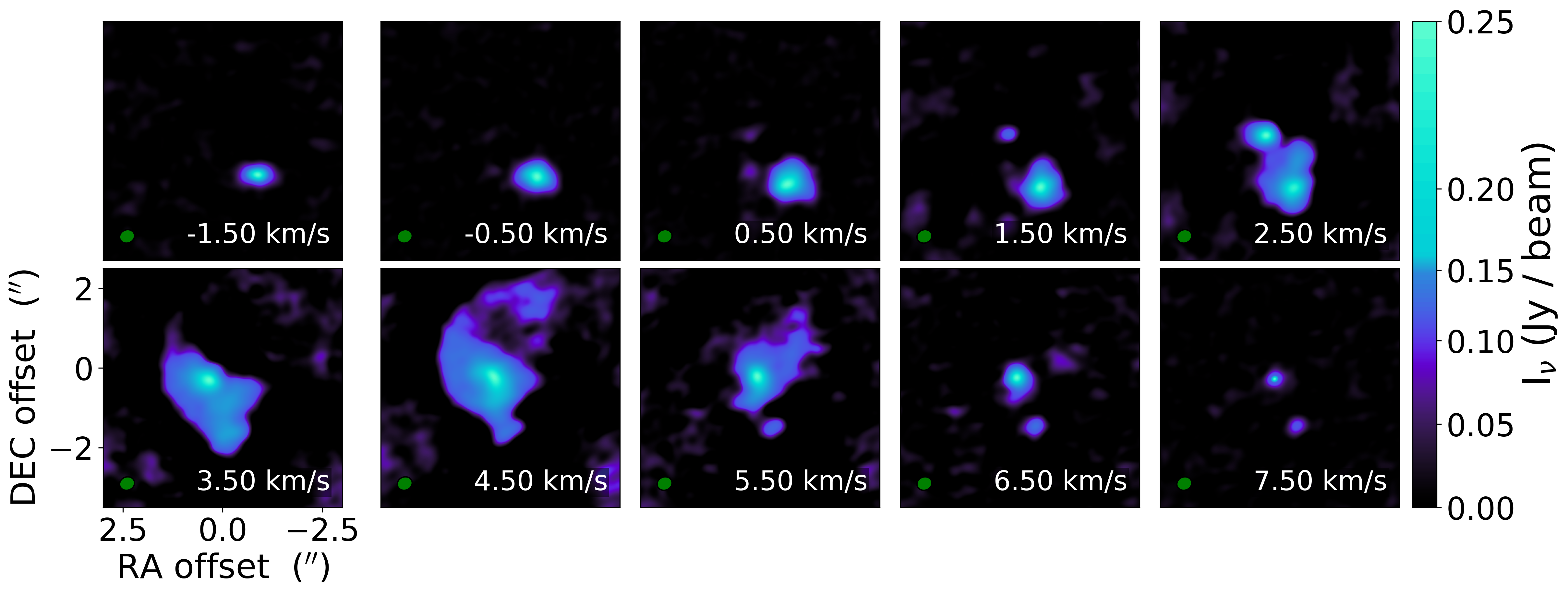}
  \vspace{-0.5cm}
  \caption{Channel maps of the binary system J16113134-1838259. A big tail of material is visible in the channel maps from $v=3.5 - 5.5$ km s$^{-1}$, showing that the two protoplanetary disks are actively interacting.}
  \vspace{-0.3cm}
  \label{fig:2MASS_J16113134-1838259}
\end{figure}

J16113134-1838259 (also known as AS 205, see e.g., \citealt{Andrews_2018}) shows what is very likely to be a fly-by (see Fig. \ref{fig:2MASS_J16113134-1838259}). For the latter, we tried to fit the two sources separately, but due to the lack of spectral resolution, we were unable to achieve a good fit. 

\begin{figure}[t]
  \includegraphics[width=\linewidth]{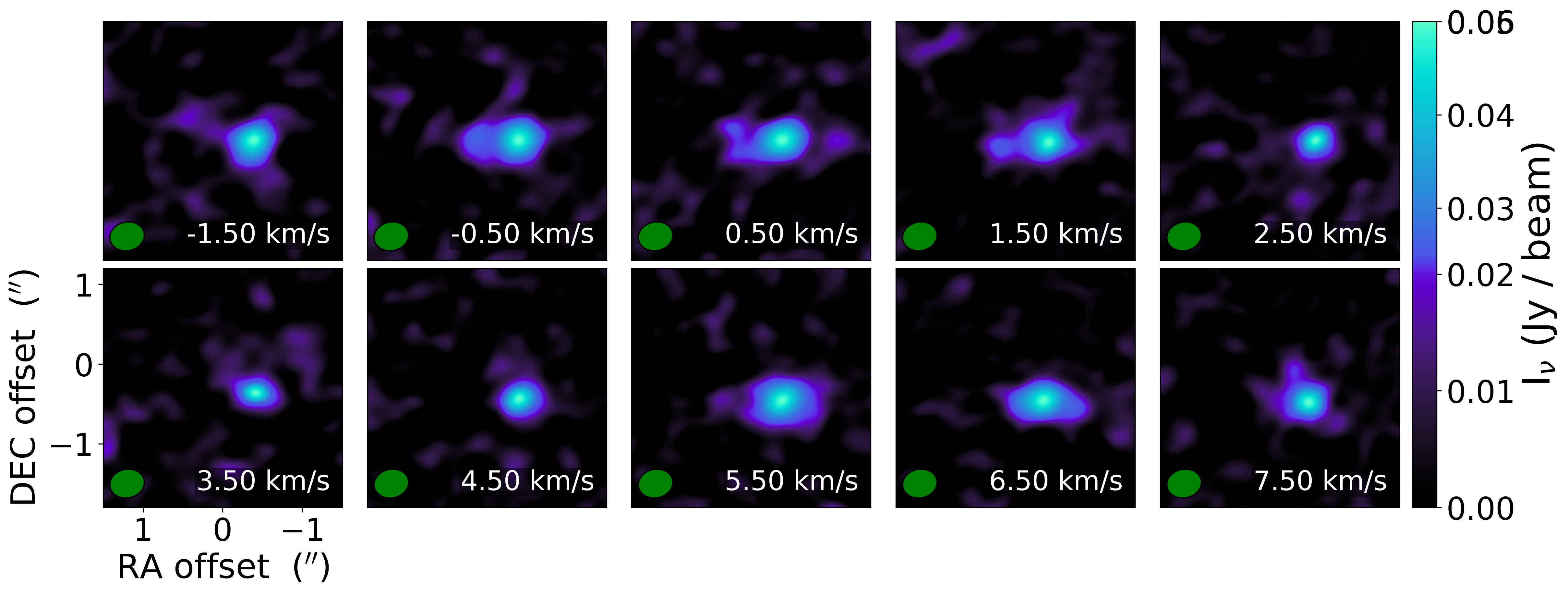}
  \vspace{-0.5cm}
  \caption{Channel maps of J16185382-2053182. Blobs of extended emission are visible in the channel maps at $v=2.3, 4, 5.7, 12.5,$ and $14.2$ km s$^{-1}$.}
  \vspace{-0.5cm}
  \label{fig:2MASS_J16185382-2053182}
\end{figure}

In Fig. \ref{fig:2MASS_J16185382-2053182}, we show the channel maps of J16185382-2053182. This system is marginally resolved, which makes it suitable for the analysis that we performed with \verb|csalt|, and we show the residuals in Fig. \ref{fig:residuals_3} (d). The results of the fit are listed in Table \ref{Table:Fits_A}.
This interesting source with a spectral type of M0 was originally removed from the sample of \cite{Carpenter2025}, as the disk classification of the star may be contaminated by another source in close proximity. We report the fact that we derived a stellar mass of $\sim2.5$ M$_{\odot}$ for this object, a result which is very discrepant with its associated spectral type and that could hint toward the presence of hidden companions.

\vspace{-0.5cm}
\section{Conclusions}
\label{sec:conclusions}
In this work, we fit the visibilities of the uniformly surveyed Upper Scorpius SFR following the sample of \cite{Carpenter2025} and using two new parametric prescriptions to extract the disk bulk properties. For this scope, we used the software \verb|csalt| (Andrews et al., in prep.) to model visibility spectra using two new parametric prescriptions. From our analysis, we derive the following conclusions.
\vspace{-0.1cm}
\begin{enumerate}[label=(\roman*)]
    \item We fit a total of 37 protoplanetary disks out of a sample of 83 $^{12}$CO $J=3-2$ detection, and we report that visibility fitting becomes difficult if the S/N of the integrated flux is $\lesssim10$.
    \item We report the correlation between the $^{12}$CO brightness temperature at a given radius and the $^{12}$CO $R_{90\%}$. We used this correlation to understand that the $^{12}$CO optically thick layer is more deeply embedded in the protoplanetary disk structure than the superheated-dust optically thin-to-stellar-irradiation layer.
    \item We extracted the $^{12}$CO $J=3 - 2$-emitting layers and $ \langle z/r \rangle $ values for "common" protoplanetary disks, deriving a median $ \langle z/r \rangle \sim 0.16$ for our sample, which is smaller than what was reported for larger and brighter disks (\citealt{Galloway2025}). Moreover, we find that the correlation between $\langle z/r \rangle$ and $R_{90\%}$ found by \cite{Law2021} for extended and bright disks still holds for the compact ones.
\end{enumerate}

%\vspace{-0.1cm}
\begin{acknowledgements}
We thank the anonymous referee for their important suggestions which improved the quality and clarity of the paper.
L.Z. is grateful to S. Facchini for some important informal suggestions for this work.
L.Z. and G.R. acknowledge support from the European Union (ERC Starting Grant DiscEvol, project number 101039651), and from Fondazione Cariplo, grant No. 2022-1217. C.F.M. is funded by the European Union (ERC, WANDA, 101039452). Views and opinions expressed are, however, those of the authors only and do not necessarily reflect those of the European Union or the European Research Council. Neither the European Union nor the granting authority can be held responsible for them. T.P.C. was supported by the Heising-Simons Foundation through a 51 Pegasi b Fellowship. Support for C.J.L. was provided by NASA through the NASA Hubble Fellowship grant No. HST-HF2-51535.001-A. Computational resources were provided by INDACO Platform, 
which is a project of High Performance Computing at the University of MILAN, and CINECA (\url{http://www.unimi.it}, \url{https://www.cineca.it}). This paper makes use of the following ALMA data: ADS/JAO.ALMA\#2011.0.00526.S, ADS/JAO.ALMA\#2012.1.00688.S, ADS/JAO.ALMA\#2013.1.00395.S, ADS/JAO.ALMA\#2018.1.00564.S. ALMA is a partnership of ESO (representing its member states), NSF (USA) and NINS (Japan), together with NRC (Canada), NSTC and ASIAA (Taiwan), and KASI (Republic of Korea), in cooperation with the Republic of Chile. The Joint ALMA Observatory is operated by ESO, AUI/NRAO and NAOJ. This paper also makes use of the following VLT/X-Shooter spectra: \#097.C-0378, \#0101.C-0866, \#105.2082.003, \#113.26NN.001, \#113.26NN.003, \#115.27XL.001.
\end{acknowledgements}

\vspace{-1cm}
\bibliographystyle{aa}
\bibliography{main}

\begin{appendix}
\section{Tables of best-fit parameters}
\label{appendix:best_fit}
In Tables \ref{Table:Fits_A}, \ref{Table:Fits_B}, we report the best-fit parameters derived from our analysis.

\section{Residual plots}
\label{appendix:residuals}
In Figs. \ref{fig:residuals_1}, \ref{fig:residuals_2}, \ref{fig:residuals_3}, \ref{fig:residuals_4}, \ref{fig:residuals_5}, \ref{fig:residuals_6}, we show the residual plots we created from the best-fit models following the procedure reported in Sect. \ref{sec:analysis}.

\section{The intrinsic dependence of the gas-disk size from its inclination and from the projected velocity field}
\label{appendix:R_vs_inc}
The disk geometrical framework of \verb|csalt| is designed for inference. The disk model is created in the disk reference frame in cylindrical coordinates and is then projected into the Cartesian plane of the sky for comparison with data. As a result, the parameters that describe the disk are generated and fit to the disk reference frame. Therefore, as described in Sec. \ref{sec:analysis}, the most convenient way to evaluate quantities such as $R_{90\%}$ is to directly integrate the best-fit peak intensity profile in cylindrical coordinates. 
This approach allows one to get rid of the `deprojection problem': as explained in Sec. \ref{sec:analysis}, the usual way to extract the gas-disk radii is based on velocity-integrated (moment 0) maps. However, the moment 0 integrated intensities use the projected velocity field as the weighting function: therefore, same disk properties will result in different integrands depending on the value of inc and $M_*$.
Moreover, protoplanetary disks are flared: as a result of this, the disk is seen to be larger than it actually is when considering high inclinations, because the back surface becomes particularly relevant at large radii. While the latter effect can be mitigated from high resolution observations where the height of the emission surface can be well constrained, this is in general not possible for surveys and for this reason existing surveys have not corrected for this factor (e.g., \citealt{Sanchis2021}, \citealt{Long2022}, \citealt{Trapman_2025}). In any case, the former effect remains unaccounted for.

\begin{figure}[t]
    \centering
    \includegraphics[width=\linewidth]{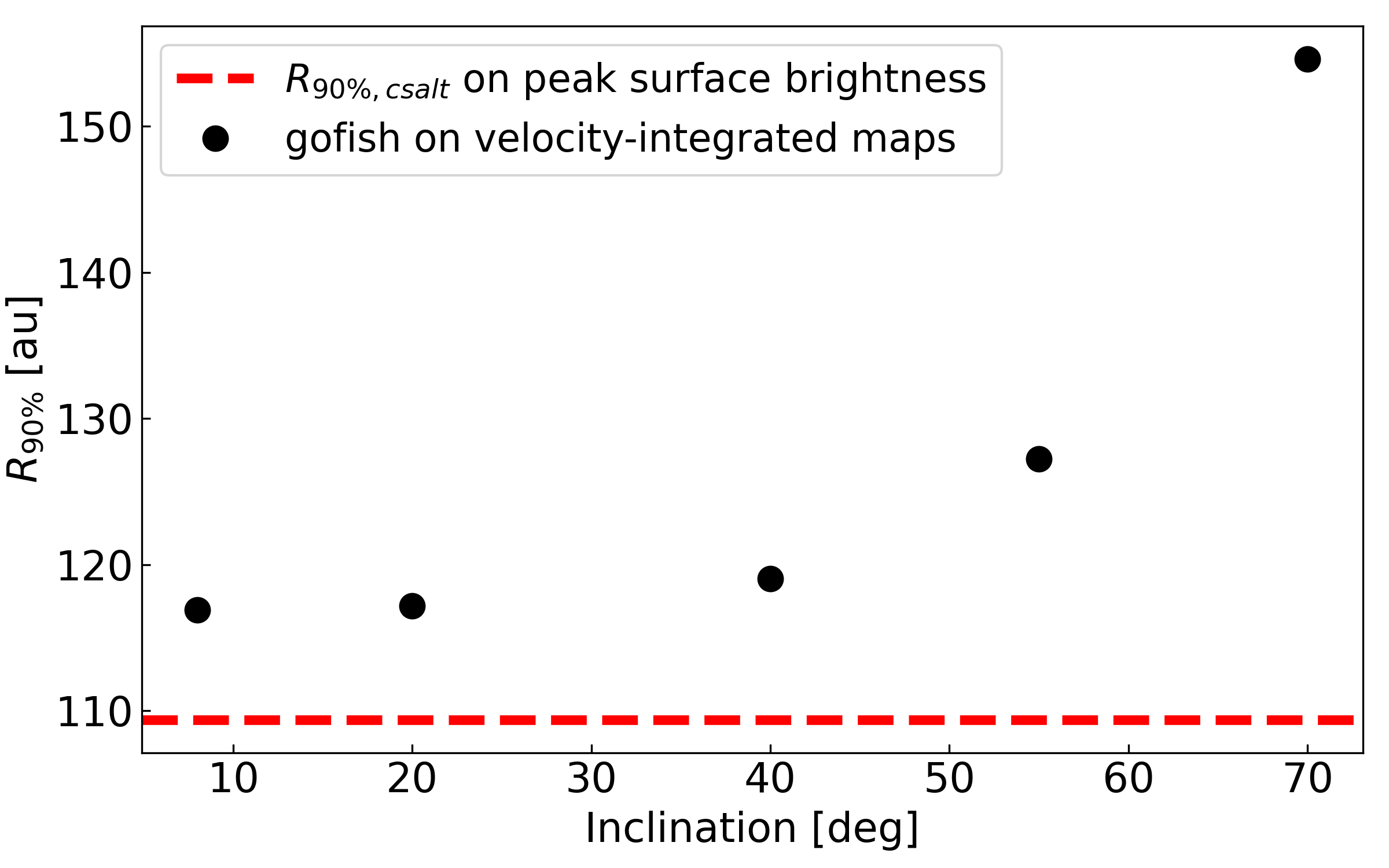}
    \caption{Illustrative example of the intrinsic dependence between the gas-disk size and its inclination when extracting gas-disk radii from velocity-integrated maps.}
    \label{fig:R_vs_inc}
\end{figure}

We present an illustrative example in Fig. \ref{fig:R_vs_inc}.
For this example, we generated five model cubes in the velocity range $[-10; 20]$ km s$^{-1}$ with a spectral resolution of $0.429$ km s$^{-1}$, to be consistent with the ALMA observations used in this work, using the dedicated framework of \verb|csalt|. The model parameters were chosen following the best-fit model parameters of J16213469-2612269 (for which we derived $ \langle z/r \rangle  \sim 0.14$), which we showed in Sec. \ref{sec:analysis}, and were kept fixed, while we set the inclination to five different values: 8°, 20°, 40°, 55°, 70°.
Then, we created the integrated moment maps of our models and convolved them with the beam of our observations, which in the case of J16213469-2612269 was $(0.4^{\prime\prime}, 0.3^{\prime\prime})$, with $\text{PA}=-81.5$°.
Then, using \texttt{gofish}, we extracted the radial profile of our convolved models and integrated it to derive $R_{90\%}$, the radius that contains $90\%$ of the total disk flux. Fig. \ref{fig:R_vs_inc} shows the effect of extracting $R_{90\%}$ on velocity-integrated channel maps without keeping into consideration the flared nature of protoplanetary disks: the radius of a very inclined disk (70°) with an aspect ratio of $0.14$ can be $\sim 40\%$ greater than the true value.

\section{Data and model-integrated fluxes}
\label{appendix:fluxes}
To extract the fluxes from the data, we first generated moment 0 maps using a bandwidth of $\pm15$ km s$^{-1}$ centered around the systemic velocity: this bandwidth allows us to include the entire disk emission for all sources. For this step, we used bettermoments (\citealt{Foreman-Mackey_2018}). Secondly, we evaluated geometric masks using the best-fit results reported in Tables \ref{Table:Fits_A}, \ref{Table:Fits_B}, and summed the emission within the mask for each moment map. The mask size was set to match $R_{90\%}+0.2 ^{\prime\prime}$. For the flux uncertainties, we moved the mask around the moment 0 maps in 8 different locations that were not enclosing any disk emission and not too far away from the image center, determined the flux in each of these locations and finally computed the standard deviation of the fluxes to obtain the integrated disk flux uncertainty. The flux from the models was computed in a similar way: we first generated datacubes using the best-fit parameters reported in Tables \ref{Table:Fits_A}, \ref{Table:Fits_B} on the $\pm15$ km s$^{-1}$ bandwidth, with the same spectral resolution as the observations. We then computed the disk fluxes using the same mask as for observations, and bootsrapped an uncertainty on the fluxes using the posterior distribution of the generated fluxes.

\section{Vertical heights}
\label{appendix:vertical_heights}
\begin{figure*}
    \centering
    \includegraphics[width=\linewidth]{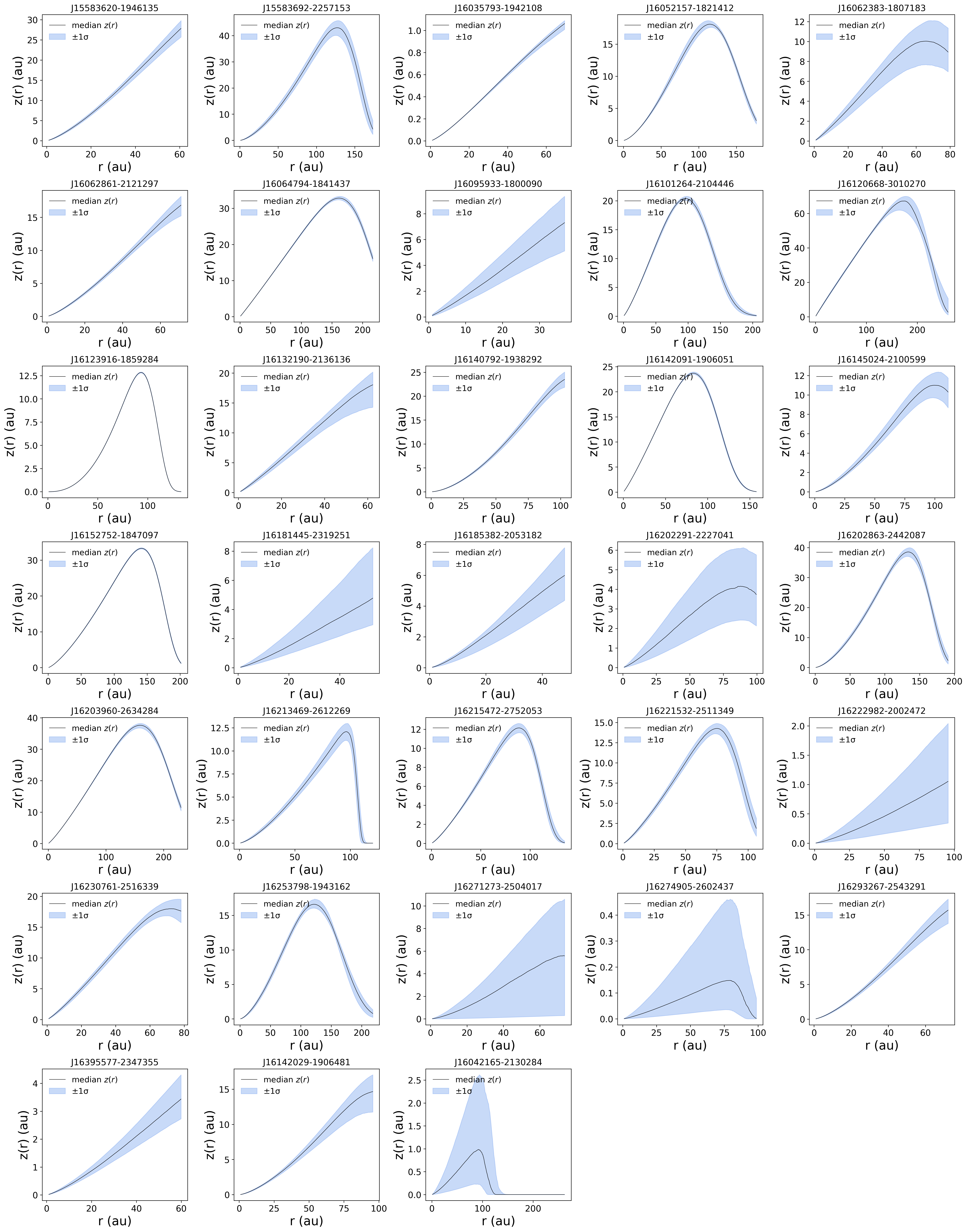}
  \caption{Vertical heights of the 33 disks fit with the prescription A. We created the profiles according to Eq. \ref{eq:z(r)} using all the random draws of the MCMC chains of $z_1$, ${\psi}_z$, $r_z$, and $\phi_z$, evaluating the associated $z(r)$ profiles, and taking their median value and the $16^{\text{th}}$ and $84^{\text{th}}$ percentiles for the associated uncertainties. }
  \label{fig:all_disks_zr}
\end{figure*}

We show in Fig. \ref{fig:all_disks_zr} shows the vertical height profiles we derived in this work using model A (see Sect. \ref{sec:analysis}) as a function of radius. We truncated each profile at a distance $R_{90\%}+0.2^{\prime\prime}$ from the central star. 

\addcontentsline{toc}{section}{Appendix A (tables)}
\renewcommand{\thetable}{A.\arabic{table}}

\clearpage
\begin{landscape}
\begin{table}
\setcounter{table}{0}
\caption{Fit results using the A parametrization prescription.}
\centering
\resizebox{\linewidth}{!}{
\begin{tabular}{l c c c c c c c c c c c c c c c c c c c c c} 
 \hline
 \hline
 \noalign{\vskip 0.03in} 
 Source (2MASS) & dist [pc] & $R_{90\%}$ [au] & $R_{68\%}$ [au] & inc [°] & PA [°] & $M_*$ [M$_{\odot}$] & $r_z$ [au] & $z_1$ [$^{\prime\prime}$]& ${\psi}_z$ & T$_{10}$ [K] & $q$ & $r_\tau$ [au] & $\phi_\tau$ & $\tau_{10}$ & ${\psi}_\tau$ & $\phi_z$ & v$_{sys}$ [ms$^{-1}$] & d$_{\text{RA}} [^{\prime\prime}]$ & d$_{\text{DEC}} [^{\prime\prime}]$ & $\log(\text{L}_*) \> [\text{L}_{\odot}]$\\ [0.4ex]
 \hline
 \hline
 \noalign{\vskip 0.03in} 
J15583620-1946135 & 155.6 & $45.6_{-0.8}^{+0.8}$ & $36.5_{-0.8}^{+0.8}$ & $45.1_{-1.5}^{+1.6}$ & $-23.3_{-1.0}^{+1.1}$ & $0.1_{-0.0}^{+0.0}$ & $112.2_{-5.5}^{+5.0}$ & $0.7_{-0.1}^{+0.1}$ & $1.3_{-0.1}^{+0.1}$ & $47.0_{-1.7}^{+1.2}$ & $-0.6_{-0.0}^{+0.0}$ & $45.0_{-1.8}^{+1.1}$ & $37.9_{-13.8}^{+8.6}$ & $9.3_{-0.9}^{+1.4}$ & $0.0_{-1.7}^{+0.8}$ & $5.9_{-2.3}^{+2.9}$ & $6768.0_{-10.5}^{+11.4}$ & $-0.3_{-0.0}^{+0.0}$ & $-0.4_{-0.0}^{+0.0}$ & $-0.9\pm0.1$ \\ [0.4ex]
J15583692-2257153 & 166.5 & $158.5_{-1.6}^{+1.6}$ & $125.2_{-1.6}^{+0.8}$ & $26.3_{-1.3}^{+1.4}$ & $168.5_{-0.4}^{+0.5}$ & $0.7_{-0.1}^{+0.1}$ & $154.6_{-3.4}^{+2.6}$ & $0.5_{-0.0}^{+0.0}$ & $1.5_{-0.1}^{+0.1}$ & $116.4_{-2.2}^{+2.1}$ & $-0.6_{-0.0}^{+0.0}$ & $133.4_{-2.2}^{+2.0}$ & $9.5_{-0.6}^{+0.3}$ & $4.9_{-0.1}^{+0.0}$ & $-0.0_{-0.1}^{+0.0}$ & $9.3_{-0.9}^{+0.5}$ & $7634.1_{-5.1}^{+5.3}$ & $-0.3_{-0.0}^{+0.0}$ & $-0.2_{-0.0}^{+0.0}$ & $0.5\pm0.1$ \\ [0.4ex]
J16035793-1942108 & 152.9 & $54.5_{-0.0}^{+0.0}$ & $44.0_{-0.0}^{+0.0}$ & $51.3_{-0.0}^{+0.0}$ & $1.9_{-0.0}^{+0.0}$ & $0.5_{-0.0}^{+0.0}$ & $155.9_{-18.2}^{+11.0}$ & $0.0_{-0.0}^{+0.0}$ & $1.3_{-0.1}^{+0.0}$ & $76.0_{-0.1}^{+0.2}$ & $-0.6_{-0.0}^{+0.0}$ & $52.0_{-0.1}^{+0.2}$ & $31.8_{-0.7}^{+1.3}$ & $17.7_{-0.0}^{+0.0}$ & $-0.0_{-0.0}^{+0.0}$ & $1.7_{-0.5}^{+0.9}$ & $7264.9_{-0.5}^{+0.7}$ & $-0.2_{-0.0}^{+0.0}$ & $-0.4_{-0.0}^{+0.0}$ & $-0.8\pm0.1$ \\ [0.4ex]
J16052157-1821412 & 148.9 & $161.4_{-1.5}^{+1.5}$ & $126.5_{-0.7}^{+0.7}$ & $69.2_{-0.1}^{+0.1}$ & $-23.6_{-0.2}^{+0.3}$ & $1.1_{-0.0}^{+0.0}$ & $145.8_{-3.5}^{+2.3}$ & $0.2_{-0.0}^{+0.0}$ & $1.5_{-0.0}^{+0.0}$ & $198.0_{-1.8}^{+1.6}$ & $-0.7_{-0.0}^{+0.0}$ & $133.3_{-3.5}^{+5.5}$ & $8.0_{-0.7}^{+1.2}$ & $4.9_{-0.1}^{+0.0}$ & $-0.8_{-0.3}^{+0.2}$ & $5.3_{-0.5}^{+0.4}$ & $5437.6_{-4.1}^{+8.3}$ & $-0.0_{-0.0}^{+0.0}$ & $-0.6_{-0.0}^{+0.0}$ & $0.0\pm0.1$\\ [0.4ex]
J16062383-1807183 & 148.6 & $63.9_{-2.2}^{+0.7}$ & $51.5_{-1.5}^{+0.7}$ & $32.5_{-1.9}^{+1.6}$ & $191.6_{-1.2}^{+1.1}$ & $0.1_{-0.0}^{+0.0}$ & $90.6_{-5.1}^{+6.1}$ & $0.2_{-0.0}^{+0.0}$ & $1.1_{-0.1}^{+0.1}$ & $43.1_{-1.0}^{+1.0}$ & $-0.5_{-0.0}^{+0.0}$ & $52.9_{-1.8}^{+1.4}$ & $9.4_{-0.8}^{+0.4}$ & $4.9_{-0.1}^{+0.1}$ & $-0.1_{-0.1}^{+0.1}$ & $5.0_{-1.9}^{+2.6}$ & $5652.5_{-6.9}^{+7.0}$ & $-0.4_{-0.0}^{+0.0}$ & $-0.5_{-0.0}^{+0.0}$ & $-$ \\ [0.4ex]
J16062861-2121297 & 139.7 & $55.9_{-0.7}^{+0.7}$ & $43.0_{-0.7}^{+0.0}$ & $33.6_{-1.4}^{+1.2}$ & $252.1_{-0.2}^{+0.2}$ & $0.5_{-0.0}^{+0.0}$ & $108.7_{-5.7}^{+6.2}$ & $0.3_{-0.0}^{+0.0}$ & $1.3_{-0.0}^{+0.0}$ & $86.0_{-1.1}^{+1.4}$ & $-0.8_{-0.0}^{+0.0}$ & $55.2_{-1.3}^{+1.2}$ & $21.4_{-3.7}^{+3.9}$ & $4.0_{-0.1}^{+0.0}$ & $-0.0_{-0.1}^{+0.0}$ & $6.5_{-2.1}^{+2.3}$ & $3614.5_{-8.6}^{+8.2}$ & $-0.3_{-0.0}^{+0.0}$ & $-0.5_{-0.0}^{+0.0}$ & $-0.7\pm0.1$ \\ [0.4ex]
J16064794-1841437 & 155.8 & $201.6_{-0.8}^{+0.8}$ & $158.2_{-0.0}^{+0.8}$ & $58.9_{-0.1}^{+0.1}$ & $199.8_{-0.1}^{+0.1}$ & $0.9_{-0.0}^{+0.0}$ & $211.9_{-1.3}^{+1.4}$ & $0.2_{-0.0}^{+0.0}$ & $1.0_{-0.0}^{+0.0}$ & $130.0_{-0.7}^{+0.6}$ & $-0.6_{-0.0}^{+0.0}$ & $170.9_{-0.9}^{+0.9}$ & $9.7_{-0.2}^{+0.2}$ & $5.0_{-0.0}^{+0.0}$ & $-0.0_{-0.0}^{+0.0}$ & $7.1_{-0.4}^{+0.5}$ & $7090.8_{-2.7}^{+2.4}$ & $-0.2_{-0.0}^{+0.0}$ & $-0.4_{-0.0}^{+0.0}$ & $-$ \\ [0.4ex]
J16095933-1800090 & 135.3 & $21.8_{-2.0}^{+5.4}$ & $15.2_{-1.3}^{+2.1}$ & $30.1_{-3.5}^{+3.3}$ & $-61.1_{-6.7}^{+5.5}$ & $0.2_{-0.0}^{+0.1}$ & $77.2_{-8.2}^{+10.4}$ & $0.3_{-0.1}^{+0.1}$ & $1.1_{-0.1}^{+0.1}$ & $77.7_{-6.4}^{+5.5}$ & $-0.8_{-0.1}^{+0.1}$ & $79.5_{-10.1}^{+9.5}$ & $3.9_{-3.0}^{+3.3}$ & $1.2_{-0.1}^{+0.2}$ & $-4.6_{-0.3}^{+0.6}$ & $5.6_{-2.9}^{+3.0}$ & $3794.1_{-91.5}^{+82.2}$ & $-0.3_{-0.0}^{+0.0}$ & $-0.3_{-0.0}^{+0.0}$ & $-1.0\pm0.1$ \\ [0.4ex]
J16101264-2104446 & 153.8 & $190.7_{-1.5}^{+0.8}$ & $150.2_{-0.8}^{+0.8}$ & $57.1_{-0.2}^{+0.1}$ & $33.9_{-0.1}^{+0.1}$ & $1.2_{-0.0}^{+0.0}$ & $129.5_{-2.7}^{+3.1}$ & $0.3_{-0.0}^{+0.0}$ & $1.2_{-0.0}^{+0.0}$ & $154.4_{-1.0}^{+1.1}$ & $-0.7_{-0.0}^{+0.0}$ & $160.2_{-3.9}^{+3.0}$ & $8.7_{-0.9}^{+0.7}$ & $4.0_{-0.0}^{+0.0}$ & $-0.0_{-0.0}^{+0.0}$ & $4.1_{-0.3}^{+0.6}$ & $6284.5_{-2.3}^{+2.4}$ & $-0.2_{-0.0}^{+0.0}$ & $-0.3_{-0.0}^{+0.0}$ & $0.0\pm0.1$ \\ [0.4ex]
J16120668-3010270 & 131.9 & $245.4_{-2.6}^{+5.9}$ & $198.4_{-1.9}^{+4.6}$ & $43.7_{-0.6}^{+0.5}$ & $43.5_{-0.3}^{+0.2}$ & $0.8_{-0.0}^{+0.0}$ & $224.4_{-4.8}^{+2.0}$ & $0.4_{-0.0}^{+0.0}$ & $1.0_{-0.0}^{+0.0}$ & $64.4_{-1.1}^{+1.0}$ & $-0.5_{-0.0}^{+0.0}$ & $199.3_{-5.3}^{+4.0}$ & $8.7_{-1.2}^{+0.8}$ & $5.0_{-0.0}^{+0.0}$ & $-0.0_{-0.0}^{+0.0}$ & $9.0_{-3.2}^{+0.7}$ & $4450.9_{-4.3}^{+4.0}$ & $-0.3_{-0.0}^{+0.0}$ & $-0.4_{-0.0}^{+0.0}$ & $-$ \\ [0.4ex]
J16123916-1859284 & 134.7 & $118.4_{-0.7}^{+0.0}$ & $87.5_{-0.0}^{+0.0}$ & $59.6_{-0.0}^{+0.0}$ & $286.2_{-0.1}^{+0.0}$ & $0.6_{-0.0}^{+0.0}$ & $107.6_{-0.0}^{+0.0}$ & $0.3_{-0.0}^{+0.0}$ & $2.4_{-0.0}^{+0.0}$ & $142.0_{-0.3}^{+0.2}$ & $-0.9_{-0.0}^{+0.0}$ & $96.6_{-0.3}^{+0.3}$ & $4.5_{-0.0}^{+0.0}$ & $1.6_{-0.0}^{+0.0}$ & $-0.0_{-0.0}^{+0.0}$ & $10.0_{-0.0}^{+0.0}$ & $4452.9_{-1.4}^{+1.1}$ & $-0.2_{-0.0}^{+0.0}$ & $-0.3_{-0.0}^{+0.0}$ & $-0.7\pm0.1$ \\ [0.4ex]
J16132190-2136136 & 144.8 & $47.4_{-1.5}^{+2.1}$ & $32.9_{-1.1}^{+1.1}$ & $39.2_{-1.9}^{+1.9}$ & $197.0_{-2.3}^{+1.5}$ & $0.6_{-0.1}^{+0.1}$ & $100.1_{-5.6}^{+5.0}$ & $0.4_{-0.0}^{+0.0}$ & $1.1_{-0.1}^{+0.1}$ & $61.2_{-2.0}^{+2.1}$ & $-0.5_{-0.0}^{+0.0}$ & $98.2_{-5.3}^{+6.2}$ & $2.8_{-1.2}^{+2.0}$ & $2.7_{-0.2}^{+0.2}$ & $-4.8_{-0.2}^{+0.3}$ & $5.8_{-3.0}^{+2.7}$ & $5612.7_{-30.2}^{+41.3}$ & $-0.1_{-0.0}^{+0.0}$ & $-0.6_{-0.0}^{+0.0}$ & $-0.9\pm0.1$ \\ [0.4ex]
J16140792-1938292 & 159.5 & $88.0_{-0.8}^{+0.8}$ & $71.7_{-0.8}^{+0.0}$ & $37.3_{-0.7}^{+0.7}$ & $-4.4_{-0.2}^{+0.2}$ & $1.2_{-0.0}^{+0.0}$ & $135.4_{-6.7}^{+6.1}$ & $0.4_{-0.0}^{+0.0}$ & $1.7_{-0.1}^{+0.0}$ & $92.9_{-1.6}^{+1.3}$ & $-0.5_{-0.0}^{+0.0}$ & $75.1_{-0.8}^{+0.5}$ & $9.8_{-0.4}^{+0.2}$ & $4.0_{-0.0}^{+0.0}$ & $-0.0_{-0.0}^{+0.0}$ & $7.8_{-1.5}^{+1.3}$ & $7235.2_{-4.4}^{+5.0}$ & $-0.1_{-0.0}^{+0.0}$ & $-0.3_{-0.0}^{+0.0}$ & $0.0\pm0.1$ \\ [0.4ex]
J16142091-1906051 & 135.3 & $142.7_{-2.0}^{+1.4}$ & $102.4_{-1.3}^{+1.3}$ & $62.6_{-0.2}^{+0.2}$ & $82.2_{-0.1}^{+0.1}$ & $0.6_{-0.0}^{+0.0}$ & $110.9_{-0.6}^{+0.5}$ & $0.4_{-0.0}^{+0.0}$ & $1.1_{-0.0}^{+0.0}$ & $163.6_{-0.9}^{+0.8}$ & $-0.9_{-0.0}^{+0.0}$ & $109.8_{-2.2}^{+2.4}$ & $6.3_{-0.3}^{+0.3}$ & $5.0_{-0.0}^{+0.0}$ & $-0.0_{-0.0}^{+0.0}$ & $5.3_{-0.1}^{+0.1}$ & $4084.2_{-6.3}^{+7.6}$ & $-0.3_{-0.0}^{+0.0}$ & $-0.6_{-0.0}^{+0.0}$ & $0.0\pm0.1$ \\ [0.4ex]
J16145024-2100599 & 138.5 & $96.0_{-1.4}^{+0.8}$ & $71.7_{-0.7}^{+0.7}$ & $43.1_{-0.7}^{+0.9}$ & $163.2_{-0.3}^{+0.3}$ & $0.7_{-0.0}^{+0.0}$ & $125.2_{-4.5}^{+5.1}$ & $0.2_{-0.0}^{+0.0}$ & $1.5_{-0.1}^{+0.1}$ & $100.3_{-2.3}^{+1.5}$ & $-0.8_{-0.0}^{+0.0}$ & $85.1_{-1.2}^{+1.1}$ & $9.6_{-0.5}^{+0.3}$ & $3.9_{-0.1}^{+0.0}$ & $-0.1_{-0.1}^{+0.0}$ & $7.7_{-1.7}^{+1.6}$ & $3886.0_{-5.8}^{+6.0}$ & $-0.1_{-0.0}^{+0.0}$ & $-0.6_{-0.0}^{+0.0}$ & $-0.6\pm0.1$ \\ [0.4ex]
J16152752-1847097 & 137.3 & $186.4_{-0.7}^{+0.7}$ & $150.2_{-0.7}^{+0.0}$ & $65.5_{-0.1}^{+0.1}$ & $162.2_{-0.1}^{+0.1}$ & $0.6_{-0.0}^{+0.0}$ & $174.2_{-0.3}^{+0.3}$ & $0.3_{-0.0}^{+0.0}$ & $1.3_{-0.0}^{+0.0}$ & $106.3_{-0.7}^{+0.4}$ & $-0.6_{-0.0}^{+0.0}$ & $157.3_{-0.7}^{+0.6}$ & $9.9_{-0.1}^{+0.1}$ & $5.0_{-0.0}^{+0.0}$ & $-0.0_{-0.0}^{+0.0}$ & $9.4_{-0.2}^{+0.1}$ & $4555.0_{-2.0}^{+2.4}$ & $-0.2_{-0.0}^{+0.0}$ & $-0.4_{-0.0}^{+0.0}$ & $-0.7\pm0.1$ \\ [0.4ex]
J16181445-2319251 & 137.7 & $38.0_{-3.0}^{+3.0}$ & $23.5_{-1.3}^{+2.0}$ & $45.2_{-1.4}^{+2.1}$ & $154.2_{-1.8}^{+1.7}$ & $0.2_{-0.0}^{+0.0}$ & $111.4_{-5.4}^{+5.4}$ & $0.1_{-0.0}^{+0.1}$ & $1.3_{-0.1}^{+0.1}$ & $68.7_{-4.7}^{+5.2}$ & $-0.6_{-0.1}^{+0.1}$ & $110.0_{-4.9}^{+5.2}$ & $3.5_{-2.1}^{+3.5}$ & $1.4_{-0.2}^{+0.2}$ & $-3.7_{-0.2}^{+0.3}$ & $4.5_{-2.4}^{+3.3}$ & $3881.5_{-39.0}^{+33.9}$ & $-0.1_{-0.0}^{+0.0}$ & $-0.5_{-0.0}^{+0.0}$ & $-1.0\pm0.1$ \\ [0.4ex]
J16185382-2053182 & 131.0 & $32.9_{-1.0}^{+1.0}$ & $22.4_{-0.6}^{+0.6}$ & $34.5_{-2.8}^{+3.2}$ & $173.2_{-1.6}^{+5.3}$ & $2.5_{-0.3}^{+0.5}$ & $105.4_{-7.5}^{+6.8}$ & $0.2_{-0.0}^{+0.1}$ & $1.4_{-0.1}^{+0.1}$ & $120.3_{-4.9}^{+6.2}$ & $-0.8_{-0.0}^{+0.0}$ & $100.1_{-6.2}^{+6.2}$ & $2.6_{-1.0}^{+1.3}$ & $2.1_{-0.1}^{+0.1}$ & $-4.8_{-0.1}^{+0.2}$ & $6.6_{-2.7}^{+2.0}$ & $8532.8_{-63.4}^{+55.0}$ & $-0.4_{-0.0}^{+0.0}$ & $-0.4_{-0.0}^{+0.0}$ & $0.0\pm0.1$ \\ [0.4ex]
J16202291-2227041 & 142.0 & $84.6_{-3.5}^{+4.3}$ & $65.9_{-2.1}^{+2.8}$ & $73.0_{-1.3}^{+1.1}$ & $72.2_{-1.0}^{+1.0}$ & $0.4_{-0.0}^{+0.0}$ & $110.5_{-5.6}^{+5.3}$ & $0.1_{-0.0}^{+0.0}$ & $1.3_{-0.1}^{+0.1}$ & $61.6_{-5.1}^{+4.8}$ & $-0.5_{-0.1}^{+0.0}$ & $86.8_{-5.3}^{+6.8}$ & $7.6_{-2.4}^{+1.5}$ & $1.6_{-0.4}^{+1.0}$ & $-1.3_{-1.2}^{+0.6}$ & $5.4_{-3.6}^{+3.0}$ & $3174.2_{-33.8}^{+36.6}$ & $-0.2_{-0.0}^{+0.0}$ & $-0.4_{-0.0}^{+0.0}$ & $-1.0\pm0.1$\\ [0.4ex]
J16202863-2442087 & 152.7 & $176.0_{-0.7}^{+0.7}$ & $140.9_{-0.7}^{+1.5}$ & $40.7_{-0.5}^{+0.5}$ & $180.6_{-0.2}^{+0.1}$ & $0.7_{-0.0}^{+0.0}$ & $163.5_{-1.4}^{+1.7}$ & $0.4_{-0.0}^{+0.0}$ & $1.6_{-0.1}^{+0.0}$ & $57.5_{-1.8}^{+1.2}$ & $-0.5_{-0.0}^{+0.0}$ & $148.5_{-0.9}^{+0.8}$ & $9.9_{-0.2}^{+0.1}$ & $5.0_{-0.0}^{+0.0}$ & $-0.0_{-0.0}^{+0.0}$ & $8.3_{-1.0}^{+1.2}$ & $3880.0_{-2.2}^{+2.3}$ & $-0.1_{-0.0}^{+0.0}$ & $-0.3_{-0.0}^{+0.0}$ & $-0.6\pm0.1$ \\ [0.4ex]
J16203960-2634284 & 148.3 & $215.1_{-0.7}^{+0.7}$ & $165.1_{-0.7}^{+0.0}$ & $62.5_{-0.1}^{+0.1}$ & $42.2_{-0.1}^{+0.1}$ & $0.6_{-0.0}^{+0.0}$ & $208.8_{-1.0}^{+1.0}$ & $0.3_{-0.0}^{+0.0}$ & $1.2_{-0.0}^{+0.0}$ & $139.2_{-0.8}^{+0.9}$ & $-0.7_{-0.0}^{+0.0}$ & $188.5_{-1.2}^{+1.0}$ & $9.6_{-0.2}^{+0.2}$ & $4.0_{-0.0}^{+0.0}$ & $-0.0_{-0.0}^{+0.0}$ & $6.1_{-0.4}^{+0.4}$ & $3945.9_{-2.6}^{+2.5}$ & $-0.2_{-0.0}^{+0.0}$ & $-0.2_{-0.0}^{+0.0}$ & $-0.7\pm0.1$ \\ [0.4ex]
J16213469-2612269 & 150.8 & $105.3_{-0.7}^{+1.5}$ & $87.6_{-1.5}^{+0.7}$ & $47.8_{-0.5}^{+0.4}$ & $44.0_{-0.2}^{+0.2}$ & $1.0_{-0.0}^{+0.0}$ & $106.6_{-1.6}^{+1.1}$ & $0.2_{-0.0}^{+0.0}$ & $1.4_{-0.1}^{+0.1}$ & $50.3_{-1.7}^{+1.4}$ & $-0.4_{-0.0}^{+0.0}$ & $92.0_{-4.2}^{+4.2}$ & $14.9_{-2.7}^{+4.8}$ & $9.0_{-0.6}^{+1.1}$ & $0.3_{-1.1}^{+0.5}$ & $34.8_{-10.2}^{+10.7}$ & $5381.6_{-5.3}^{+4.3}$ & $-0.0_{-0.0}^{+0.0}$ & $-0.4_{-0.0}^{+0.0}$ & $-$ \\ [0.4ex]
J16215472-2752053 & 159.3 & $119.8_{-0.8}^{+0.8}$ & $94.9_{-0.8}^{+0.0}$ & $65.4_{-0.2}^{+0.2}$ & $206.5_{-0.2}^{+0.2}$ & $0.8_{-0.0}^{+0.0}$ & $110.4_{-1.8}^{+1.6}$ & $0.2_{-0.0}^{+0.0}$ & $1.2_{-0.0}^{+0.1}$ & $121.5_{-1.4}^{+1.3}$ & $-0.7_{-0.0}^{+0.0}$ & $100.1_{-2.2}^{+1.5}$ & $9.2_{-0.7}^{+0.5}$ & $5.0_{-0.0}^{+0.0}$ & $-0.0_{-0.0}^{+0.0}$ & $9.2_{-1.5}^{+0.5}$ & $7427.7_{-4.5}^{+4.5}$ & $-0.1_{-0.0}^{+0.0}$ & $-0.5_{-0.0}^{+0.0}$ & $-0.6\pm0.1$ \\ [0.4ex]
J16221532-2511349 & 139.0 & $91.6_{-0.7}^{+0.7}$ & $74.7_{-0.7}^{+0.7}$ & $54.6_{-0.4}^{+0.4}$ & $15.6_{-0.3}^{+0.3}$ & $0.5_{-0.0}^{+0.0}$ & $94.5_{-2.9}^{+2.6}$ & $0.3_{-0.0}^{+0.0}$ & $1.2_{-0.1}^{+0.1}$ & $56.5_{-1.1}^{+0.9}$ & $-0.5_{-0.0}^{+0.0}$ & $76.4_{-1.0}^{+0.9}$ & $9.7_{-0.4}^{+0.2}$ & $5.0_{-0.0}^{+0.0}$ & $-0.0_{-0.0}^{+0.0}$ & $8.2_{-1.4}^{+1.2}$ & $3516.4_{-5.4}^{+4.6}$ & $-0.1_{-0.0}^{+0.0}$ & $-0.5_{-0.0}^{+0.0}$ & $-0.9\pm0.1$ \\ [0.4ex]
J16222982-2002472 & 126.0 & $80.6_{-0.6}^{+0.6}$ & $64.6_{-0.6}^{+0.0}$ & $17.4_{-0.9}^{+0.9}$ & $148.6_{-0.3}^{+0.2}$ & $1.4_{-0.1}^{+0.1}$ & $161.8_{-4.4}^{+5.7}$ & $0.0_{-0.0}^{+0.0}$ & $1.3_{-0.1}^{+0.1}$ & $101.1_{-0.9}^{+1.1}$ & $-0.6_{-0.0}^{+0.0}$ & $3.7_{-0.5}^{+1.5}$ & $1.1_{-0.0}^{+0.1}$ & $17.6_{-0.8}^{+0.5}$ & $-4.7_{-0.2}^{+0.5}$ & $21.2_{-14.4}^{+22.2}$ & $2268.3_{-3.5}^{+2.7}$ & $-0.0_{-0.0}^{+0.0}$ & $-0.5_{-0.0}^{+0.0}$ & $-0.4\pm0.1$ \\ [0.4ex]
J16230761-2516339 & 136.9 & $63.2_{-1.8}^{+1.7}$ & $48.1_{-0.7}^{+0.7}$ & $63.2_{-1.2}^{+1.3}$ & $88.3_{-0.6}^{+0.6}$ & $0.4_{-0.0}^{+0.0}$ & $96.7_{-6.8}^{+7.1}$ & $0.4_{-0.0}^{+0.0}$ & $1.2_{-0.1}^{+0.1}$ & $72.7_{-2.7}^{+2.9}$ & $-0.4_{-0.0}^{+0.0}$ & $70.3_{-5.4}^{+4.6}$ & $4.1_{-1.0}^{+2.1}$ & $2.8_{-0.4}^{+0.2}$ & $-3.6_{-0.3}^{+0.4}$ & $5.3_{-1.3}^{+1.3}$ & $3825.6_{-16.6}^{+17.2}$ & $-0.1_{-0.0}^{+0.0}$ & $-0.5_{-0.0}^{+0.0}$ & $-1.3\pm0.1$ \\ [0.4ex]
J16253798-1943162 & 150.8 & $203.2_{-1.5}^{+2.2}$ & $156.8_{-0.7}^{+1.5}$ & $73.1_{-0.2}^{+0.2}$ & $71.6_{-0.1}^{+0.1}$ & $0.7_{-0.0}^{+0.0}$ & $154.1_{-3.1}^{+4.3}$ & $0.2_{-0.0}^{+0.0}$ & $1.5_{-0.1}^{+0.1}$ & $105.1_{-1.5}^{+1.6}$ & $-0.7_{-0.0}^{+0.0}$ & $177.2_{-2.3}^{+2.0}$ & $9.7_{-0.4}^{+0.2}$ & $4.0_{-0.0}^{+0.0}$ & $-0.0_{-0.0}^{+0.0}$ & $4.2_{-0.5}^{+0.9}$ & $7667.4_{-5.0}^{+5.6}$ & $-0.2_{-0.0}^{+0.0}$ & $-0.5_{-0.0}^{+0.0}$ & $-0.7\pm0.1$ \\ [0.4ex]
J16271273-2504017 & 134.7 & $58.5_{-6.7}^{+8.6}$ & $37.5_{-3.9}^{+4.2}$ & $40.5_{-1.3}^{+1.6}$ & $140.6_{-3.3}^{+4.1}$ & $1.1_{-0.1}^{+0.1}$ & $99.7_{-8.2}^{+7.7}$ & $0.1_{-0.1}^{+0.1}$ & $1.4_{-0.1}^{+0.1}$ & $73.2_{-8.8}^{+12.6}$ & $-0.7_{-0.1}^{+0.1}$ & $98.2_{-8.4}^{+7.0}$ & $14.4_{-6.0}^{+5.2}$ & $2.1_{-0.4}^{+0.7}$ & $-3.4_{-0.7}^{+0.7}$ & $7.9_{-4.8}^{+5.8}$ & $3336.4_{-49.2}^{+61.2}$ & $-0.1_{-0.0}^{+0.0}$ & $-0.5_{-0.0}^{+0.0}$ & $-0.7\pm0.1$ \\ [0.4ex]
J16274905-2602437 & 130.9 & $83.7_{-0.6}^{+0.6}$ & $64.6_{-0.6}^{+0.6}$ & $38.5_{-0.5}^{+0.7}$ & $49.0_{-0.3}^{+0.2}$ & $0.2_{-0.0}^{+0.0}$ & $91.2_{-6.0}^{+6.2}$ & $0.0_{-0.0}^{+0.0}$ & $1.3_{-0.1}^{+0.1}$ & $70.2_{-1.5}^{+1.2}$ & $-0.7_{-0.0}^{+0.0}$ & $72.3_{-3.7}^{+3.4}$ & $13.9_{-2.6}^{+3.6}$ & $13.6_{-0.7}^{+1.1}$ & $-1.0_{-1.7}^{+1.2}$ & $28.1_{-19.5}^{+15.5}$ & $2468.4_{-3.5}^{+4.0}$ & $-0.2_{-0.0}^{+0.0}$ & $-0.6_{-0.0}^{+0.0}$ & $-0.6\pm0.1$\\ [0.4ex]
J16293267-2543291 & 142.8 & $57.2_{-0.7}^{+0.7}$ & $46.7_{-0.7}^{+0.1}$ & $38.9_{-1.8}^{+1.3}$ & $248.8_{-0.4}^{+0.5}$ & $0.5_{-0.0}^{+0.0}$ & $116.1_{-5.7}^{+5.8}$ & $0.3_{-0.0}^{+0.0}$ & $1.4_{-0.1}^{+0.1}$ & $64.7_{-1.8}^{+1.9}$ & $-0.5_{-0.0}^{+0.0}$ & $48.3_{-1.2}^{+1.3}$ & $8.9_{-1.0}^{+0.8}$ & $4.3_{-0.5}^{+0.4}$ & $-0.9_{-0.9}^{+0.6}$ & $4.6_{-1.0}^{+1.2}$ & $4067.7_{-6.3}^{+6.7}$ & $-0.1_{-0.0}^{+0.0}$ & $-0.4_{-0.0}^{+0.0}$ & $-0.8\pm0.1$ \\ [0.4ex]
J16395577-2347355 & 150.6 & $44.9_{-1.5}^{+1.5}$ & $33.8_{-0.7}^{+0.7}$ & $50.1_{-1.4}^{+1.2}$ & $241.0_{-0.8}^{+0.8}$ & $0.4_{-0.0}^{+0.0}$ & $120.4_{-5.9}^{+6.8}$ & $0.1_{-0.0}^{+0.0}$ & $1.3_{-0.1}^{+0.1}$ & $92.5_{-4.5}^{+4.4}$ & $-0.5_{-0.1}^{+0.1}$ & $45.1_{-5.7}^{+4.6}$ & $3.2_{-0.9}^{+1.6}$ & $2.1_{-0.5}^{+0.4}$ & $-3.1_{-0.6}^{+0.9}$ & $6.3_{-3.0}^{+2.6}$ & $3547.7_{-21.3}^{+17.3}$ & $-0.0_{-0.0}^{+0.0}$ & $-0.4_{-0.0}^{+0.0}$ & $-0.7\pm0.1$ \\ [0.4ex]
J16142029-1906481 & 138.8 & $80.7_{-1.4}^{+2.0}$ & $60.3_{-0.7}^{+1.4}$ & $60.5_{-0.9}^{+1.0}$ & $187.0_{-0.7}^{+0.6}$ & $1.2_{-0.0}^{+0.0}$ & $123.6_{-7.0}^{+9.2}$ & $0.2_{-0.0}^{+0.0}$ & $1.5_{-0.1}^{+0.1}$ & $147.2_{-3.6}^{+5.2}$ & $-0.3_{-0.0}^{+0.0}$ & $89.2_{-7.5}^{+6.2}$ & $3.0_{-0.7}^{+1.2}$ & $3.3_{-0.3}^{+0.2}$ & $-3.8_{-0.1}^{+0.3}$ & $7.3_{-2.3}^{+1.9}$ & $3828.5_{-18.3}^{+16.9}$ & $-0.3_{-0.0}^{+0.0}$ & $-0.2_{-0.0}^{+0.0}$ & $-$\\ [0.4ex]
J16042165-2130284 & 144.6 & $247.1_{-0.0}^{+0.7}$ & $205.5_{-0.7}^{+0.0}$ & $6.2_{-0.6}^{+1.1}$ & $258.7_{-0.2}^{+0.3}$ & $1.2_{-0.3}^{+0.3}$ & $112.3_{-9.7}^{+10.0}$ & $0.0_{-0.0}^{+0.0}$ & $1.3_{-0.2}^{+0.2}$ & $90.6_{-0.6}^{+0.6}$ & $-0.4_{-0.0}^{+0.0}$ & $114.6_{-2.7}^{+3.0}$ & $3.3_{-0.1}^{+0.1}$ & $5.5_{-0.0}^{+0.1}$ & $1.0_{-0.0}^{+0.0}$ & $23.5_{-16.0}^{+17.7}$ & $4539.9_{-1.0}^{+0.9}$ & $-0.2_{-0.0}^{+0.0}$ & $-0.4_{-0.0}^{+0.0}$ & $-$ \\ [0.4ex]

 \hline
 \hline
\end{tabular}
}
\tablefoot{The distances were taken from the Gaia DR3 collaboration (\citealt{Gaia_DR3_2023}). The stellar luminosities were taken from Empey et al., in prep. The full table is available on \texttt{GitHub}\protect\hyperref[footnote:github]{\textsuperscript{\ref*{footnote:github}}}.}
\label{Table:Fits_A}
%\vspace{-0.5cm}
\end{table}

\begin{table}
\setcounter{table}{1}
\caption{Fit results using the B parametrization prescription.}
\centering
\resizebox{\linewidth}{!}{
\begin{tabular}{l c c c c c c c c c c c c c c c c c c c} 
 \hline
 \hline
 \noalign{\vskip 0.03in} 
 Source (2MASS) & dist [pc] & $R_{90\%}$ [au] & $R_{68\%}$ [au] & inc [°] & PA [°] & $M_*$ [M$_{\odot}$] & R$_{\text{out}}$ [au] & T$_{10}$ [K] & T$_{q}$ & $\tau_{10}$ & v$_{sys}$ [ms$^{-1}$] & d$_{\text{RA}} [^{\prime\prime}]$ & d$_{\text{DEC}} [^{\prime\prime}]$ & $\log(\text{L}_*) \> [\text{L}_{\odot}]$\\ [0.4ex]
 \hline
 \hline
\noalign{\vskip 0.03in} 

J16012268-2408003 & 140.3 & $48.0_{-1.4}^{+1.4}$ & $34.9_{-0.7}^{+0.7}$ & $21.6_{-0.5}^{+0.7}$ & $-6.9_{-1.8}^{+1.4}$ & $1.9_{-0.1}^{+0.0}$ & $54.0_{-1.7}^{+1.7}$ & $99.8_{-1.4}^{+1.4}$ & $-1.0_{-0.0}^{+0.0}$ & $4.9_{-0.1}^{+0.0}$ & $3539.2_{-37.2}^{+41.5}$ & $-0.2_{-0.0}^{+0.0}$ & $-0.6_{-0.0}^{+0.0}$ & $-0.9 \pm0.1$\\ [0.4ex]
J16145026-2332397 & 144.0 & $36.6_{-0.7}^{+1.5}$ & $27.4_{-0.7}^{+0.7}$ & $15.0_{-2.6}^{+2.3}$ & $148.1_{-1.4}^{+1.4}$ & $0.5_{-0.1}^{+0.2}$ & $40.9_{-1.4}^{+1.5}$ & $71.3_{-0.7}^{+0.7}$ & $-0.9_{-0.0}^{+0.0}$ & $5.0_{-0.1}^{+0.0}$ & $4774.8_{-13.1}^{+13.4}$ & $-0.1_{-0.0}^{+0.0}$ & $-0.3_{-0.0}^{+0.0}$ & $-1.1\pm0.1$ \\ [0.4ex]
J16020757-2257467 & 139.6 & $42.9_{-2.0}^{+2.8}$ & $34.1_{-1.4}^{+2.0}$ & $57.0_{-4.0}^{+3.1}$ & $84.7_{-2.4}^{+2.4}$ & $0.6_{-0.1}^{+0.1}$ & $47.4_{-2.8}^{+3.0}$ & $106.5_{-5.7}^{+6.1}$ & $-0.7_{-0.1}^{+0.1}$ & $3.9_{-0.9}^{+0.8}$ & $3795.6_{-74.0}^{+60.3}$ & $-0.2_{-0.0}^{+0.0}$ & $-0.5_{-0.0}^{+0.0}$ & $-1.2\pm0.1$  \\ [0.4ex]
J16121242-1907191 & 138.2 & $38.5_{-1.3}^{+2.0}$ & $30.4_{-1.3}^{+0.8}$ & $36.7_{-2.5}^{+2.7}$ & $152.0_{-2.2}^{+2.0}$ & $0.03_{-0.01}^{+0.01}$ & $41.9_{-1.6}^{+1.8}$ & $40.2_{-0.9}^{+0.9}$ & $-0.6_{-0.0}^{+0.0}$ & $4.9_{-0.1}^{+0.0}$ & $3310.6_{-13.4}^{+12.0}$ & $-0.2_{-0.0}^{+0.0}$ & $-0.5_{-0.0}^{+0.0}$ & $-1.2\pm0.1$\\ [0.4ex]

 \hline
 \hline
\end{tabular}
} 
\tablefoot{The distances were taken from the Gaia DR3 collaboration (\citealt{Gaia_DR3_2023}). The stellar luminosities were taken from Empey et al., in prep. The full table is available on \texttt{GitHub}\protect\hyperref[footnote:github]{\textsuperscript{\ref*{footnote:github}}}.}
\label{Table:Fits_B}
\end{table}
\end{landscape}
\clearpage

\addcontentsline{toc}{section}{Appendix B (figures)}
\setcounter{figure}{0}
\renewcommand{\thefigure}{B.\arabic{figure}}

\begin{landscape}
\begin{figure}[t!]
    \centering
    \subfloat[Residual plot of J15583620-1946135.]{\includegraphics[width=0.67\textwidth]{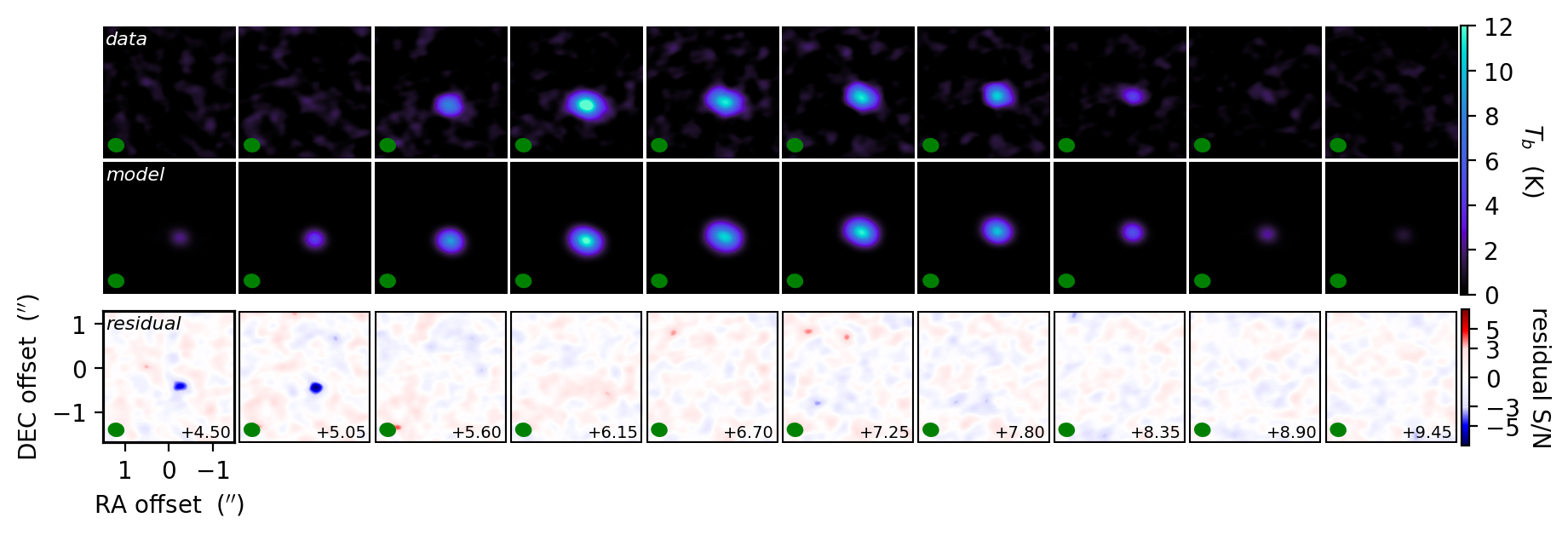}}
    \subfloat[Residual plot of J15583692-2257153.]{\includegraphics[width=0.67\textwidth]{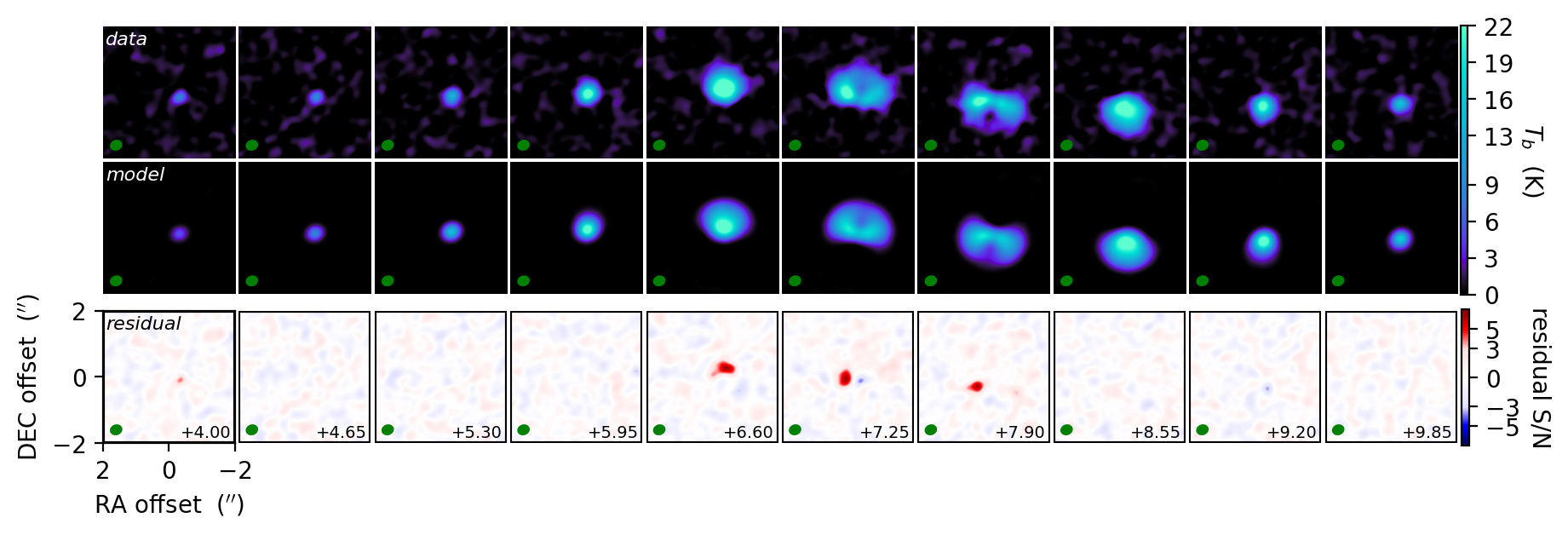}}\\[1ex]
    \subfloat[Residual plot of J16035793-1942108.]{\includegraphics[width=0.67\textwidth]{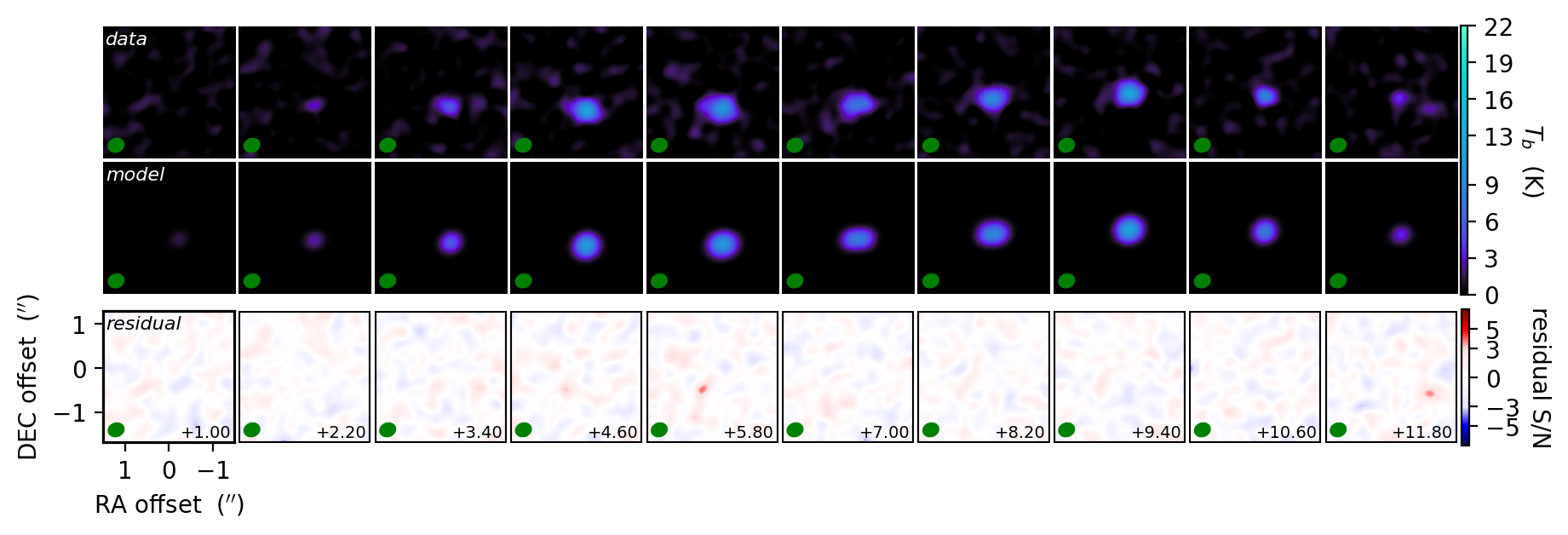}}
    \subfloat[Residual plot of J16052157-1821412.]{\includegraphics[width=0.67\textwidth]{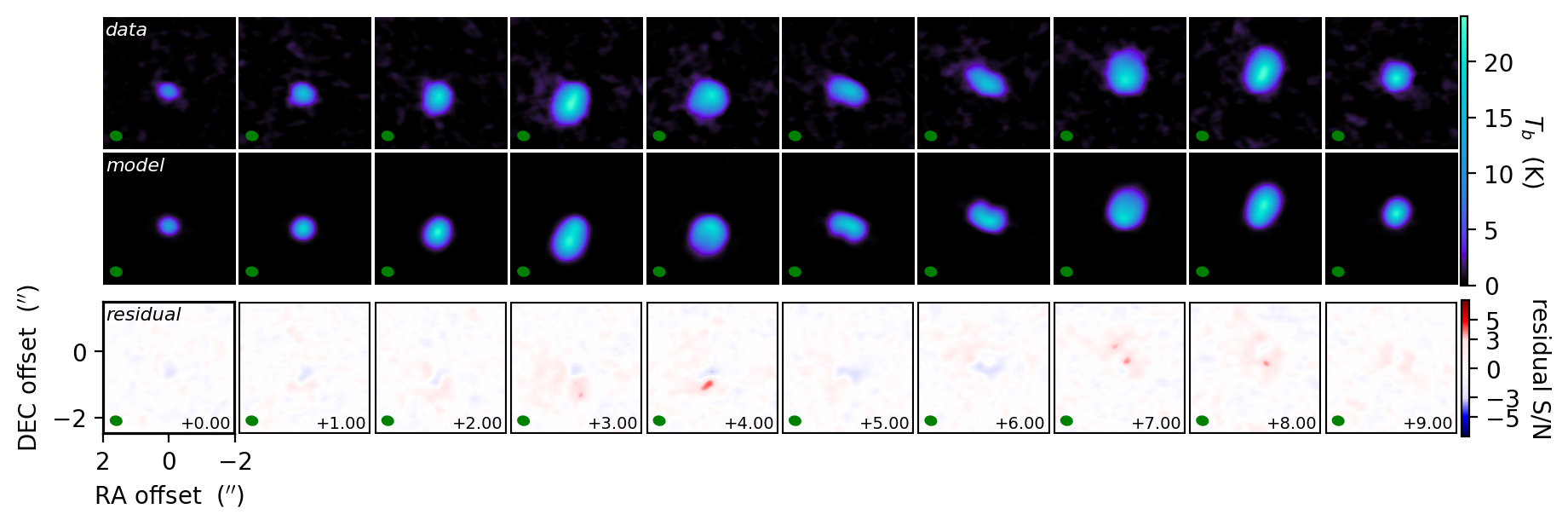}}\\[1ex]
    \subfloat[Residual plot of J16062383-1807183.]{\includegraphics[width=0.67\textwidth]{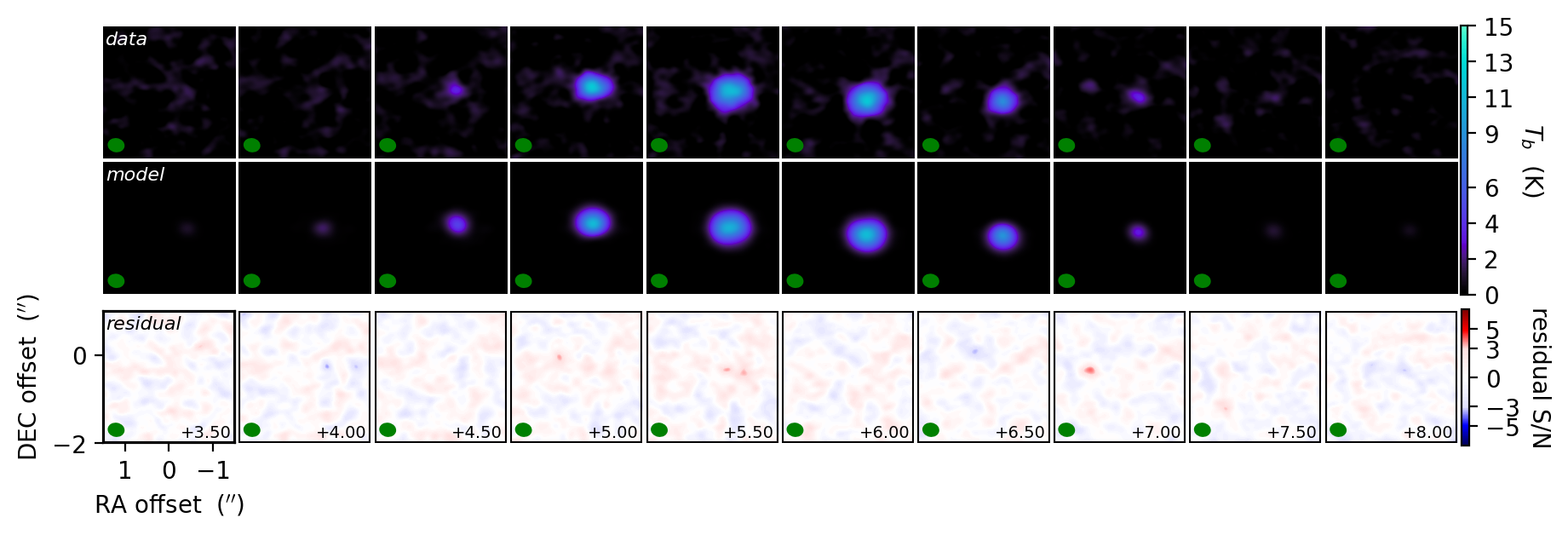}}
    \subfloat[Residual plot of J16064794-1841437.]{\includegraphics[width=0.67\textwidth]{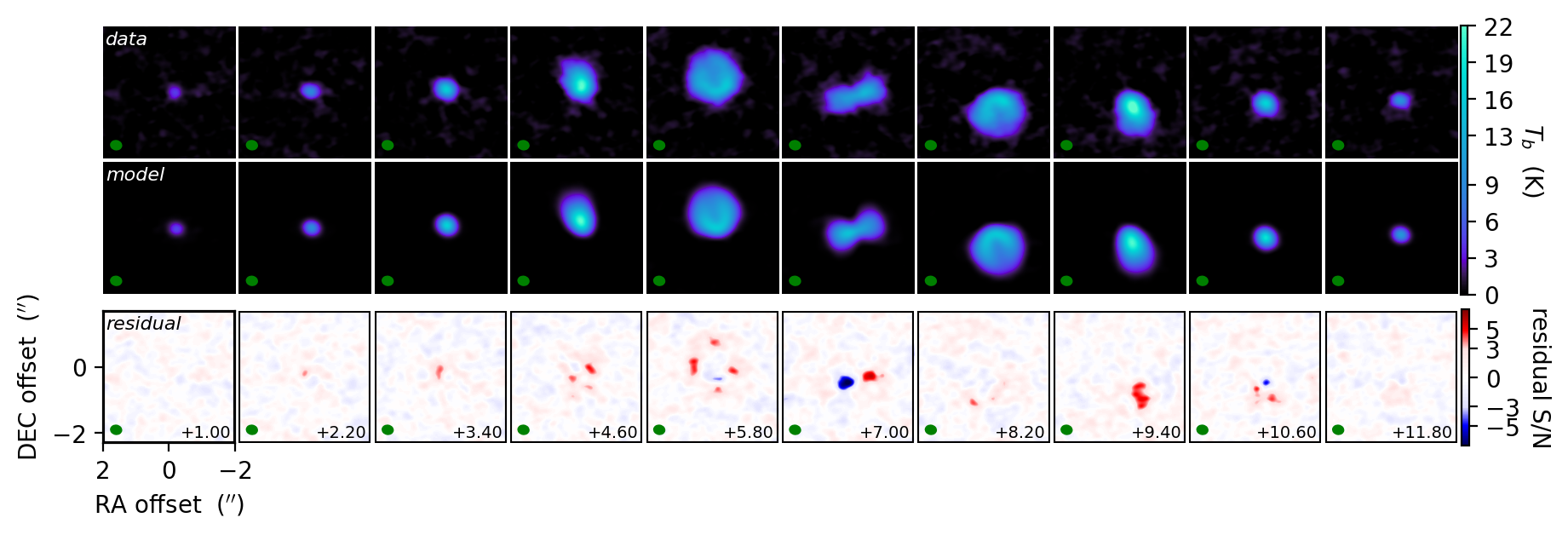}}
    \caption{Residual plots of J15583620-1946135 (a), J15583692-2257153 (b), J16035793-1942108 (c), J16052157-1821412 (d), J16062383-1807183 (e), J16064794-1841437 (f). The residual plots for all the disks analyzed in this work and their corresponding corner plots can be found on \texttt{GitHub}\protect\hyperref[footnote:github]{\textsuperscript{\ref*{footnote:github}}}.}
    \label{fig:residuals_1}
\end{figure}
\end{landscape}

\begin{landscape}
\begin{figure}[t!]
    \centering
    \subfloat[Residual plot of J16095933-1800090.]{\includegraphics[width=0.67\textwidth]{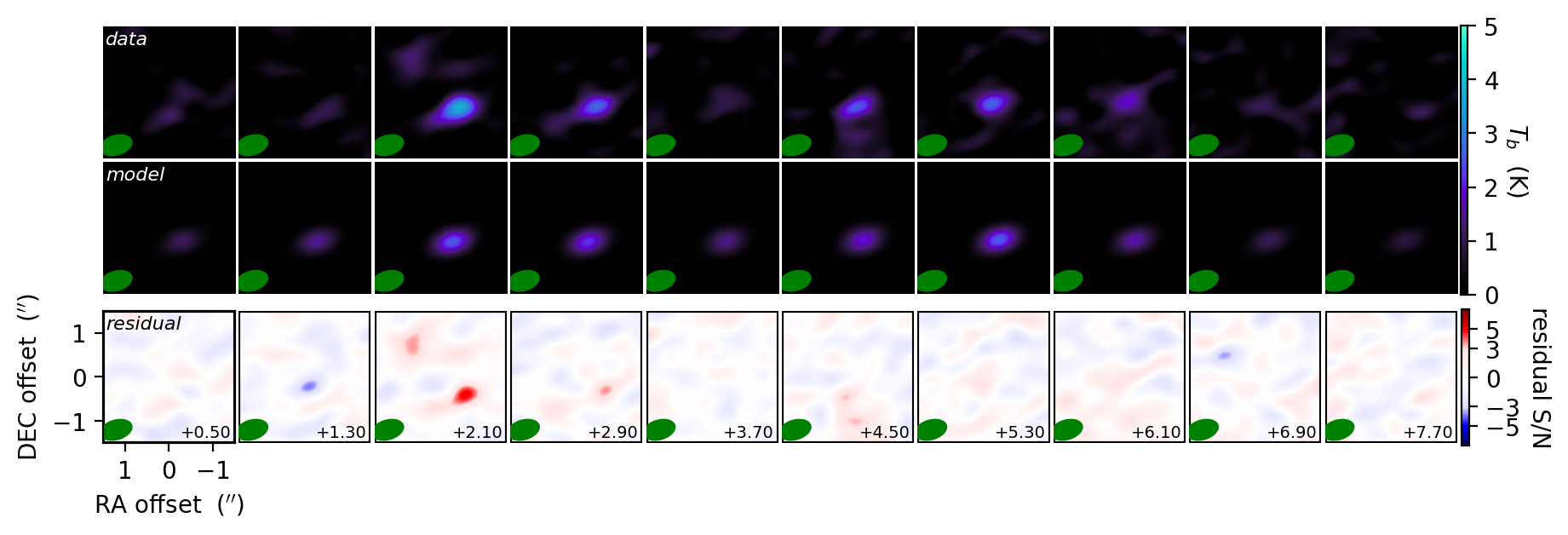}}
    \subfloat[Residual plot of J16101264-2104446.]{\includegraphics[width=0.67\textwidth]{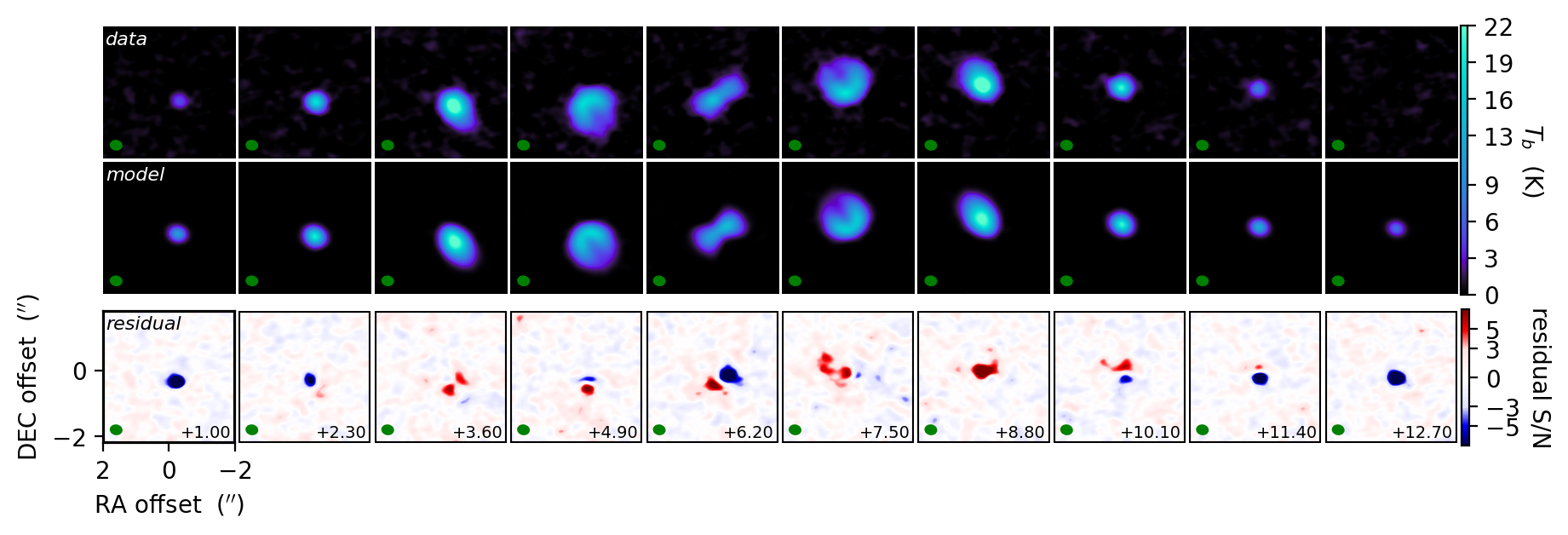}}\\[1ex]
    \subfloat[Residual plot of J16120668-3010270.]{\includegraphics[width=0.67\textwidth]{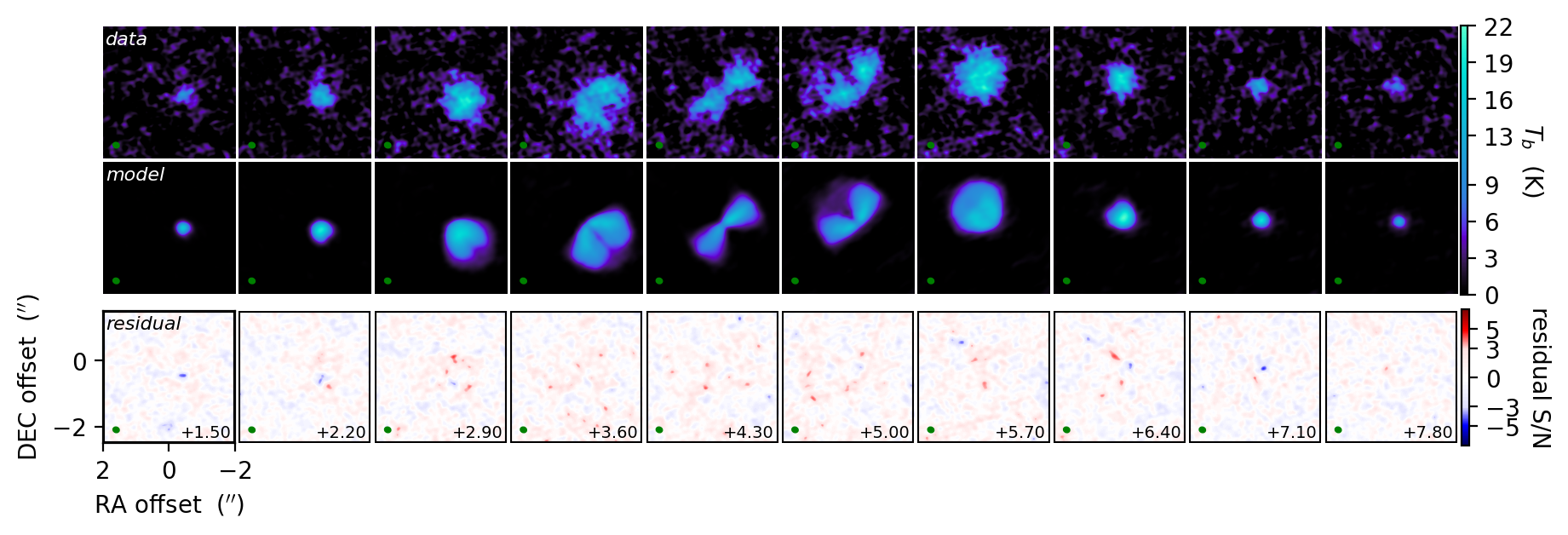}}
    \subfloat[Residual plot of J16132190-2136136.]{\includegraphics[width=0.67\textwidth]{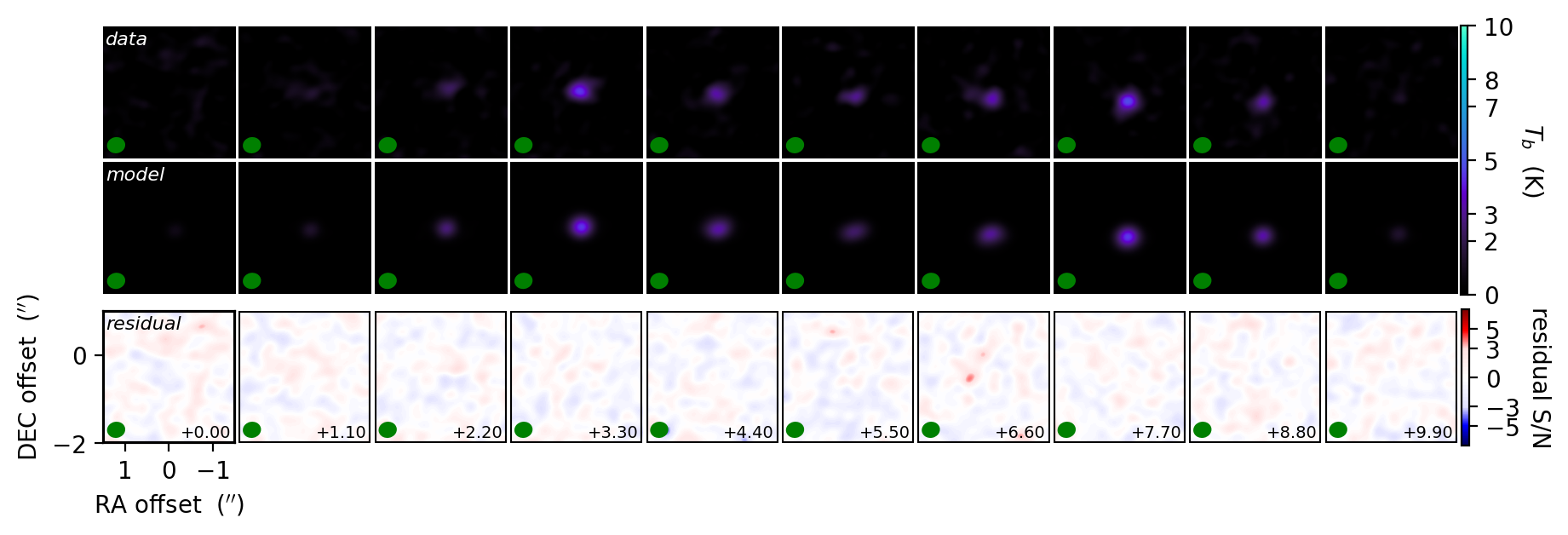}}\\[1ex]
    \subfloat[Residual plot of J16140792-1938292.]{\includegraphics[width=0.67\textwidth]{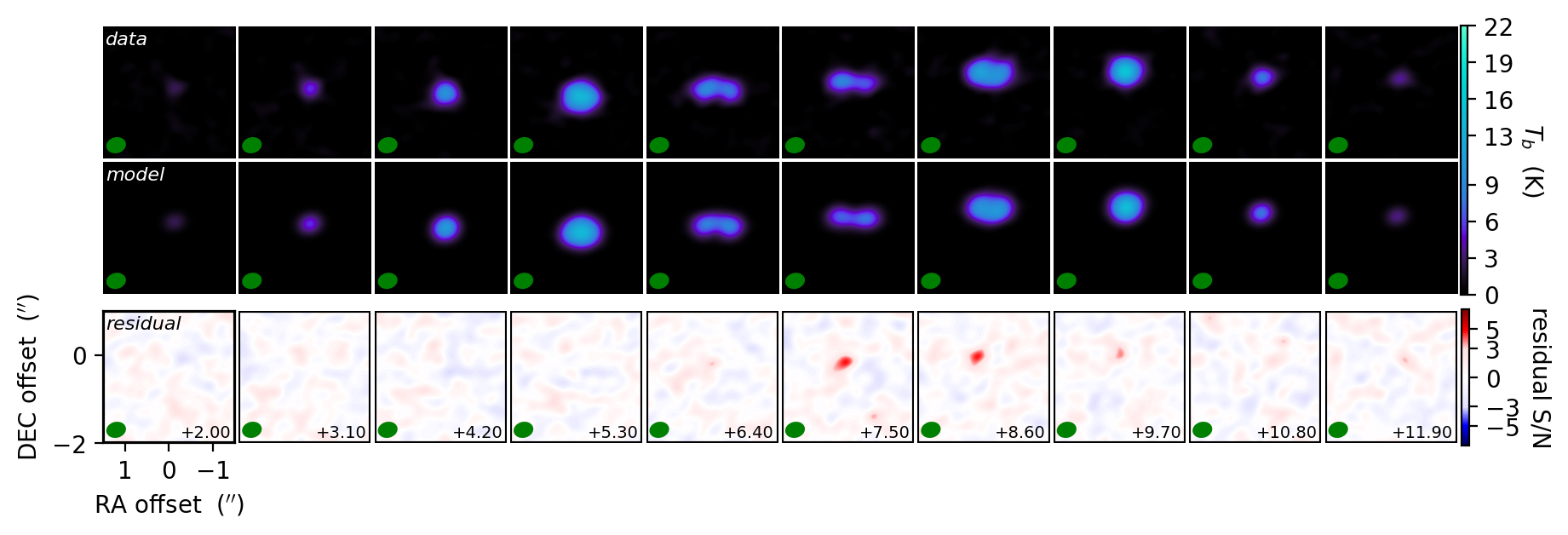}}
    \subfloat[Residual plot of J16142091-1906051.]{\includegraphics[width=0.67\textwidth]{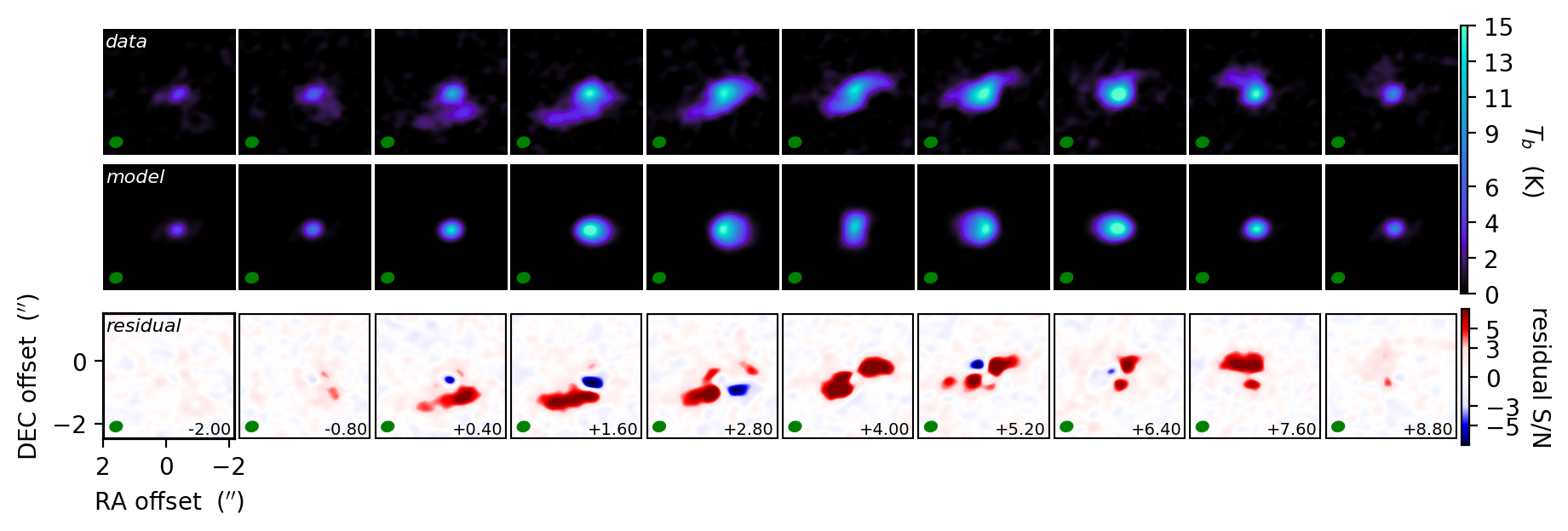}}
    \caption{Residual plots of J16095933-1800090 (a), J16101264-2104446 (b), J16120668-3010270 (c), J16132190-2136136 (d), J16140792-1938292 (e), J16142091-1906051 (f). The residual plots for all the disks analyzed in this work and their corresponding corner plots can be found on \texttt{GitHub}\protect\hyperref[footnote:github]{\textsuperscript{\ref*{footnote:github}}}.}
    \label{fig:residuals_2}
\end{figure}
\end{landscape}

\begin{landscape}
\begin{figure}[t!]
    \centering
    \subfloat[Residual plot of J16145024-2100599.]{\includegraphics[width=0.67\textwidth]{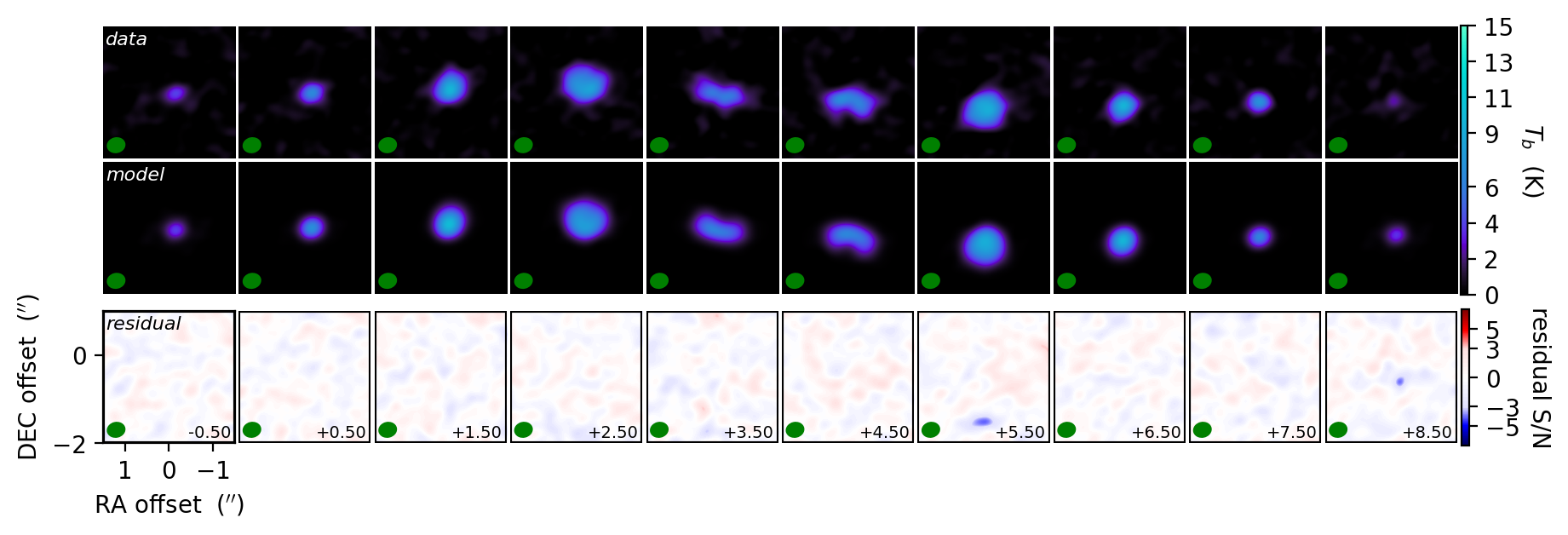}}
    \subfloat[Residual plot of J16152752-1847097.]{\includegraphics[width=0.67\textwidth]{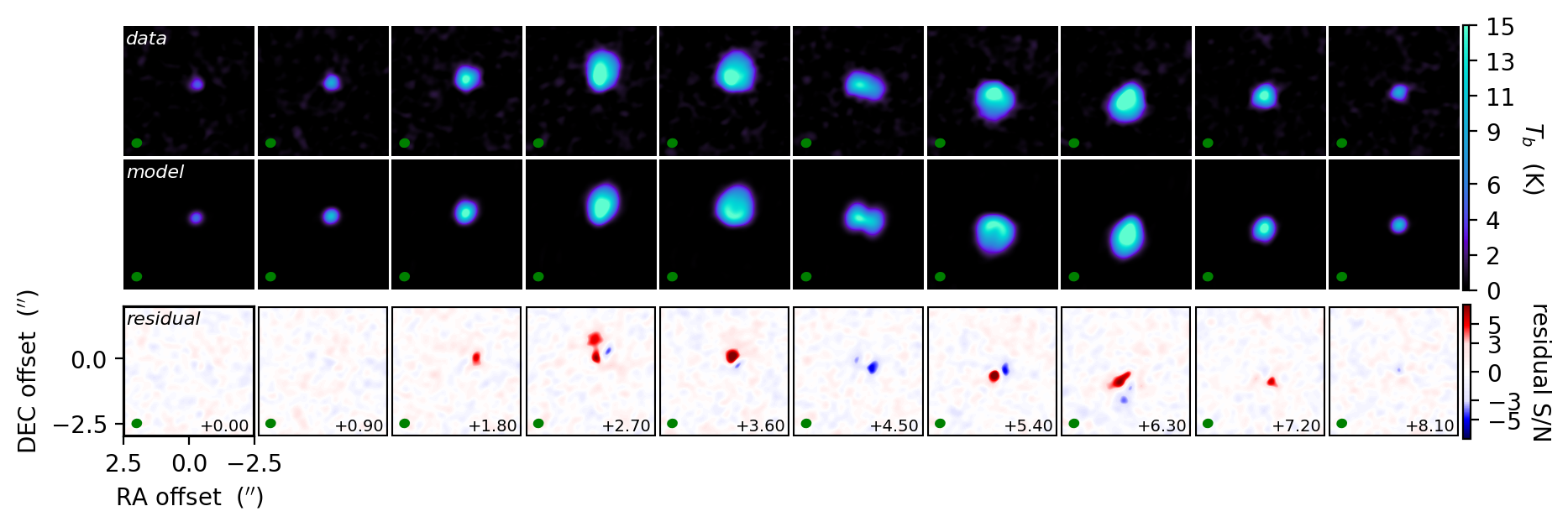}}\\[1ex]
    \subfloat[Residual plot of J16181445-2319251.]{\includegraphics[width=0.67\textwidth]{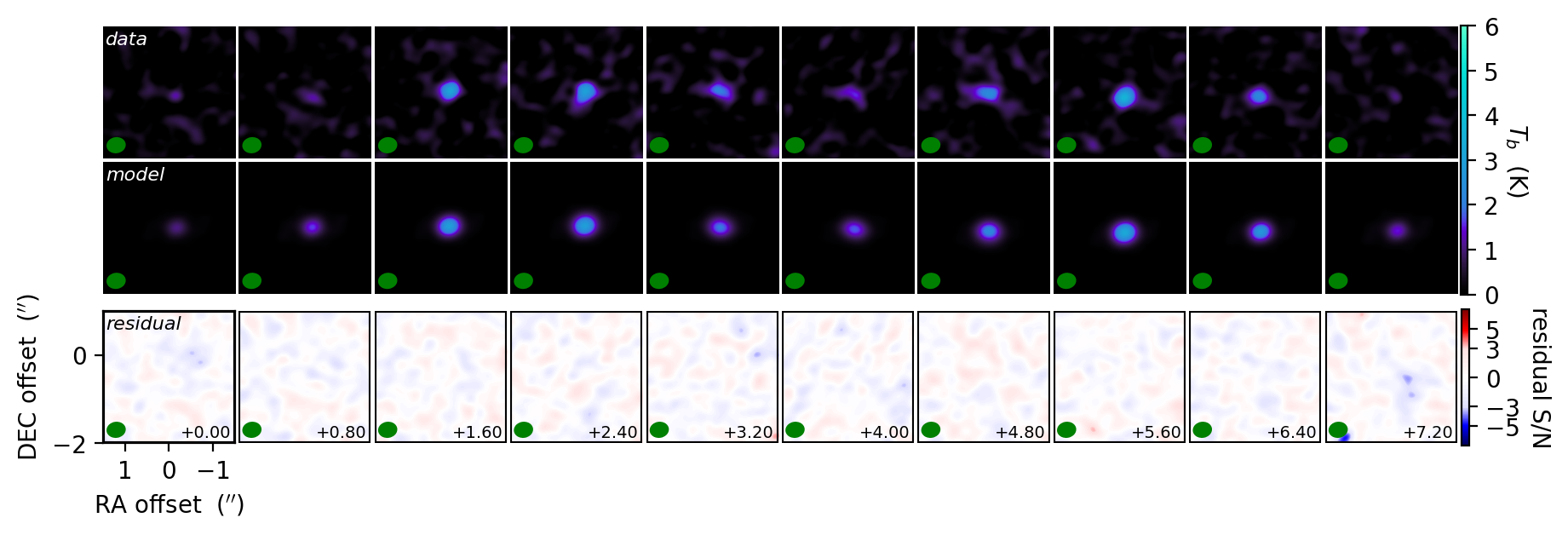}}
    \subfloat[Residual plot of J16185382-2053182.]{\includegraphics[width=0.67\textwidth]{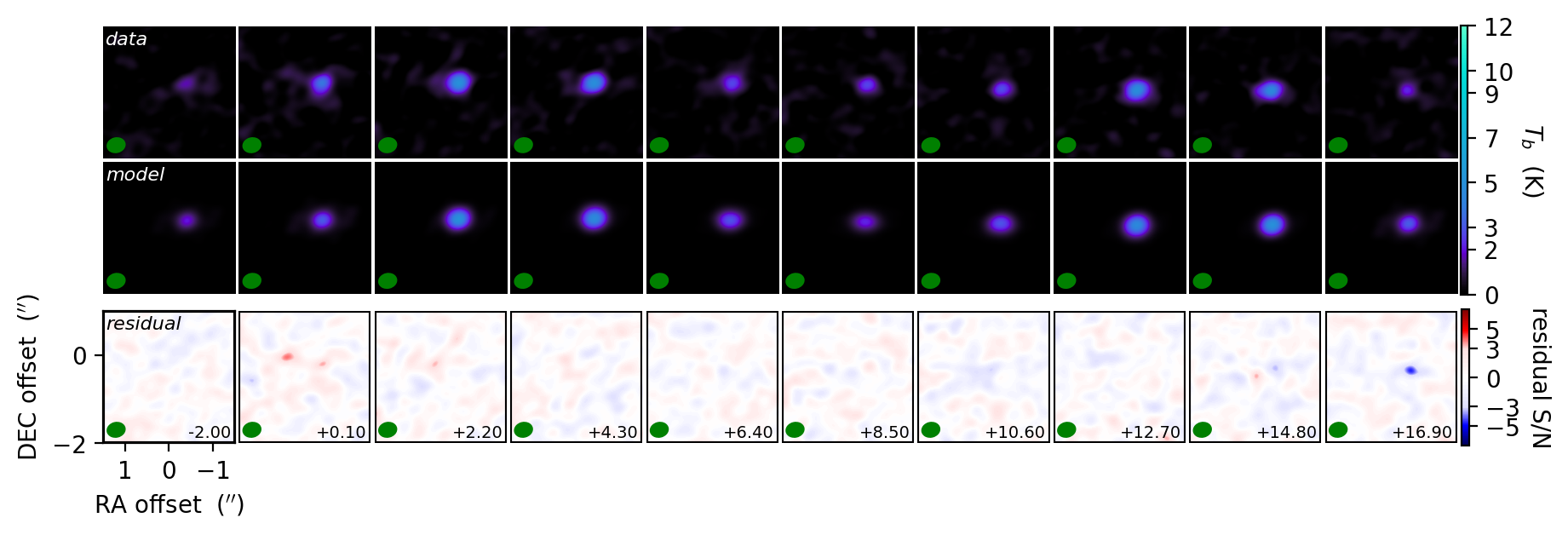}}\\[1ex]
    \subfloat[Residual plot of J16202291-2227041.]{\includegraphics[width=0.67\textwidth]{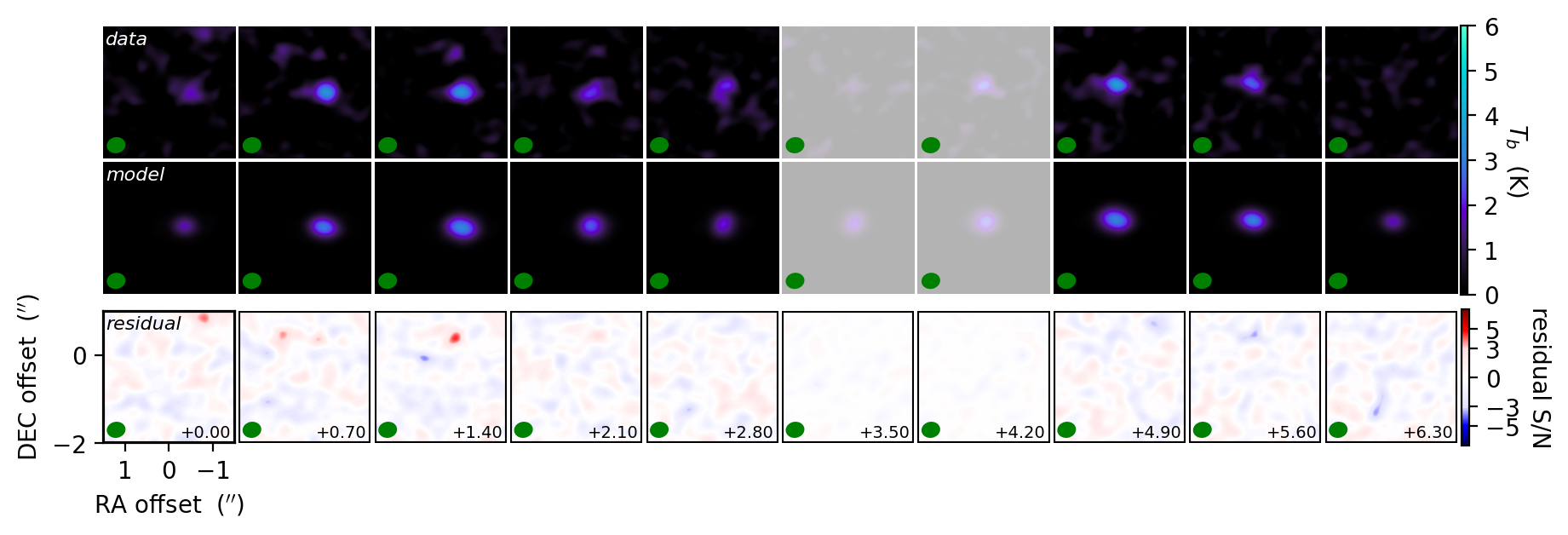}}
    \subfloat[Residual plot of J16202863-2442087.]{\includegraphics[width=0.67\textwidth]{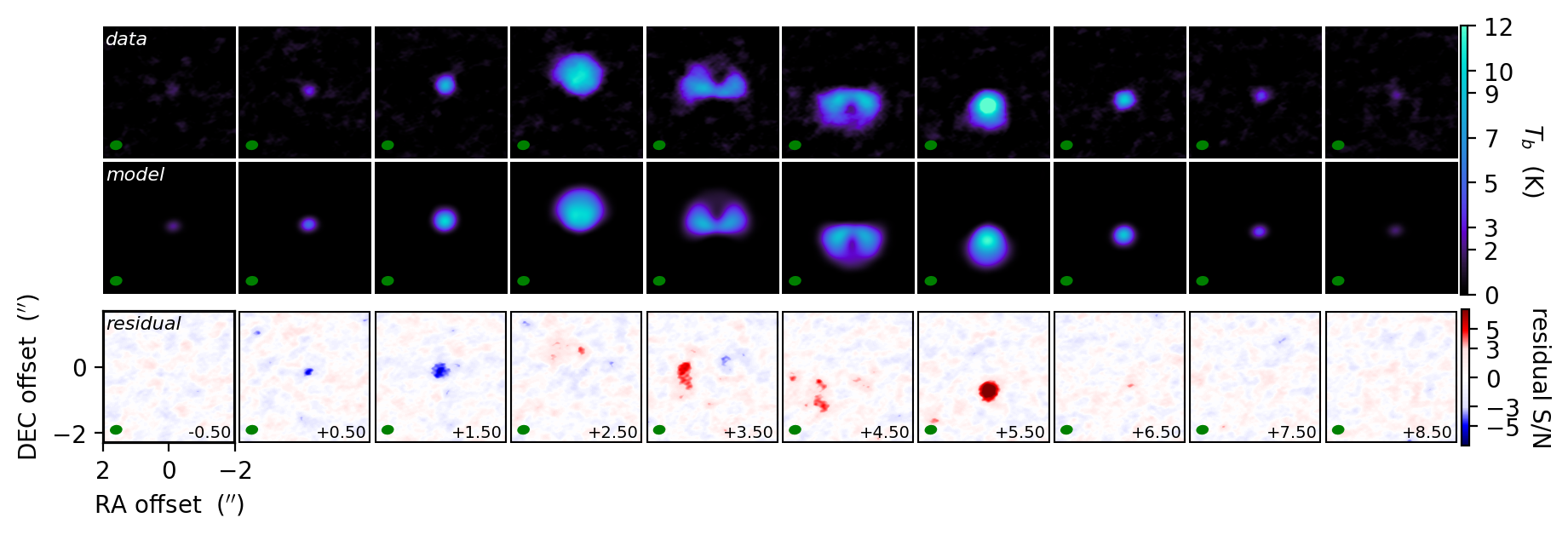}}
    \caption{Residual plots of J16145024-2100599 (a), J16152752-1847097 (b), J16181445-2319251 (c), J16185382-2053182 (d), J16202291-2227041 (e), J16202863-2442087 (f). The residual plots for all the disks analyzed in this work and their corresponding corner plots can be found on \texttt{GitHub}\protect\hyperref[footnote:github]{\textsuperscript{\ref*{footnote:github}}}.}
    \label{fig:residuals_3}
\end{figure}
\end{landscape}

\begin{landscape}
\begin{figure}[t!]
    \centering
    \subfloat[Residual plot of J16203960-2634284.]{\includegraphics[width=0.67\textwidth]{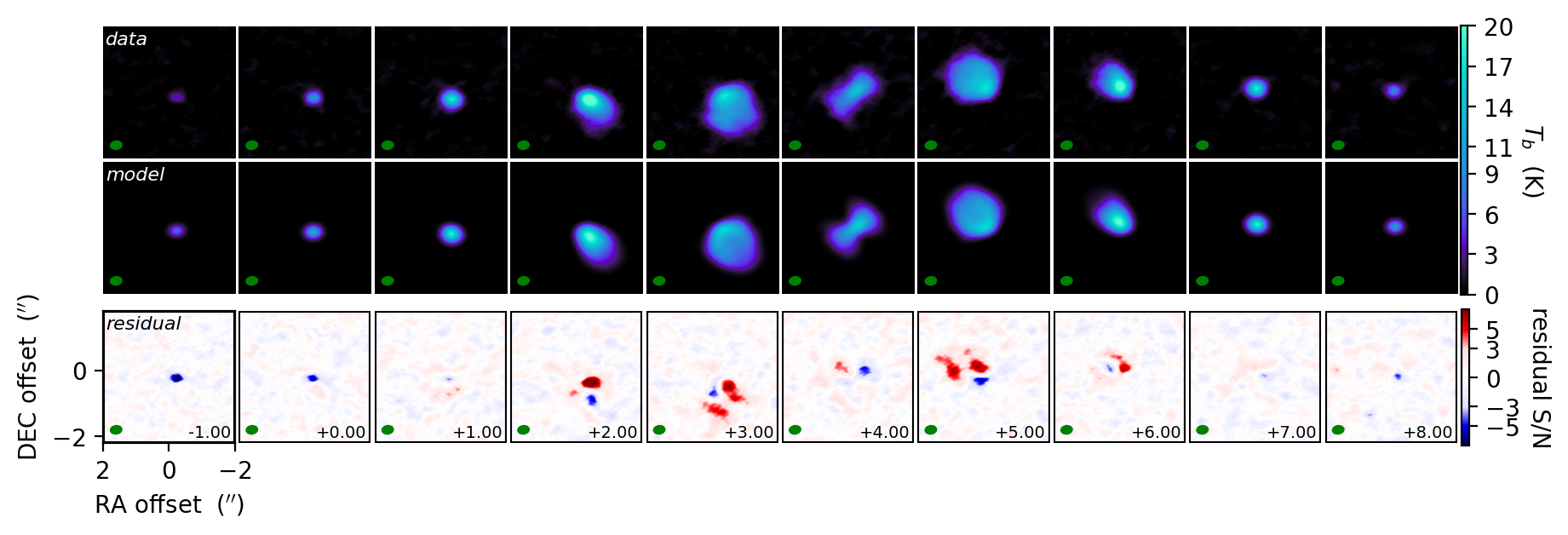}}
    \subfloat[Residual plot of J16215472-2752053.]{\includegraphics[width=0.67\textwidth]{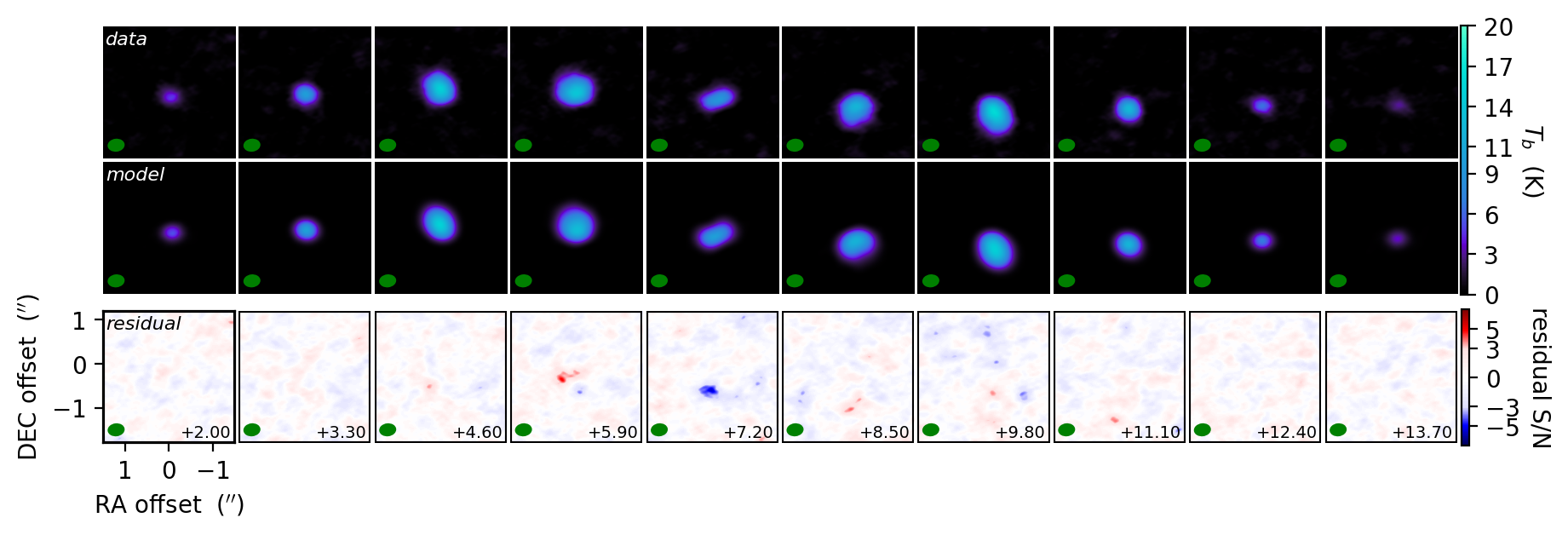}}\\[1ex]
    \subfloat[Residual plot of J16221532-2511349.]{\includegraphics[width=0.67\textwidth]{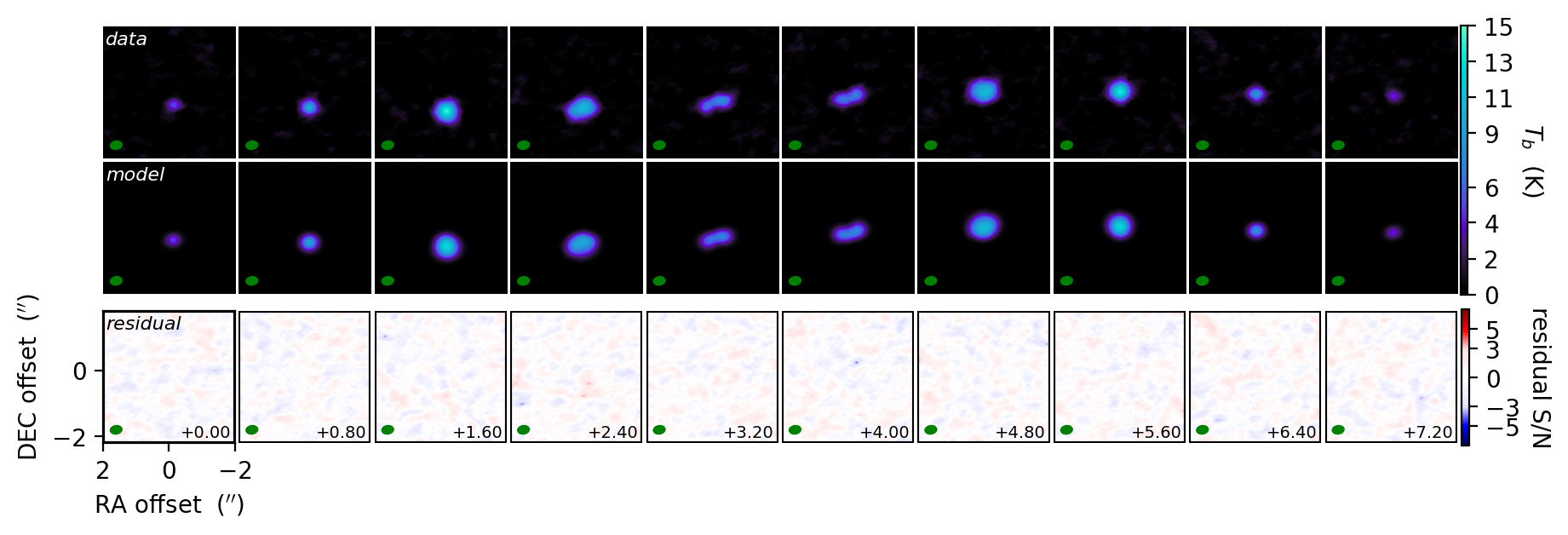}}
    \subfloat[Residual plot of J16222982-2002472.]{\includegraphics[width=0.67\textwidth]{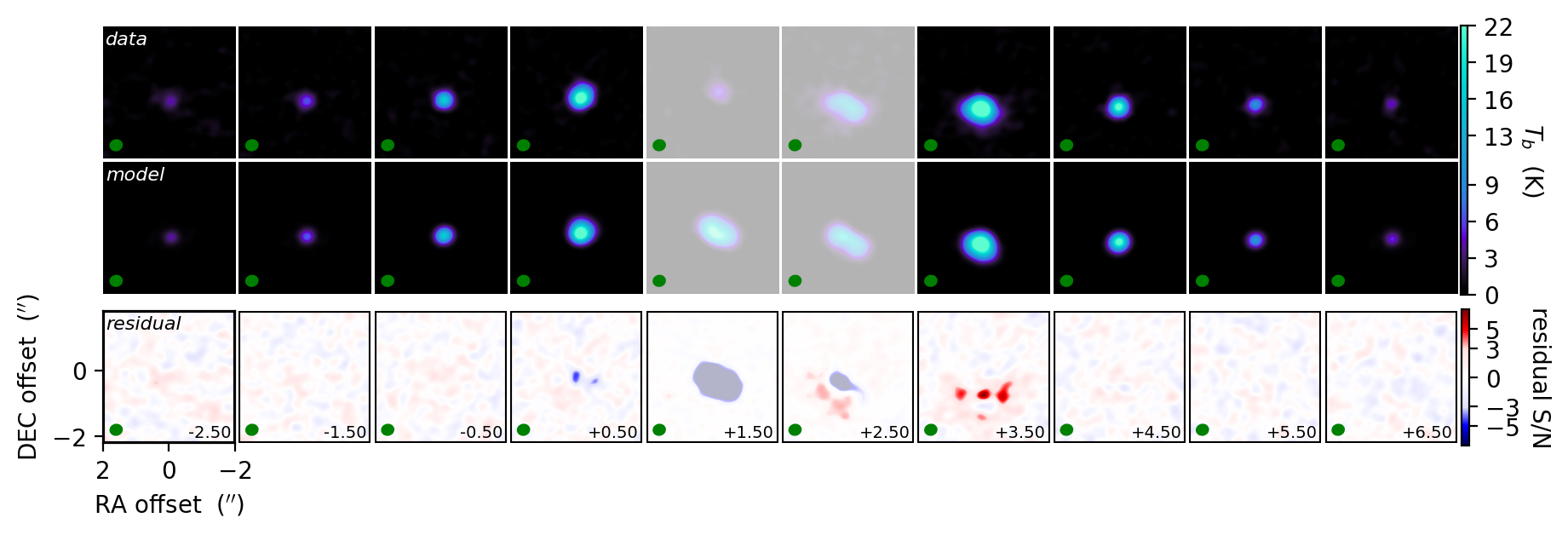}}\\[1ex]
    \subfloat[Residual plot of J16230761-2516339.]{\includegraphics[width=0.67\textwidth]{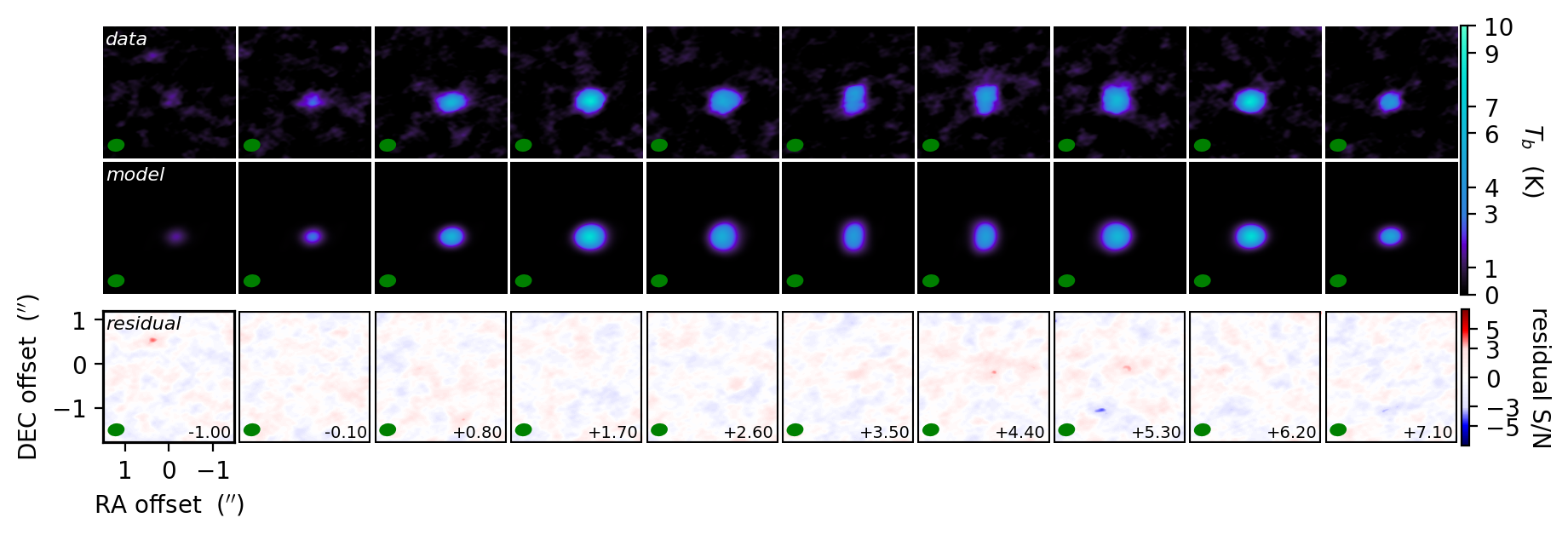}}
    \subfloat[Residual plot of J16253798-1943162.]{\includegraphics[width=0.67\textwidth]{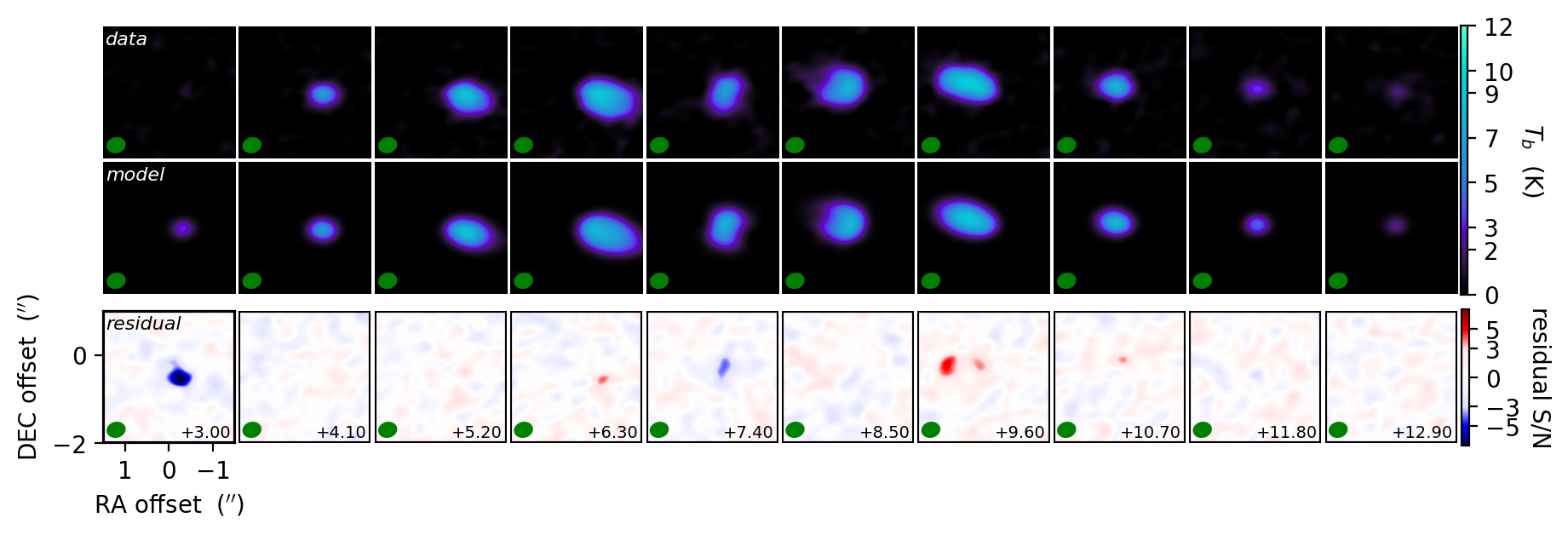}}
    \caption{Residual plots of J16203960-2634284 (a), J16215472-2752053 (b), J16221532-2511349 (c), J16222982-2002472 (d), J16230761-2516339 (e), J16253798-1943162 (f). The residual plots for all the disks analyzed in this work and their corresponding corner plots can be found on \texttt{GitHub}\protect\hyperref[footnote:github]{\textsuperscript{\ref*{footnote:github}}}.}
    \label{fig:residuals_4}
\end{figure}
\end{landscape}

\begin{landscape}
\begin{figure}[t!]
    \centering
    \subfloat[Residual plot of J16271273-2504017.]{\includegraphics[width=0.67\textwidth]{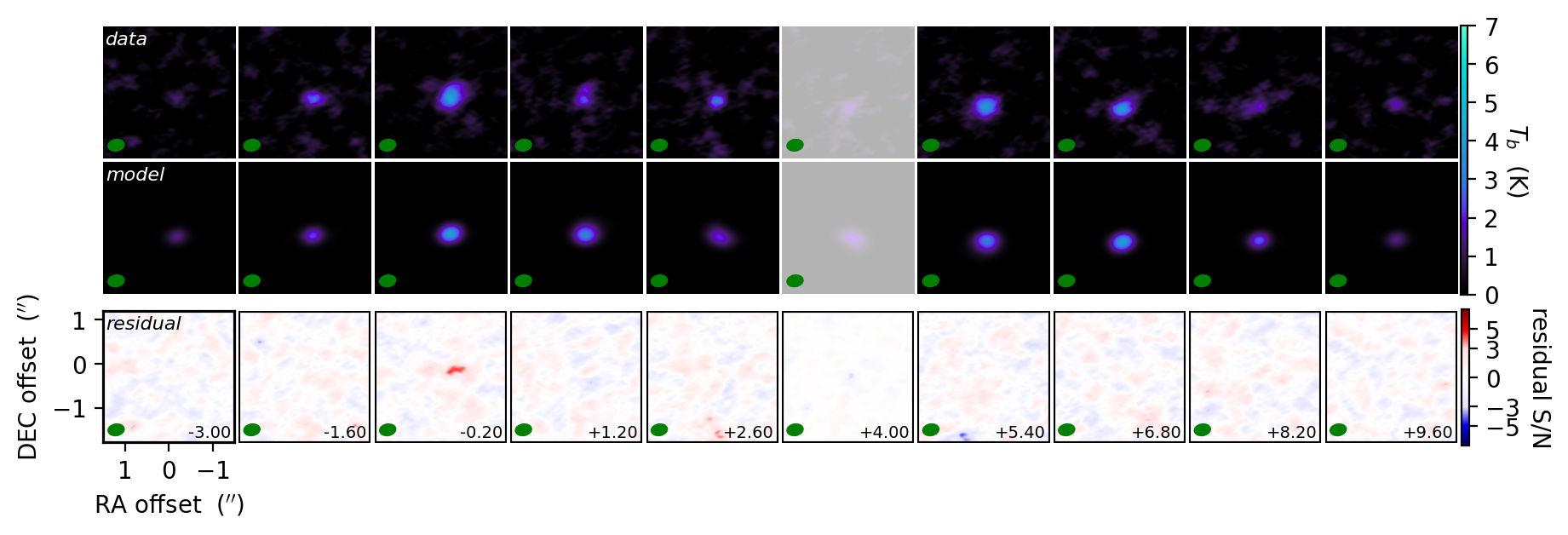}}
    \subfloat[Residual plot of J16274905-2602437.]{\includegraphics[width=0.67\textwidth]{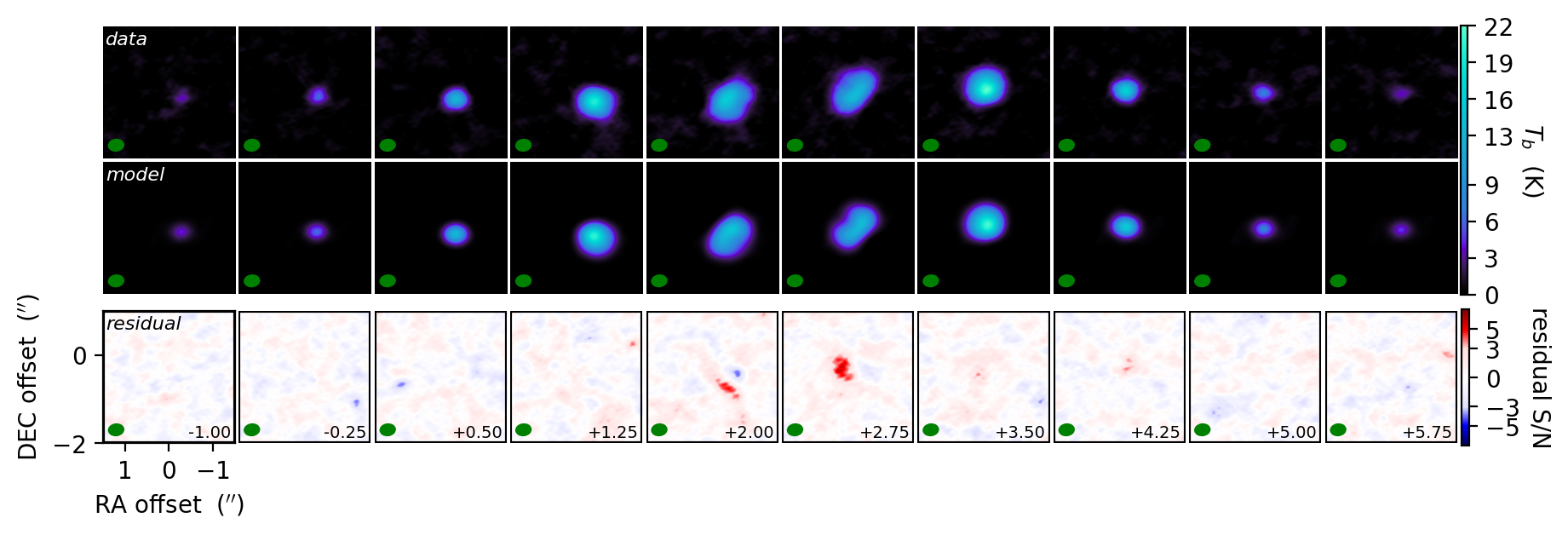}}\\[1ex]
    \subfloat[Residual plot of J16293267-2543291.]{\includegraphics[width=0.67\textwidth]{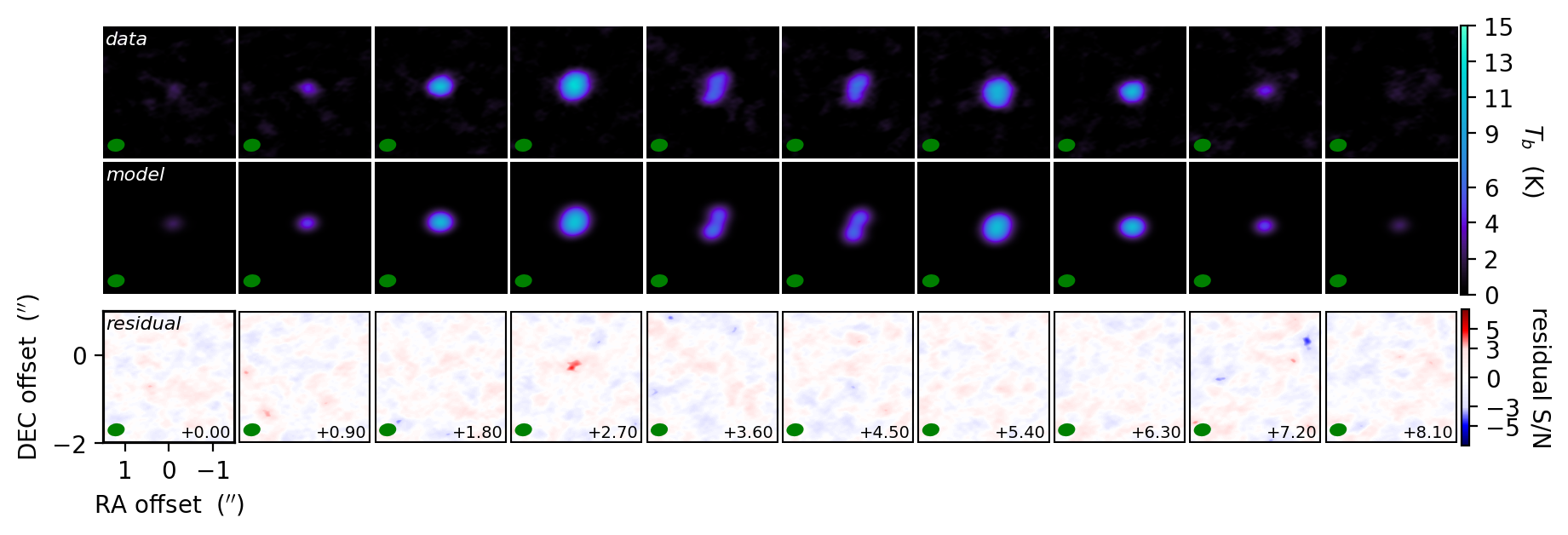}}
    \subfloat[Residual plot of J16395577-2347355.]{\includegraphics[width=0.67\textwidth]{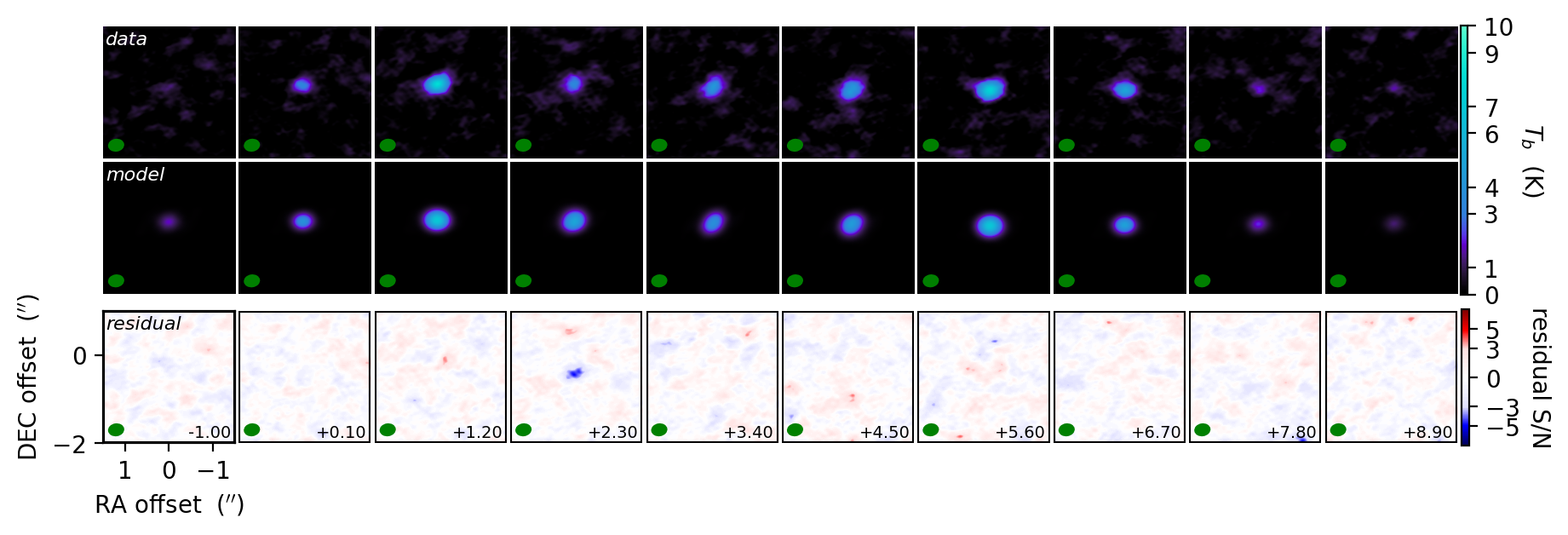}}\\[1ex]
    \subfloat[Residual plot of J16142029-1906481.]{\includegraphics[width=0.67\textwidth]{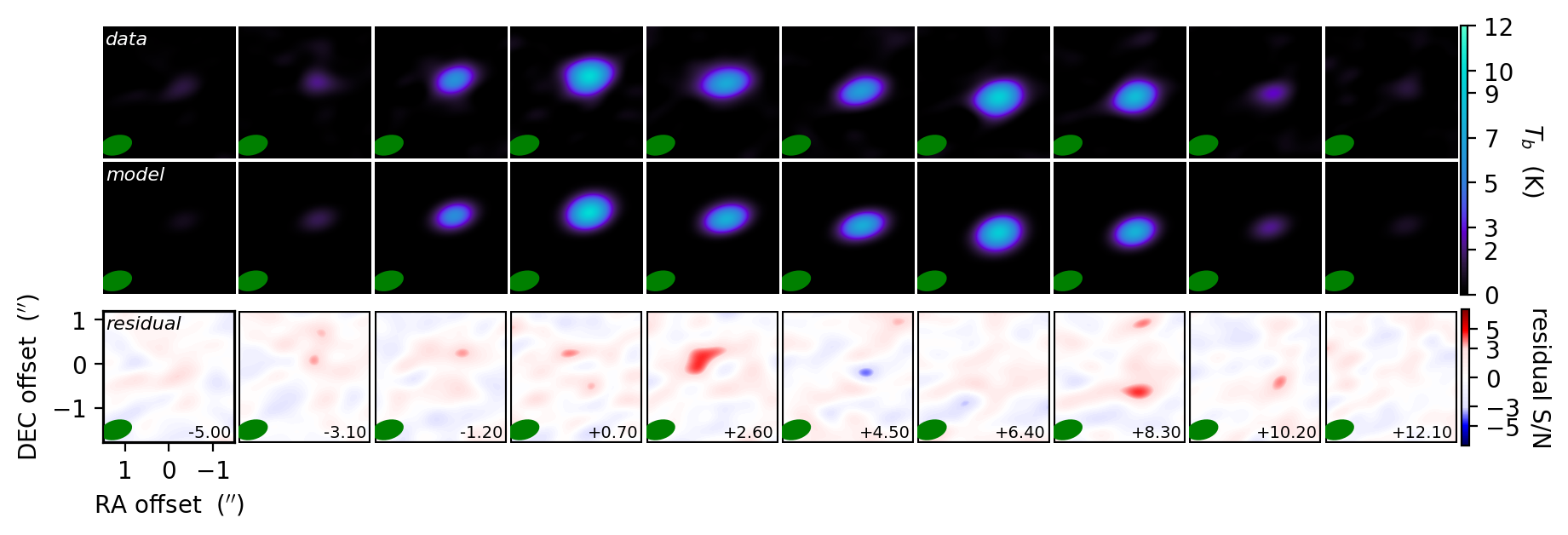}}
    \subfloat[Residual plot of J16042165-2130284.]{\includegraphics[width=0.67\textwidth]{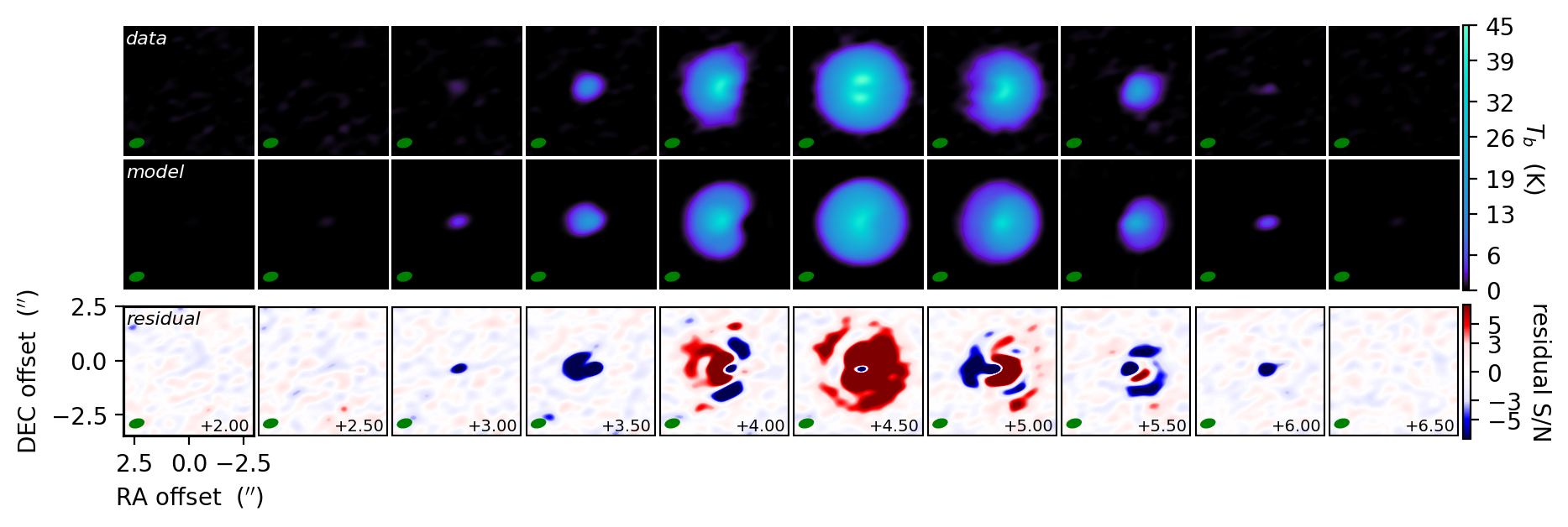}}
    \caption{Residual plots of J16271273-2504017 (a), J16274905-2602437 (b), J16293267-2543291 (c), J16395577-2347355 (d), J16142029-1906481 (e), J16042165-2130284 (f). The residual plots for all the disks analyzed in this work and their corresponding corner plots can be found on \texttt{GitHub}\protect\hyperref[footnote:github]{\textsuperscript{\ref*{footnote:github}}}.}
    \label{fig:residuals_5}
\end{figure}
\end{landscape}

\begin{landscape}
\begin{figure}[t!]
    \centering
    \subfloat[Residual plot of J16012268-2408003.]{\includegraphics[width=0.67\textwidth]{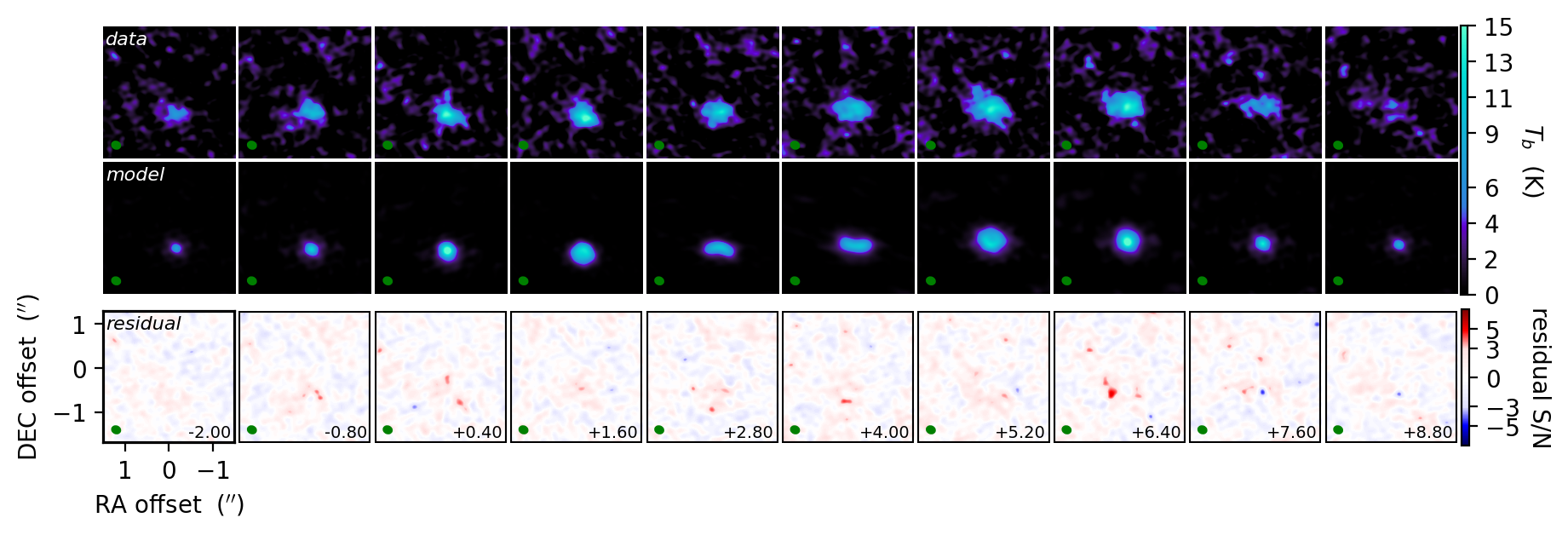}}
    \subfloat[Residual plot of J16145026-2332397.]{\includegraphics[width=0.67\textwidth]{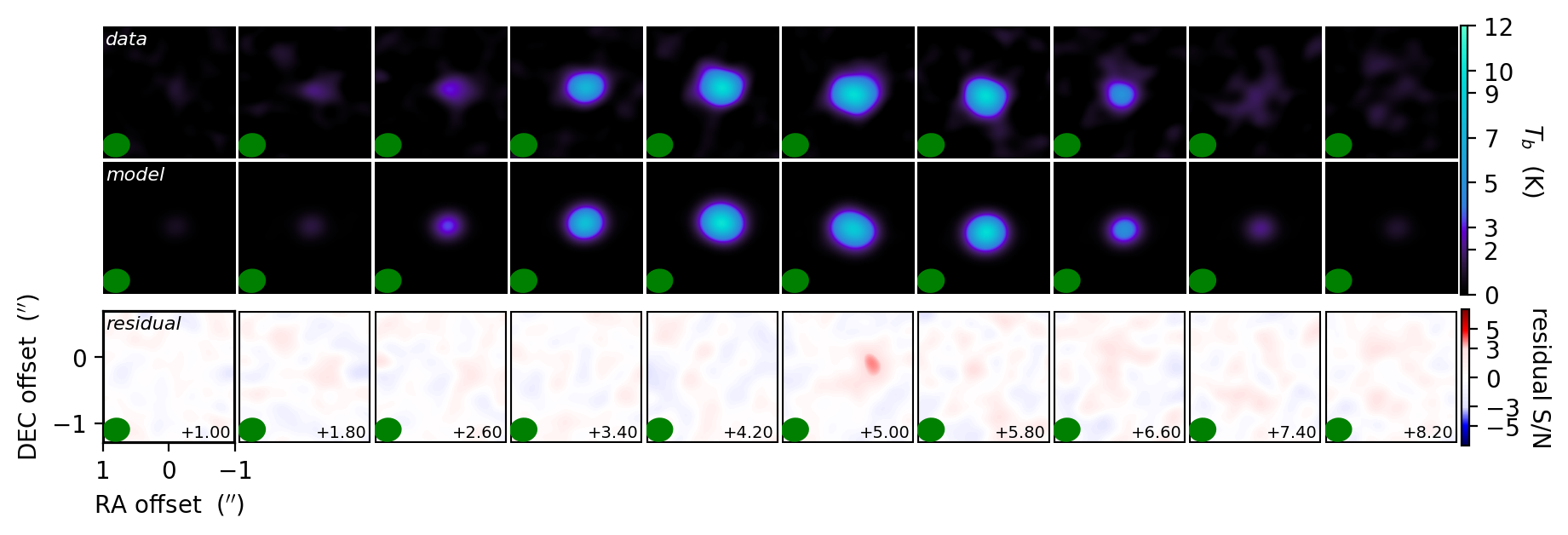}}\\[1ex]
    \subfloat[Residual plot of J16020757-2257467.]{\includegraphics[width=0.67\textwidth]{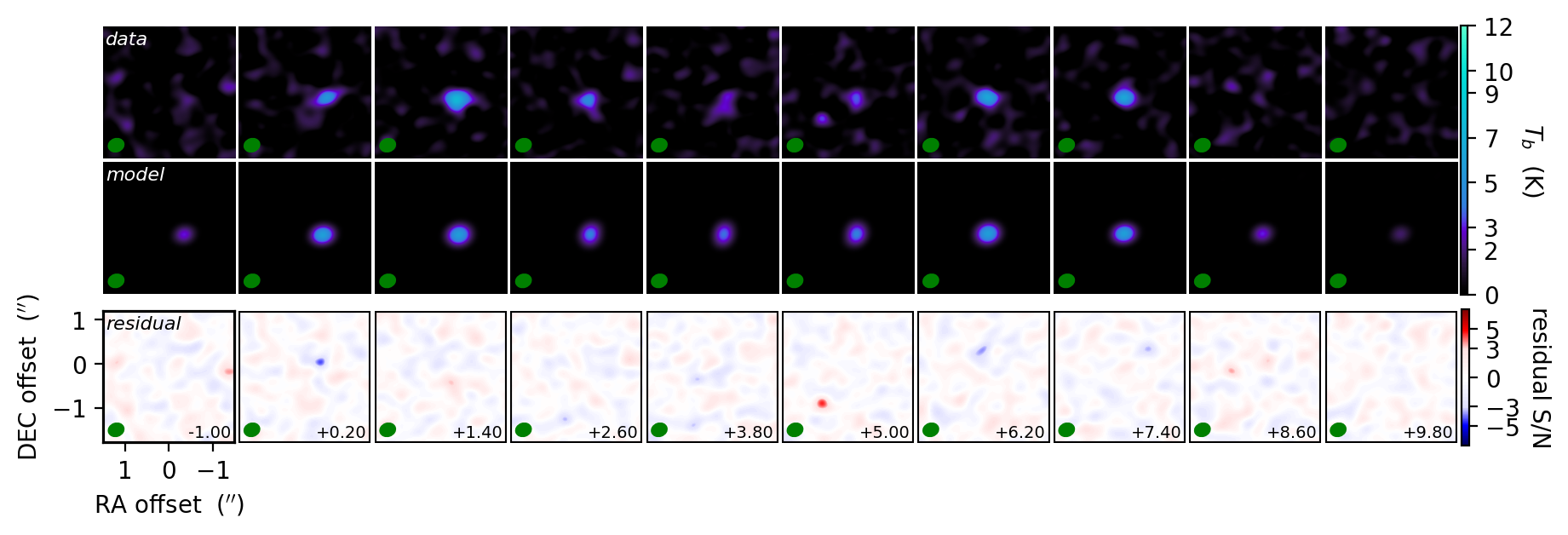}}
    \subfloat[Residual plot of J16121242-1907191.]{\includegraphics[width=0.67\textwidth]{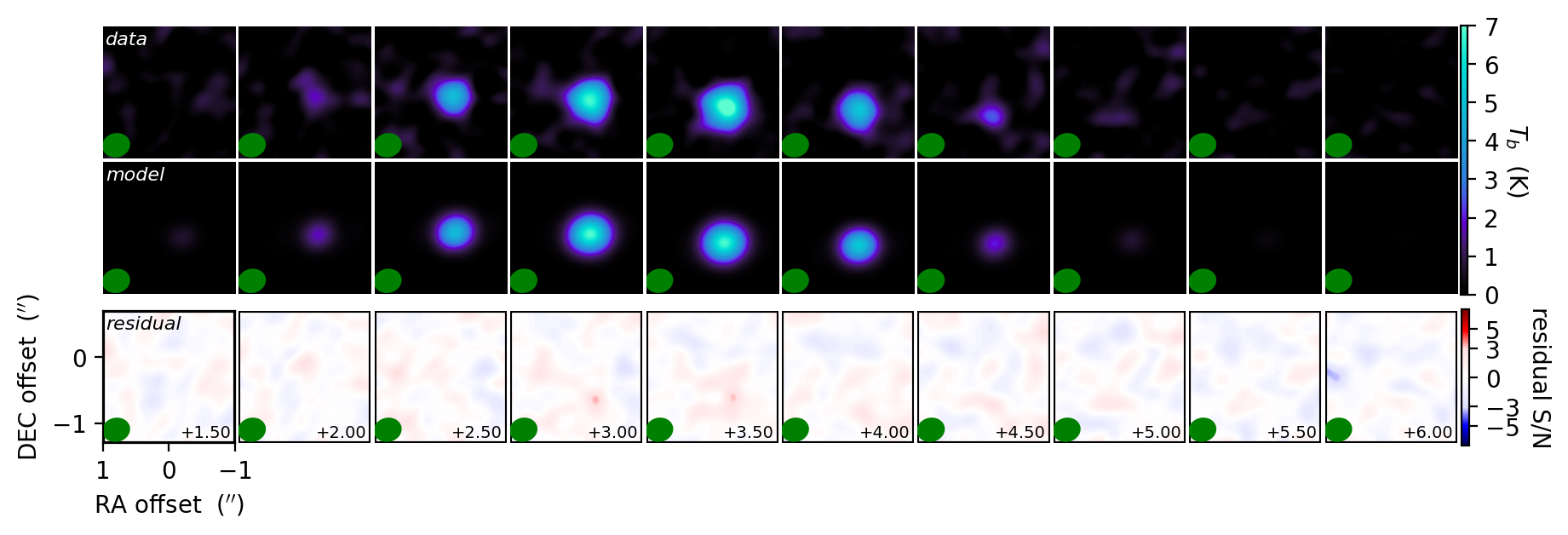}}\\[1ex]
    \caption{Residual plots of J16012268-2408003 (a), J16145026-2332397 (b), J16020757-2257467 (c), J16121242-1907191 (d). The residual plots for all the disks analyzed in this work and their corresponding corner plots can be found on \texttt{GitHub}\protect\hyperref[footnote:github]{\textsuperscript{\ref*{footnote:github}}}.}
    \label{fig:residuals_6}
\end{figure}
\end{landscape}

\end{appendix}

\end{document}